\documentclass[11pt]{article}
\usepackage{geometry}
\usepackage{amsmath}
\usepackage{amssymb}
\usepackage{latexsym}
\usepackage{epsfig}
\usepackage[vcentermath]{youngtab}
\usepackage{slashbox}
\usepackage[usenames,dvipsnames]{pstricks}

\usepackage{amsmath,amssymb,amsbsy}
\usepackage{latexsym,layout,amsfonts,amssymb,fontenc,xcolor,amsthm,multirow,amsfonts,bm,textcomp}
\usepackage{hyperref}
\usepackage{empheq}
\usepackage{accents}

\newsavebox{\uuunit}
\sbox{\uuunit}
    {\setlength{\unitlength}{0.825em}
     \begin{picture}(0.6,0.7)
        \thinlines
        \put(0,0){\line(1,0){0.5}}
        \put(0.15,0){\line(0,1){0.7}}
        \put(0.35,0){\line(0,1){0.8}}
       \multiput(0.3,0.8)(-0.04,-0.02){12}{\rule{0.5pt}{0.5pt}}
     \end {picture}}

\def\2{\frac12}
\def\4{\frac14}

\def\equationautorefname~#1\null{eq.~(#1)\null
}

\hypersetup{
    bookmarks=false,         
    unicode=false,          
    pdftoolbar=true,        
    pdfmenubar=true,        
    pdffitwindow=false, 
    pdfstartview={FitH},    
    pdftitle={P fluxes and exotic branes},    
    pdfauthor={Davide M. Lombardo, Fabio Riccioni, Stefano Risoli},     
    pdfsubject={},   
    pdfnewwindow=true,      
    colorlinks=false,       
    linkcolor=blue,          
    citecolor=blue,        
    filecolor=blue,      
    urlcolor=blue           
    linkbordercolor={blue},
    citebordercolor={purple},
    urlbordercolor={gray}
}

\begin{document}

\begin{titlepage}
\begin{center}

\hfill UUITP-08/18\\

\vskip 1.2cm

{\Large \bf
Space-filling branes \& gaugings
}

\vskip 1.1cm

{\bf  Giuseppe Dibitetto\,$^1$,  Fabio Riccioni\,$^2$  and Stefano Risoli\,$^{2,3}$}

\vskip 30pt

{\em $^1$ \hskip -.1truecm
Department of Physics and Astronomy, Uppsala University,\\ Box 516, SE-751 20 Uppsala, Sweden
 \vskip 25pt }


{\em $^2$ \hskip -.1truecm
 INFN Sezione di Roma,   Dipartimento di Fisica, Universit\`a di Roma ``La Sapienza'',\\ Piazzale Aldo Moro 2, 00185 Roma, Italy
 \vskip 25pt }


{\em $^3$ \hskip -.1truecm  Dipartimento di Fisica, Universit\`a di Roma ``La Sapienza'',\\ Piazzale Aldo Moro 2, 00185 Roma, Italy
  }

\vskip 0.8cm

{email addresses: {\tt giuseppe.dibitetto@physics.uu.se},  {\tt fabio.riccioni@roma1.infn.it}, {\tt stefano.risoli@roma1.infn.it}  \vskip 25pt}

\vskip .4cm

\end{center}

\vskip 0.4cm

\begin{center} {\bf ABSTRACT}\\[3ex]
\end{center}

We consider in any dimension the supersymmetric $\mathbb{Z}_2$ truncations of the maximal supergravity theories.  In each dimension and for each truncation we determine all the sets of 1/2-BPS space-filling branes that preserve the supersymmetry of the truncated theory and the representations of the symmetry of such theory to which they belong. We show that in any dimension below eight these sets always contain exotic branes, that are objects that do not have a ten-dimensional origin. We repeat the same analysis for half-maximal theories and for the quarter-maximal theories in four and three dimensions. We then discuss all the possible gaugings of these theories as described in terms of the embedding tensor. In general, the truncation acts on the quadratic constraints of the embedding tensor in such a way that some representations survive the truncation although they are not required by the supersymmetry of the truncated theory. We show that for any theory, among these representations, the highest-dimensional ones are precisely those of the 1/2-BPS space-filling branes that preserve the same supersymmetry of the truncated theory, and we interpret this result as the fact that these quadratic constraints after the truncation become tadpole conditions for such branes.

\end{titlepage}

\newpage
\setcounter{page}{1} \tableofcontents


\setcounter{page}{1} \numberwithin{equation}{section}

\section{Introduction}

It is well known that the ${\rm SO(32)}$ type-I string theory in ten
dimensions is obtained from the type-IIB  theory by performing the
orientifold projection \cite{orientifolds}. In the closed sector,
the projection is due to the O9-plane, while the open sector arises due to
the presence of D9-branes \cite{Polchinski:1995mt}, and RR and
NSNS tadpole cancellations correspond to the fact that the charge
and tension of the O9-plane are cancelled by those of the D9-branes.
In the low-energy theory, the projection in the closed sector acts
as a $\mathbb{Z}_2$ truncation to ${\cal N}=1$ supergravity, in
which the spinors are halved and,  among the gauge potentials, the
NSNS 2-form $B_2$  and the RR 4-form $C_4$ are projected out, while
the RR 2-form $C_2$ survives.

From the point of view of supergravity, there is another consistent
supersymmetric  $\mathbb{Z}_2$ truncation,  in which all RR fields
are projected out, leading to the gravity sector of the heterotic
theory. The two truncations are related by S-duality. Denoting with
$\psi_\mu $ the gravitino of the IIB theory, which is a doublet of
Majorana-Weyl spinors of the same chirality, using the conventions
of \cite{Bergshoeff:1999bx} the gravitino-dependent part of the
supersymmetry transformations of $B_2$, $C_2$ and $C_4$ can be
schematically written in the string frame as
\begin{eqnarray}
& & \delta C_{\mu\nu}   = i e^{-\phi} \bar{\epsilon} \gamma_{[\mu}
\sigma_1 \psi_{\nu ]} +...\nonumber \\
& & \delta B_{\mu \nu} = i \bar{\epsilon} \gamma_{[\mu}
\sigma_3\psi_{\nu]}+... \\
& & \delta C_{\mu\nu\rho\sigma} =  e^{-\phi} \bar{\epsilon}
\gamma_{[\mu\nu\rho} \sigma_2 \psi_{\sigma ]}+... \quad , \nonumber
\end{eqnarray}
where the Pauli matrices act on the doublets of spinors. The O9
truncation is then  realised in the spinor sector as the projection
\begin{equation}
{\rm O9}: \Psi = \pm \sigma_1 \Psi
\end{equation}
while the S-dual truncation, which we label SO9, acts as
\begin{equation}
{\rm SO9}: \Psi = \pm  \sigma_3 \Psi \quad ,
\end{equation}
 where with $\Psi$ we denote any
spinor in the theory \cite{Bergshoeff:1999bx}.

In the low-energy theory, the occurrence of D9-branes is signalled
by the fact that one can consistently introduce a RR 10-form in the
supersymmetry algebra, whose transformation contains the Pauli
matrix $\sigma_1$ consistently with the fact that the field survives
the O9 truncation \cite{Bergshoeff:1999bx}. Analogously, one can
consider the S-dual of the RR 10-form potential, and write an
effective action for the 1/2-BPS brane that is charged under it
\cite{Bergshoeff:2006ic}. The tension of such brane scales like
$g_s^{-4}$ \cite{Hull:1997kt,Hull:1998he}, and  the presence of the Pauli matrix $\sigma_3$ 
in the
supersymmetry variation of the potential  signals that
it survives the SO9 truncation.\footnote{In \cite{Hull:1998he} it
was conjectured that the ${\rm SO(32)}$ heterotic theory can be
obtained from type-IIB by performing  the S-dual of the orientifold
projection, and the charge and tension of the S-dual of the O9-plane
are cancelled by these branes, that are the  S-duals of the
D9-branes and are defined as end-points of D-strings. We will not
discuss this issue in this paper.} The presence of two space-filling
1/2-BPS branes, each of the two surviving each of the two
truncations, is also signalled by the presence of the doublet of
central charges $Z_{\mu}^a$, $a=1,2$, in the supersymmetry algebra.
Indeed, if $\mu$ is along the time direction, this can be dualised
to $Z_{i_1 ...i_9}^a$, where the $i$'s are space indices, which is
a doublet of 9-brane central charges \cite{Hull:1997kt}. For each
$\mathbb{Z}_2$ truncation, the supersymmetry preserved by the
1/2-BPS 9-brane that survives the projection is exactly the
supersymmetry of the truncated theory.

The 10-forms that couple to the 9-branes in IIB belong to a
quadruplet ({\it i.e.}  a spin-3/2 representation) of the global
symmetry ${\rm SL(2,\mathbb{R})}$ of IIB supergravity
\cite{Bergshoeff:2005ac}, and are more precisely the spin $3/2$ and
$-3/2$ components ({\it i.e.} the longest weights) of that
representation \cite{Bergshoeff:2006ic,Bergshoeff:2006gs}. The same
applies  to maximal theories in lower dimensions: in any dimension
$D$ one can determine the representation of the global symmetry
group $G$ to which the RR $D$-form potentials belong
\cite{Riccioni:2007au,Bergshoeff:2007qi}, and the space-filling
1/2-BPS branes turn out to correspond to the long weights of that
representation \cite{Bergshoeff:2013sxa}. The analysis of
\cite{Riccioni:2007au,Bergshoeff:2007qi} was performed by suitably
decomposing the very-extended Kac-Moody algebra ${\rm E_{11}}$
\cite{West:2001as}, and we will especially make use of the results
of \cite{Riccioni:2007au}, where the representations of the
potentials in the lower-dimensional theories were shown to arise
from the dimensional reduction of both standard potentials and
mixed-symmetry potentials in ten dimensions, that follow from the
decomposition of the ${\rm E_{11}}$ algebra
\cite{Kleinschmidt:2003mf}.

A crucial result that applies to all the maximal theories in
dimension less than ten is the fact that the 1/2-BPS condition for
space-filling branes is degenerate, which means that different
branes can preserve the same supersymmetry
\cite{Bergshoeff:2013sxa}.\footnote{The same applies to 1/2-BPS
defect branes \cite{Bergshoeff:2011se} and domain walls
\cite{Bergshoeff:2012pm} of the maximal theories.} This degeneracy
was determined in \cite{Bergshoeff:2013sxa} by simply observing that
the number of 1/2-BPS space-filling branes, that are the long
weights of the representation of the $D$-forms, is always a multiple
of the dimension of the R-symmetry group of the vector central
charge $Z_\mu$.  As in the IIB theory, we can associate to each
space-filling brane a $\mathbb{Z}_2$ truncation to the
half-supersymmetric theory, and given that the degenerate branes all
preserve the supersymmetry of the same truncation, we arrive at the
obvious conclusion that the number of different supersymmetric
$\mathbb{Z}_2$ truncations is precisely the dimension of the
representation of the central charge $Z_\mu$. The first result of
this paper will be to identify these truncations, and for each
truncation to identify the branes that are not projected out, {\it
i.e.} the branes that preserve the same supersymmetry of the
truncated theory. We give in Table \ref{numberofbranesmax} the
number of branes and the corresponding degeneracy in any dimension,
as well as the dimension of the vector central charge, which gives
the number of different supersymmetric $\mathbb{Z}_2$ truncations.

\begin{table}[t!]
\begin{center}
\begin{tabular}{|c||c|c|c||c|c||c|}
\hline \rule[-1mm]{0mm}{6mm} $D$ & $G$  & repr. & branes & R-symmetry & $Z_\mu$ & deg.\\
\hline
\hline \rule[-1mm]{0mm}{6mm} 8 & ${\rm SL(3,\mathbb{R})\times SL(2,\mathbb{R})}$ & $({\bf 15,1})$ & 6 & ${\rm U(2)}$ & ${\bf 3}$ & 2 \\
\hline \rule[-1mm]{0mm}{6mm} 7 & ${\rm SL(5,\mathbb{R})}$ & ${\bf 70}$ & 20 & ${\rm USp(4)}$ & ${\bf 5}$ & 4 \\
\hline \rule[-1mm]{0mm}{6mm} \multirow{2}{*}{6} & \multirow{2}{*}{${\rm SO(5,5)}$}
& ${\bf 320}$ & 80 & \multirow{2}{*}{${\rm USp(4)\times USp(4)}$ }& $({\bf 5,1})+({\bf 1,5})$ & 8 \\
& & ${\bf \overline{126}}$ & 16 & & $({\bf 1,1})$ & 16\\
\hline \rule[-1mm]{0mm}{6mm} 5 & ${\rm E_{6(6)}}$ & ${\bf \overline{1728}}$ & 432 & ${\rm USp(8)}$ & ${\bf 27}$ & 16 \\
\hline \rule[-1mm]{0mm}{6mm} 4 & ${\rm E_{7(7)}}$ & ${\bf {8645}}$ & 2016 & ${\rm SU(8)}$ & ${\bf 63}$ & 32 \\
\hline \rule[-1mm]{0mm}{6mm} 3 & ${\rm E_{8(8)}}$ & ${\bf {147250}}$ & 17280 & ${\rm SO(16)}$ & ${\bf 135}$ & 128 \\
\hline

\end{tabular}
\end{center}
  \caption{\sl The 1/2-BPS space-filling branes of the maximal theories in any dimension
  and their degeneracy \cite{Bergshoeff:2013sxa}. The number of branes is given in the fourth column,
  while the third column contains the representation of the corresponding $D$-form potential.
  The sixth column contains the representation of the central charge and in the last column we list
  the degeneracy, which is simply the ratio between the number of branes and the dimension of the
  representation of the central charge. In six dimensions the first line corresponds to branes supporting
  a vector multiplet, and the second line to branes supporting a tensor multiplet.
  \label{numberofbranesmax}}
\end{table}

We will start considering explicitly the  eight-dimensional
case.\footnote{In $D=9$ the dimension of the central charge and the
degeneracy of the space-filling branes are identical to the IIB
case.} We will determine the supersymmetry transformations of all
the fields in a manifestly ${\rm SL}(3,\mathbb{R}) \times {\rm
SL}(2,\mathbb{R}) $-covariant notation,\footnote{As far as we know,
this result was not available in the literature.} and we will use
this to show that there are three different $\mathbb{Z}_2$
truncations to minimal supergravity coupled to two vector
multiplets.  The R-symmetry of the theory is ${\rm SO(3) \times
SO(2)}$, and we will therefore introduce two sets of Pauli matrices:
the matrices $\sigma_i$, $i=1,2,3$, generate the ${\rm SO(3)}$
Clifford algebra, while $\tau_a$, $a=1,2$, which numerically are
equal to the first two Pauli matrices, generate the ${\rm SO(2)}$
Clifford algebra, and $\tau_3$, which is the third Pauli matrix, is
the ${\rm SO(2)}$ chirality matrix. There are three 2-forms in the
theory, coming from $B_2$, $C_2$  and the compactified $C_4$ in IIB,
and we will find that the gravitino-dependent part of their
supersymmetry transformations can be schematically written in the
string frame as
\begin{eqnarray}
& & \delta C_{\mu\nu}   = i e^{-\phi} \bar{\epsilon} \gamma_{[\mu}
\sigma_1 \tau_3 \psi_{\nu ]} +...\nonumber \\
& & \delta B_{\mu \nu} = i \bar{\epsilon} \gamma_{[\mu}
\sigma_3 \tau_3 \psi_{\nu]}+... \label{2formstransfintro}  \\
& & \delta C_{\mu\nu \, x^1 x^2} = i e^{-\phi} \bar{\epsilon}
\gamma_{[\mu} \sigma_2 \tau_3 \psi_{\nu ]}+... \quad , \nonumber
\end{eqnarray}
where $x^i$ ($i=1,2$) are the two compact directions. It is easy to identify the three $\mathbb{Z}_2$ truncations as
\begin{eqnarray}
& & {\rm O9}: \ \ \Psi = \pm \sigma_1 \tau_3 \Psi \nonumber \\
& & {\rm SO9}: \, \Psi = \pm \sigma_3 \tau_3  \Psi \label{truncationsD=8}\\
& & {\rm O7}: \ \ \Psi = \pm \sigma_2 \tau_3 \Psi \nonumber \quad ,
\end{eqnarray}
and only one 2-form survives each truncation. We will  then write
down the variation of the 8-forms in the $({\bf 15,1})$, and show
that for each truncation in eq. \eqref{truncationsD=8} there are two
space-filling branes that survive.

The fact that there are two space-filling branes preserved by each
truncation is not surprising if one considers in particular the O7
truncation. Indeed, we know that the D7-brane and its S-dual
preserve the same supersymmetry
\cite{Greene:1989ya,Bergshoeff:2006jj}. Performing T-dualities in
$x^1$ and $x^2$, the D7-brane is mapped to the D9-brane, while the
S-dual of the D7-brane is mapped to an {\it exotic} space-filling
brane, {\it i.e.} a brane charged with respect to an 8-form
potential whose IIB origin is a mixed-symmetry potential.  These branes survive the O9 truncation in
eight dimensions. Similarly, by S-duality one obtains the branes
that survive the SO9 truncation. All these arguments can then be
repeated in all lower dimensions, and by multiple T and S-duality
transformations one obtains all the different truncations and all
the branes that preserve the same supersymmetry of each truncation.
Most of these branes are exotic, and we identify them with the
corresponding components of the mixed-symmetry potential using the
universal T-duality rules derived in \cite{Lombardo:2016swq}.

The $\mathbb{Z}_2$ truncation of the maximal theory in $D$
dimensions gives half-maximal supergravity coupled to $d=10-D$
vector multiplets. This theory has global symmetry
$\mathbb{R}^+\times {\rm SO}(d,d)$ in dimension higher than four,
${\rm SL(2,\mathbb{R})}\times {\rm SO(6,6)}$ in four dimensions and
${\rm SO(8,8)}$ in three dimensions, and we identify in all
dimensions the irreducible representations of these groups that
contain the branes preserving the same supersymmetry of the
truncated theory. The particular truncation such that
${\mathbb{R}}^+$ is identified with the string dilaton scaling (and
therefore ${\rm SO}(d,d)$ is T-duality) is always the SO9
truncation.

Starting from six dimensions, apart from the branes that preserve
the same supersymmetry of the truncation, there are additional
space-filling branes surviving the truncation which are 1/2-BPS
states of the truncated theory. As in the maximal case, one can
determine the vector central charge $Z_\mu$ as a representation of
the R-symmetry of the half-maximal theory, and relate it to the
number of space-filling branes to determine their degeneracy
\cite{Bergshoeff:2012jb}. We list in Table
\ref{numberofbraneshalfmax} the number of 1/2-BPS space-filling
branes, the central charge and the degeneracy for the truncated
theories. In the table we denote with 6A the ${\cal N}=(1,1)$ theory
and with 6B the ${\cal N}=(2,0)$ theory, and the latter case
corresponds to IIB compactified on $T^4/\mathbb{Z}_2$, so that the
truncation is geometric. Exactly as in the maximal theory, the
number of vector central charges gives the number of $\mathbb{Z}_2$
truncations to quarter-maximal theories, and the degeneracy gives
the number of space-filling branes that preserve the same
supersymmetry of the truncation. We will be able to show that in all
cases the branes that preserve the same supersymmetry of  a given
truncation of the half-maximal theory are the union of two different
sets of degenerate branes of the maximal theory.

\begin{table}[t!]
\begin{center}
\begin{tabular}{|c||c|c|c||c|c||c|}
\hline \rule[-1mm]{0mm}{6mm} $D$  & $G$  & repr. & branes & R-symmetry & $Z_\mu$ & deg.\\
\hline
\hline \rule[-1mm]{0mm}{6mm} \multirow{3}{*}{6A} & \multirow{3}{*}{$\mathbb{R}^+ \times {\rm SO(4,4)}$}
& ${\bf 35}_{\rm V}$ &  8 & \multirow{3}{*}{${\rm USp(2)\times USp(2)}$} & \multirow{3}{*}{$({\bf 1,1})$} & 8 \\
  &  & ${\bf 35}_{\rm S} $ &  8 &  &  & 8 \\
  &  & ${\bf 35}_{\rm C}$ &  8 &  &  & 8 \\
\hline \rule[-1mm]{0mm}{6mm} 6B & ${\rm SO(5,5)}$ & ${\bf 320}$ &  80 & ${\rm USp(4)}$ & ${\bf 5}$ & 16 \\
\hline \rule[-1mm]{0mm}{6mm} \multirow{2}{*}{5} & \multirow{2}{*}{$\mathbb{R}^+ \times {\rm SO(5,5)}$}
& ${\bf 320}$ & 80 & \multirow{2}{*}{${\rm USp(4)}$} & \multirow{2}{*}{${\bf 5}$} & 16 \\
& & ${\bf 210}$ & 80 & &  & 16 \\
\hline \rule[-1mm]{0mm}{6mm} \multirow{2}{*}{4}  & \multirow{2}{*}{${\rm SL(2,\mathbb{R})}
\times {\rm SO(6,6)}$} & $({\bf 3,495}) $ &480 & \multirow{2}{*}{${\rm U(4)}$} & \multirow{2}{*}{${\bf 15}$} & 32 \\
 &  & $ ({\bf 1,2079})$ &480 &  & & 32 \\
\hline \rule[-1mm]{0mm}{6mm} 3 & ${\rm SO(8,8)}$ & ${\bf 60060}$ & 8960 & ${\rm SO(8)}$ & ${\bf 35}$ & 256 \\
\hline

\end{tabular}
\end{center}
  \caption{\sl The number and the degeneracy of the 1/2-BPS space-filling branes of the
  half-maximal theories which arise as  $\mathbb{Z}_2$ truncations of the maximal ones \cite{Bergshoeff:2012jb}.
  In the 6A theory, 8 of the branes support tensor multiplets and the remaining 16 hypermultiplets.
  In the 6B theory the branes support vector multiplets. In five and four dimensions half of the branes support
  vector multiplets and the other half support hypermultiplets \cite{Bergshoeff:2012jb}.
  \label{numberofbraneshalfmax}}
\end{table}

The analysis can be further extended to consider the $\mathbb{Z}_2$
truncation of the quarter-maximal theories. Indeed, starting from
four dimensions, apart from the branes of the half-maximal theories
that preserve the same supersymmetry of the truncation, there are
additional space-filling branes surviving the truncation which are
1/2-BPS states of the truncated quarter-maximal theory. In
\cite{Bergshoeff:2014lxa} the number of space-filling branes of the
quarter-maximal theories in four and three dimensions was determined
and then compared to the number of vector central charges to obtain
the degeneracy. We list the results in Table
\ref{numberofbranesquartermax}. Again, the number of vector central
charges gives the number of $\mathbb{Z}_2$ truncations to theories
with four supercharges, {\it i.e.} ${\cal N}=1$ in four dimensions,
and the degeneracy gives the number of space-filling branes that
preserve the same supersymmetry of the truncation. We will  show
that the branes that preserve the same supersymmetry of a given
truncation of the quarter-maximal theory are the union of four
different sets of degenerate branes of the maximal theory.

\begin{table}[t!]
\begin{center}
\begin{tabular}{|c||c|c|c||c|c||c|}
\hline \rule[-1mm]{0mm}{6mm} $D$  & $G$  & repr. & branes & R-symmetry & $Z_\mu$ & deg.\\
\hline
\hline \rule[-1mm]{0mm}{6mm} \multirow{4}{*}{4} & \multirow{4}{*}{${\rm SL(2,\mathbb{R})}^3 \times {\rm SO(4,4)}$} & $({\bf 1,1,1,350})$ & 96 & \multirow{4}{*}{${\rm U(2)}$} & \multirow{4}{*}{${\bf 3}$} & 32 \\
 &  & $({\bf 1,3,3,28})$ & 96 &  &  & 32 \\
 &  & $({\bf 3,1,3,28})$ & 96 &  &  & 32 \\
 &  & $({\bf 3,3,1,28})$ & 96 &  &  & 32 \\
\hline \rule[-1mm]{0mm}{6mm} \multirow{2}{*}{3} & \multirow{2}{*}{${\rm SO(4,4)} \times {\rm SO(4,4)}$} & $({\bf 28,350})$ & 2304 &\multirow{2}{*}{${\rm SU(2) \times SU(2)}$} & \multirow{2}{*}{${\bf (3,3)}$} & 256 \\
&  & $({\bf 350,28})$ & 2304 & & & 256 \\
\hline

\end{tabular}
\end{center}
  \caption{\sl The number and the degeneracy of 1/2-BPS space-filling branes of the quarter-maximal theories
  resulting from $\mathbb{Z}_2$ truncations \cite{Bergshoeff:2014lxa} (see also tables 7 and 8 of \cite{Pradisi:2014fqa}).
  \label{numberofbranesquartermax}}
\end{table}

The truncation of the maximal theory to the half-maximal one can
also be performed in the presence of gaugings. In particular, the
truncation of ${\cal N}=8$ gauged supergravity to ${\cal N}=4$
gauged supergravity coupled to six vector multiplets was studied in
\cite{Dibitetto:2011eu} using the embedding tensor formalism
\cite{Nicolai:2000sc,Nicolai:2001ac,deWit:2002vt,deWit:2005ub}. Decomposing the
embedding tensor of the maximal theory \cite{deWit:2007kvg} under
${\rm SL(2,\mathbb{R})}\times {\rm SO(6,6)}$ and projecting out  the
representations that are odd under ${\mathbb{Z}}_2$, one is left
with the embedding tensor of the half-maximal theory
\cite{Schon:2006kz}. On the other hand, by projecting out the
representations of the quadratic constraints that are odd under
${\mathbb{Z}}_2$, one is left with more than the quadratic
constraints of the half-maximal theory. Among the representations of
the quadratic constraints that survive the ${\mathbb{Z}}_2$
truncation but are not required by supersymmetry, the
highest-dimensional one contains space-filling branes that preserve
the same supersymmetry of the ${\mathbb{Z}}_2$ truncation. The fact
that this quadratic constraint is not required in  ${\cal N}=4$
although it is not projected out has therefore the natural
interpretation that it becomes a tadpole condition for the
corresponding brane \cite{Dibitetto:2011eu}.

Using the results of the first part of this paper, we will
generalise this to any maximal theory. All the space-filling branes
that preserve the same supersymmetry of the ${\mathbb{Z}}_2$
truncation belong to the representation of the symmetry of the
half-maximal theory which is the highest-dimensional representation
of the quadratic constraint which survives the  truncation but is
not required by the supersymmetry of  the truncated theory.  We will
also show that exactly the same applies for the truncation from the
half-maximal to the quarter-maximal theories, using the quadratic
constraints of the embedding tensor of ${\cal N}=2$ theories
discussed in \cite{deWit:2005ub}. The truncation of the
four-dimensional ${\cal N}=2$ theory whose symmetry appears in Table
\ref{numberofbranesquartermax} gives the ${\cal N}=1$ theory with
${\rm SL}(2,\mathbb{R})^7$ global symmetry. Minimal supersymmetry
does not require any quadratic constraint for the embedding tensor,
and consistently we find that all the highest-dimensional
representations of the quadratic constraints of the ${\cal N}=2$
theory that survive the ${\mathbb{Z}}_2$ truncation coincide with
the representations of the space-filling branes which preserve the
same supersymmetry of the truncation. To obtain this result, we will
use the analysis of \cite{Lombardo:2017yme}, where the space-filling
branes of the ${\rm SL}(2,\mathbb{R})^7$ ${\cal N}=1$ model that
arises from the IIB O3/O7 $T^6/(\mathbb{Z}_2 \times \mathbb{Z}_2)$
orientifold  were derived.

Finally, we will discuss the truncations of the gauged theories with
lower supersymmetry from the point of view of the maximal theories.
Considering again the IIB O3/O7 $T^6/(\mathbb{Z}_2 \times
\mathbb{Z}_2)$ orientifold, the embedding tensor of the
four-dimensional theory arises from geometric and non-geometric IIB
fluxes. These fluxes satisfy Bianchi identities, and we will show
that these Bianchi identities are in the same representations as 
the space-filling branes that preserve the same supersymmetry of the
orbifold. Again, the ${\rm SL}(2,\mathbb{R})^7$ analysis
performed in \cite{Lombardo:2017yme} will be crucial to get this
result. The result  also applies to $T^4/\mathbb{Z}_2 \times T^n$
orientifolds.

The plan of the paper is as follows. In section 2 we derive the
supersymmetry transformations of the fields of maximal supergravity
in a manifestly ${\rm SL}(3,\mathbb{R}) \times {\rm
SL}(2,\mathbb{R}) $-covariant notation, and we use this to derive
the three independent $\mathbb{Z}_2$ truncations to the half-maximal
theory coupled to two vector multiplets. We determine the
space-filling branes that for each truncation preserve the same
supersymmetry of the truncated theory. In section 3 we generalise
this result to any dimension and any supersymmetry. In section 4 we
discuss gauged supergravities, and we show that in general the
highest-dimensional representations of the quadratic constraint that
survive the $\mathbb{Z}_2$ truncation but are not required by the
supersymmetry of the truncated theory precisely coincide with the
representations containing the space-filling branes that preserve
the same supersymmetry of the truncation. This is also done for the
truncation of theories with lower supersymmetry. In section 5 we
discuss the particular case of the IIB O3/O7 $T^6/(\mathbb{Z}_2
\times \mathbb{Z}_2)$ orientifold, and we show that the Bianchi
identities are in the representations of the space-filling branes
that preserve the same supersymmetry of the orbifold truncation.
Finally, section 6 contains our conclusions. The paper also contains
an appendix, in which the details of the $D=8$ notations and conventions used in
section 2 are explained.

\section{$D=8$ supergravity and its truncations}

The ${\rm SU}(2)$ gauged maximal $D=8$ supergravity was originally
constructed in \cite{Salam:1984ft} by dimensional reduction from
eleven dimensions on an SU(2) group manifold. This was later
generalised in \cite{Bergshoeff:2003ri} to include more general
gaugings. The supersymmetry transformations in the ungauged case can be 
recovered from these papers, but they are not suitable for our
purposes, because we need them in a formulation which is manifestly
covariant under ${\rm SL}(3,\mathbb{R}) \times {\rm
SL}(2,\mathbb{R})$. In the first subsection we will derive these
transformations imposing the closure of the supersymmetry algebra,
and in particular we will write down the gravitino-dependent part of
the supersymmetry transformation of the 8-form potentials. In the
second subsection we will show that the theory admits three
different $\mathbb{Z}_2$ truncations to the half-maximal theory
coupled to two vector multiplets, and by considering the action of
these projections on the 8-forms we will determine the space-filling
branes that are not projected out in each truncation.

\subsection{Supersymmetry algebra}

We first introduce the notation. We work with a mostly-minus space-time signature, and we denote the curved space-time indices with Greek letters $\mu,\nu,...$, while the tangent-space indices are $\alpha,\beta,\dots$.
We denote with upstairs indices
$M=1,2,3$ and $A=1,2$ the fundamentals of ${\rm SL}(3,\mathbb{R})$
and ${\rm SL}(2,\mathbb{R})$, and with $m=1,2,3$ and $a=1,2$ the
vector indices of their maximal compact subgroups ${\rm SO}(3)$ and
${\rm SO}(2)$.\footnote{Repeated $m$ and $a$ indices, regardless of
whether they are up or down, are meant to be contracted by
$\delta^{mn}$ and $\delta^{ab}$.} The seven scalars in the theory
parametrise the  coset-space  ${\rm SL}(3,\mathbb{R})/{\rm
SO}(3)\otimes {\rm SL}(2,\mathbb{R})/{\rm SO}(2)$. We describe them
introducing the matrices $L_M^m$ and $V_A^a$, together with the
inverse matrices $\tilde{L}_m^M$ and $\tilde{V}_a^A$, satisfying the
identities
\begin{eqnarray}
& & {L}_M^m \tilde{L}^{Nn} \delta_{mn} = \delta_M^N \qquad {L}_M^m \tilde{L}^{Mn} = \delta^{mn}
\qquad L_M^m L_N^n L_P^p \epsilon_{mnp} = \epsilon_{MNP}\nonumber \\
& & {V}_A^a \tilde{V}^{Bb} \delta_{ab} = \delta_A^B \quad \quad \ \ \
{V}_A^a \tilde{V}^{Ab} = \delta^{ab}  \ \ \ \qquad V_A^a V_B^b  \epsilon_{ab} = \epsilon_{AB} \quad . \label{scalaridentities}
\end{eqnarray}
We  define the Maurer-Cartan forms as
\begin{eqnarray}
& & \tilde{L}^M_m \partial_\mu {L}_{Mn} = Q_{\mu \, mn} + P_{\mu \, mn} \nonumber \\
& & \tilde{V}^A_a \partial_\mu {V}_{Ab} = Q_{\mu \, ab} + P_{\mu \, ab} \quad ,\label{MaurerCartanforms}
\end{eqnarray}
where the ${\rm SO}(3)$ and ${\rm SO}(2)$ connections $Q_{\mu \, mn}$ and
$Q_{\mu \, ab}$ are antisymmetric while $P_{\mu \, mn}$ and $P_{\mu
\, ab} $ are symmetric and traceless. The other bosonic fields are
the vielbein $e_{\mu}{}^\alpha$, the 1-form $A_{\mu \, MA}$ in the
${\bf (\overline{3},2)}$, the 2-form $A_{\mu \nu}^M$ in the ${\bf
({3},1)}$ and the 3-forms $A_{\mu \nu \rho}^A$ in the ${\bf
({1},2)}$. The field-strengths of the 3-forms satisfy a self-duality
condition.

We now move to discuss the fermionic  sector. The eight-dimensional
chirality matrix $\gamma_9$ is defined in terms of the gamma
matrices $\gamma_\mu$ as
\begin{equation}
\gamma_9 = - \tfrac{i}{8!} \epsilon_{\mu_1 ...\mu_8} \gamma^{\mu_1 ...\mu_8} \quad .\label{definitiongammanine}
\end{equation}
We also introduce the Pauli matrices $\sigma_m$ which act on ${\rm SO}(3)$
spinor indices. Similarly, we introduce the matrices $\tau_a$ acting
on the spinor indices of ${\rm SO}(2)$. Numerically, $\tau_1$ and $\tau_2$
coincide with $\sigma_1$ and $\sigma_2$. We will also need the
${\rm SO}(2)$ chirality matrix $\tau_3$, which coincides numerically with
$\sigma_3$. The eight-dimensional fermions  are the
gravitino $\psi_\mu$ and the spinors $\chi_m$ and $\chi_a$, while we denote with $\epsilon$ the supersymmetry parameter. They all have also
spinor
indices of ${\rm SO}(3) \times {\rm SO}(2)$, and satisfy a chirality condition
with respect to $\gamma_9 \tau_3$. In particular
\begin{equation}
\gamma_9 \tau_3 \psi_\mu= \psi_\mu \qquad \gamma_9 \tau_3 \chi_m= -\chi_m
\qquad \gamma_9 \tau_3 \chi_a= \chi_a \qquad\gamma_9 \tau_3 \epsilon= \epsilon \quad ,\label{sortofchirality}
\end{equation}
and thus $\chi_m$ has opposite `chirality' with respect to all the
other fermions.
All the fermions also 
satisfy the `symplectic' Majorana condition
\begin{equation}
\Psi = C \overline{\Psi}^T \quad , \label{symplMajorana}
\end{equation}
where $C$ is defined as
\begin{equation}
C = C_8 \sigma_2 \tau_1 \label{definitionofC}
\end{equation}
and the eight-dimensional Majorana matrix $C_8$ is symmetric and satisfies
\begin{equation}
C_8^\dagger \gamma_\mu C_8 = - \gamma_\mu^T \quad .\label{CdaggergammaCD=8}
\end{equation}
It can be shown that the symplectic Majorana condition of eq. \eqref{symplMajorana} is compatible with the chirality conditions defined in eq. \eqref{sortofchirality}.
Finally, the spinors $\chi_m$ and $\chi_a$ also satisfy the
irreducibility conditions
\begin{equation}
\sigma_m \chi_m = \tau_a \chi_a =0 \label{constraintsonfermionsD=8}
\quad .
\end{equation}
The number of on-shell degrees of freedom that
these fermions propagate match those of the bosons. We discuss in more detail the fermionic sector in Appendix A, where  we also
derive the properties of the various bilinears under Majorana flip.

The way we proceed  to derive the supersymmetry transformations of
the fields is by imposing that the supersymmetry algebra closes. We
first write down the final outcome of our analysis, and then we
discuss in more detail how the algebra closes on the various fields.
The supersymmetry transformations of the fermionic fields are
\begin{eqnarray}
& & \delta \psi_\mu = D_\mu \epsilon - \tfrac{1}{48} F^{\nu \rho}_{MA}
\tilde{L}^M_m \tilde{V}^A_a \gamma_{\mu \nu \rho} \sigma_m \tau_a \epsilon + \tfrac{5}{24}
F_{\mu \nu \, MA} \tilde{L}^M_m \tilde{V}^A_a \gamma^\nu \sigma_m \tau_a \epsilon\nonumber \\
&& \qquad -\tfrac{1}{36} F^{\nu\rho \sigma \, M} {L}_{Mm} \gamma_{\mu\nu\rho\sigma} \sigma_m \tau_3 \epsilon + \tfrac{1}{6}
F_{\mu \nu\rho}^M {L}_{Mm} \gamma^{\nu\rho} \sigma_m \tau_3 \epsilon \nonumber \\
& & \qquad -\tfrac{i}{16} F_{\mu\nu\rho\sigma}^A {V}_{Aa} \gamma^{\nu\rho\sigma} \tau_a \epsilon\nonumber \\
& & \delta \chi_m = -\tfrac{i}{2} P_{\mu \, mn} \gamma^\mu \sigma_n \tau_3 \epsilon + \tfrac{i}{12}
F_{\mu\nu\, MA} \tilde{L}^M_m \tilde{V}^A_a \gamma^{\mu \nu } \tau_a \tau_3 \epsilon + \tfrac{1}{24}
F_{\mu\nu \, MA}  \tilde{L}^M_n \tilde{V}^A_a \epsilon_{mnp}  \gamma^{\mu \nu } \sigma_p \tau_a \tau_3 \epsilon \nonumber\\
& & \qquad + \tfrac{i}{18} F_{\mu\nu\rho}^M {L}_{Mm} \gamma^{\mu\nu\rho} \sigma_m  \epsilon +
\tfrac{1}{36} F_{\mu\nu\rho}^M {L}_{Mn} \epsilon_{mnp} \gamma^{\mu\nu\rho} \sigma_p  \epsilon\nonumber \\
& &\delta \chi_a = -\tfrac{i}{2} P_{\mu \, ab} \gamma^\mu \tau_b \epsilon - \tfrac{i}{16} F_{\mu\nu\, MA}
\tilde{L}^M_m \tilde{V}^A_a \gamma^{\mu\nu} \sigma_m \epsilon
- \tfrac{1}{16} F_{\mu\nu \, MA}  \tilde{L}^M_m \tilde{V}^A_b\epsilon_{ab} \gamma^{\mu\nu} \sigma_m \tau_3 \epsilon \nonumber \\
& & \qquad + \tfrac{1}{64} F_{\mu\nu\rho\sigma}^A {V}_{Aa} \gamma^{\mu\nu\rho\sigma} \epsilon \quad .
\end{eqnarray}
In the transformation of the gravitino, the derivative $D_\mu$ is covariant with respect to local Lorentz, local ${\rm SO}(3)$ and local ${\rm SO}(2)$, that is
\begin{equation}
D_\mu \epsilon = \partial_\mu \epsilon + \tfrac{1}{4} \omega_{\mu \alpha\beta} \gamma^{\alpha \beta}
\epsilon+ \tfrac{i}{4} Q_{\mu \, mn} \epsilon_{mnp} \sigma_{p} \epsilon + \tfrac{i}{4} Q_{\mu \, ab} \epsilon_{ab} \tau_3 \epsilon \quad .
\end{equation}
The field-strengths $F_{\mu\nu\, MA}$, $F_{\mu\nu\rho}^M$ and
$F_{\mu\nu\rho\sigma}^A$ are defined as
\begin{eqnarray}
& & F_{\mu\nu\, MA} = 2 \partial_{[\mu} A_{\nu ] \, MA} \nonumber \\
& & F_{\mu\nu\rho}^M = 3 \partial_{[\mu } A_{\nu\rho] }^M +\tfrac{3}{8} \epsilon^{MNP} \epsilon^{AB} A_{[\mu\, NA } F_{\nu\rho] \, PB} \nonumber \\
& & F_{\mu\nu\rho\sigma}^A = 4 \partial_{[\mu} A_{\nu\rho\sigma]}^A -\tfrac{8}{9}
\epsilon^{AB} A_{[\mu\, MB} F_{\nu\rho\sigma ]}^M -\tfrac{8}{9} \epsilon^{AB} A_{[\mu\nu}^M F_{\rho\sigma]\, MB}
\end{eqnarray}
and they are invariant with respect to the gauge transformations
\begin{eqnarray}
& & \delta A_{\mu\, MA} = \partial_\mu \Lambda_{MA} \nonumber \\
& & \delta A_{\mu\nu}^M = 2 \partial_{[\mu} \Sigma_{\nu]}^M -\tfrac{1}{8}
\epsilon^{MNP} \epsilon^{AB} \Lambda_{NA} F_{\mu\nu \, PB}\nonumber\\
& & \delta A_{\mu\nu\rho}^A = 3  \partial_{[\mu} \Xi_{\nu\rho]}^A +\tfrac{2}{9}
\epsilon^{AB} \Lambda_{MB} F_{\mu\nu \rho}^M + \tfrac{4}{3} \epsilon^{AB} \Sigma_{[\mu}^M F_{\nu\rho]\, MB} \quad .
\end{eqnarray}
The supersymmetry transformations of the bosons are
\begin{eqnarray}
& & \delta e_\mu{}^{\alpha} = -i \bar{\epsilon} \gamma^\alpha \psi_\mu \nonumber \\
& & \delta L_{M m } = L_{Mn} \bar{\epsilon} \sigma_n \tau_3 \chi_m \nonumber \\
& &  \delta V_{A a } = V_{Ab} \bar{\epsilon} \tau_b \chi_a \nonumber \\
& &  \delta A_{\mu \, MA} = L_{Mm} V_{Aa} \left( i \bar{\epsilon} \sigma_m \tau_a
\psi_\mu - \bar{\epsilon} \gamma_\mu \sigma_m \chi_a - \bar{\epsilon} \gamma_\mu \tau_a \tau_3\chi_m \right) \nonumber \\
& & \delta A_{\mu\nu}^M = \tilde{L}^M_m \left(i\bar{\epsilon} \gamma_{[\mu} \sigma_m \tau_3
\psi_{\nu]} + \tfrac{1}{2} \bar{\epsilon} \gamma_{\mu\nu} \chi_m \right)+ \tfrac{1}{4} \epsilon^{MNP} \epsilon^{AB} A_{[\mu \, NA} \delta A_{\nu] \, PB}\nonumber \\
& & \delta A_{\mu\nu\rho}^A = \tilde{V}^A_a \left(\bar{\epsilon} \gamma_{[\mu\nu} \tau_a \psi_{\rho]} - \tfrac{i}{3} \bar{\epsilon} \gamma_{\mu\nu\rho} \chi_a\right) - \tfrac{2}{3} \epsilon^{AB} A_{[\mu \, MB} \delta A_{\nu  \rho]}^M \nonumber \\
& & \qquad + \tfrac{4}{3} \epsilon^{AB} A_{[\mu\nu}^M \delta A_{\rho]\, MB} + \tfrac{1}{6} \epsilon^{AB} \epsilon^{CD} \epsilon^{MNP}A_{[\mu \, MB} A_{\nu \, NC} \delta A_{\rho ]\, PD}  \quad .
\end{eqnarray}

We now discuss in some detail how the analysis of the closure of the
supersymmetry  algebra was performed. We have computed the
commutator of two supersymmetry transformations of parameters
$\epsilon_2$ and $\epsilon_1$ on the bosonic fields, and we have
imposed that this closes on all the local symmetries of the theory.
In particular, on the vielbein one obtains
\begin{equation}
[ \delta_1 , \delta_2 ] e_\mu{}^\alpha = \partial_\mu \xi^\nu e_\nu{}^\alpha+ \xi^\nu \partial_\nu e_\mu{}^\alpha  + \Lambda^{\alpha\beta} e_{\mu \beta} \quad ,
\end{equation}
where the general coordinate transformation parameter is
\begin{equation}
\xi_\mu = -i \bar{\epsilon}_2 \gamma_\mu \epsilon_1
\end{equation}
and the local Lorentz parameter is
\begin{eqnarray}
& & \Lambda_{\alpha\beta} = \xi^\nu \omega_{\nu \alpha\beta} + \tilde{L}^M_m \tilde{V}^A_a \left( \tfrac{i}{24} F^{\mu\nu}_{MA} \bar{\epsilon}_2 \gamma_{\alpha\beta\mu\nu} \sigma_m \tau_a \epsilon_1+ \tfrac{5i}{12} F_{\alpha\beta\, MA} \bar{\epsilon}_2 \sigma_m \tau_a \epsilon_1 \right) \nonumber \\
& & \qquad + L_{Mm} \left(\tfrac{i}{18}F^{\mu\nu\rho\, M} \bar{\epsilon}_2 \gamma_{\alpha\beta\mu\nu\rho} \sigma_m \tau_3 \epsilon_1 + \tfrac{2i}{3} F_{\alpha\beta \mu}  \bar{\epsilon}_2 \gamma^\mu \sigma_m \tau_3 \epsilon_1 \right) \nonumber \\
& & \qquad + \tfrac{3}{8} F_{\alpha \beta \mu\nu}^A V_{Aa} \bar{\epsilon}_2 \gamma^{\mu\nu} \tau_a \epsilon_1 \quad .
\end{eqnarray}
All the other fields also transform correctly under general
coordinate transformations. One can show that on top of this, on the
scalars one produces  local  $SO(3)$ and $SO(2)$ transformations. To
prove that the supersymmetry algebra closes of the vector $A_{\mu\,
MA}$ one needs the identities
\begin{eqnarray}
& & D_\mu L_{Mm} = \partial_\mu L_{Mm} + Q_{\mu \, mn} L_{Mn} = P_{\mu \, mn} L_{Mn} \qquad D_\mu \tilde{L}^M_m = - P_{\mu mn} \tilde{L}^M_n\nonumber \\
& &  D_\mu V_{Aa} = \partial_\mu V_{Aa} + Q_{\mu \, ab} V_{Ab} = P_{\mu \, ab} V_{Ab} \qquad D_\mu \tilde{V}^A_a = - P_{\mu ab} \tilde{V}^A_b
\end{eqnarray}
which follow from the definitions given in eq.
\eqref{MaurerCartanforms}.  The final result is that the algebra
produces a gauge transformation of parameter
\begin{equation}
\Lambda_{MA} = \Lambda^{\rm susy}_{MA} - \xi^\mu A_{\mu \, MA}
\quad ,
\end{equation}
where
\begin{equation}
\Lambda^{\rm susy}_{MA} = i L_{Mm } V_{Aa} \bar{\epsilon}_2 \sigma_m \tau_a \epsilon_1 \quad .
\end{equation}
The gauge transformation parameter of the 2-forms is
\begin{equation}
\Sigma_\mu^M = \Sigma_\mu^{{\rm susy} M} - \xi^{\nu} A_{\nu\mu}^M -\tfrac{1}{8}
\epsilon^{MNP} \epsilon^{AB} A_{\mu\, NA} \Lambda_{PB}^{\rm susy}
\quad ,
\end{equation}
where
\begin{equation}
\Sigma^{{\rm susy} M}_\mu = -\tfrac{i}{2} \tilde{L}^M_m \bar{\epsilon}_2 \gamma_\mu \sigma_m \tau_3 \epsilon_1 \quad .
\end{equation}
Finally, the gauge parameter of the 3-forms is
\begin{equation}
\Xi_{\mu\nu}^A = \Xi_{\mu\nu}^{{\rm susy}A} -\xi^{\rho} A_{\rho\mu\nu}^A + \tfrac{4}{9}
\epsilon^{AB} A_{[\mu\, MB} \Sigma_{\nu]}^{{\rm susy}M} + \tfrac{4}{9} \epsilon^{AB} A_{\mu\nu}^M \Lambda^{\rm susy}_{MB} \quad ,
\end{equation}
where
\begin{equation}
\Xi^{{\rm susy} A}_{\mu\nu} = \tfrac{1}{3} \tilde{V}^A_a \bar{\epsilon}_2 \gamma_{\mu\nu} \tau_a \epsilon_1 \quad .
\end{equation}
A crucial ingredient to prove the closure of the supersymmetry
algebra on the  3-form doublet is the self-duality relation
\begin{equation}
F_{\mu_1 ...\mu_4}^A V_{Aa} = - \tfrac{1}{4!} \epsilon_{\mu_1 ...\mu_4 \nu_1 ...\nu_4} \epsilon_{ab} F^{\nu_1 ...\nu_4 \, A} V_{Ab}
\quad .\label{selfdualityF4}
\end{equation}
We refer to Appendix A for more details on the self-duality
properties in eight  dimensions.

Following \cite{Bergshoeff:2005ac}, one can proceed and derive the supersymmetry transformations of the higher-rank forms by imposing the closure of the supersymmetry algebra, provided that the first-order duality conditions are imposed. In particular, the algebra closes on the 4-forms $A_{4 \, M}$ in the ${\bf (\overline{3},1)}$ that are dual to the 2-forms, on the 5-forms $A_{5}^{MA}$  in the ${\bf ({3},2)}$ that are dual to the 1-forms, and on the 6-forms $A_{6 \, M}{}^N$ in the ${\bf ({8},1)}$ and $A_{6 \, AB}$ in the ${\bf ({1},3)}$, that are dual to the scalars. Moreover, the algebra closes on the non-propagating 7-forms in the ${\bf (6,2)} \oplus {\bf (\overline{3},2)}$ and 8-forms in the ${\bf (15,1)} \oplus {\bf (3,3)} \oplus {\bf (3,1)}\oplus {\bf (3,1)}$ \cite{Riccioni:2007au,Bergshoeff:2007qi}. In particular, we are interested in the highest-dimensional representation of the 8-forms, which is the  ${\bf (15,1)}$. This is the irreducible representation with two symmetric indices up and one down. To determine how the field $A_8^{MN}{}_P$ behaves with respect to the different $\mathbb{Z}_2$ truncations, we only need the gravitino-dependent part of its supersymmetry transformation, which is 
\begin{equation}
\delta A_{\mu_1 ...\mu_8}{}^{MN}{}_P = \tilde{L}^M_m \tilde{L}^N_m L_{Pn} \bar{\epsilon} \gamma_{[ \mu_1 ...\mu_7} \sigma_n \psi_{\mu_8 ]} + ... \quad . \label{susytransf8forms}
\end{equation}
  
The 1/2-BPS space-filling branes correspond to the components of the 8-form potential that satisfy the highest-weight constraint. These in general are all the long weights of the representation, which in the case of the ${\bf 15}$ of $\rm{SL}(3,\mathbb{R})$ are the six components of $A_{8}{}^{MN}{}_P$ such that $M=N$ and  $M \neq P$ \cite{Bergshoeff:2013sxa}.  
 In the next subsection we will determine the three different $\mathbb{Z}_2$ truncations to the half-maximal theory coupled to two vector multiplets, and we will show that for each truncation there are two space-filling branes that survive the projection. These are the branes that preserve the supersymmetry of the truncated theory.

\subsection{$\mathbb{Z}_2$ truncations to half-supersymmetry and space-filling branes}

\begin{figure}[h]
 \centering
\scalebox{1} 
{
\begin{pspicture}(0,-2.8992147)(9.81,0.5364839)
\psline[linewidth=0.06cm,arrowsize=0.05291667cm
4.0,arrowlength=1.4,arrowinset=0.0]{->}(4.8,-0.08921477)(3.38,2.3507853)
\psline[linewidth=0.06cm,arrowsize=0.05291667cm
4.0,arrowlength=1.4,arrowinset=0.0,linecolor=black]{->}(4.8,-0.08921477)(1.98,-0.08921477)
\psline[linewidth=0.06cm,arrowsize=0.05291667cm
4.0,arrowlength=1.4,arrowinset=0.0,linecolor=black]{->}(4.8,-0.08921477)(6.22,2.3507853)
\psline[linewidth=0.06cm,arrowsize=0.05291667cm
4.0,arrowlength=1.4,arrowinset=0.0,linecolor=black]{->}(4.8,-0.08921477)(3.38,-2.5292149)
\psline[linewidth=0.06cm,arrowsize=0.05291667cm
4.0,arrowlength=1.4,arrowinset=0.0,linecolor=black]{->}(4.8,-0.08921477)(7.62,-0.08921477)
\psline[linewidth=0.06cm,arrowsize=0.05291667cm
4.0,arrowlength=1.4,arrowinset=0.0,linecolor=black]{->}(4.8,-0.08921477)(6.22,-2.5292149)
\usefont{T1}{ppl}{m}{n}
\rput{55.888874}(0.58888847,-3.6979384){\rput(3.7434375,-1.2742147){\large $-\alpha_{1}-\alpha_{2}$}}
\usefont{T1}{ppl}{m}{n}
\rput{60.39754}(3.8652787,-3.9893463){\rput(5.3234377,1.3457853){\large $\alpha_{1}+\alpha_{2}$}}
\usefont{T1}{ppl}{m}{n}
\rput(6.7434373,0.10578523){\large $\alpha_{1}$}
\usefont{T1}{ppl}{m}{n}
\rput{-56.242115}(3.9211056,4.189283){\rput(5.8434377,-1.5542147){\large $-\alpha_{2}$}}
\usefont{T1}{ppl}{m}{n}
\rput(3.1634376,0.12578523){\large $-\alpha_{1}$}
\usefont{T1}{ppl}{m}{n}
\rput{-59.27895}(0.5316899,4.2266393){\rput(3.9434376,1.6657852){\large $\alpha_{2}$}}
\pscircle[linewidth=0.02,dimen=outer,fillstyle=solid](4.8,-0.08921477){0.2}
\end{pspicture}
}

\caption{\label{the8ofsl3}\sl The roots of ${\rm SL}(3,\mathbb{R})$.}
\end{figure}
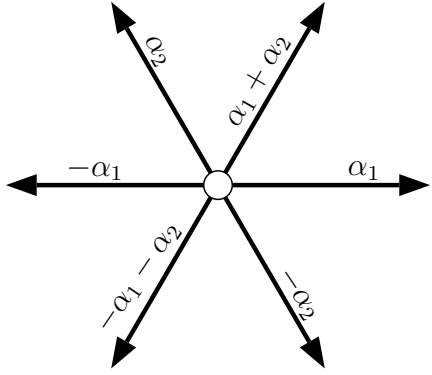

The maximal theory in $D=8$ can be truncated to half-maximal supergravity coupled to two vector multiplets. The resulting theory has global continuous symmetry $\mathbb{R}^+ \times {\rm SL}(2,\mathbb{R}) \times {\rm SL}(2,\mathbb{R})$, and therefore there are three independent truncations because there are three different ways of embedding $\mathbb{R}^+ \times {\rm SL}(2,\mathbb{R})$ inside ${\rm SL}(3,\mathbb{R})$. The three different embeddings can be easily visualised by looking at the root diagram of ${\rm SL}(3,\mathbb{R})$, which we draw in fig. \ref{the8ofsl3}. Each of the three ${\rm SL}(2,\mathbb{R})$'s are generated by one positive root, the corresponding negative root and the corresponding Cartan generator. 

We first discuss the  scalar sector. The scalars $V_A^a$ are obviously not projected out because the truncation does not act on the  ${\rm SL}(2,\mathbb{R})$ factor of the maximal theory. The index $M$ in the fundamental of ${\rm SL}(3,\mathbb{R})$ splits as $ M = (\sharp, \dot{A})$,  where ${\dot{A}} =1,2$ is the index of the fundamental of the ${\rm SL}(2,\mathbb{R})$ inside ${\rm SL}(3,\mathbb{R})$,  and similarly the SO(3) index $m$ splits as $m = (\sharp ,  \dot{a})$,  where $\dot{a}$ is the vector index of the maximal compact subgroup SO(2) of ${\rm SL}(2,\mathbb{R})$.
The scalars $L_{Mm}$ are truncated to
\begin{equation}
L_{Mm} \rightarrow ( e^{\Phi} , e^{-\Phi/2} V_{\dot{A}\dot{a}} ) \quad ,
\end{equation}
where the dilaton $\Phi$ parametrises $\mathbb{R}^+$ and the matrix  $V_{\dot{A}\dot{a}} $ satisfies the same identities as $V_{Aa}$  given in eq. \eqref{scalaridentities}.

We then derive how the truncation acts on the fermions. The gravity multiplet of the truncated theory contains a Majorana gravitino and a Majorana spinor, while each vector multiplet contains a single Majorana spinor. We find that the truncation (up to an overall sign) is 
\begin{eqnarray}
& & \psi_\mu = \sigma_\sharp \tau_3 \psi_\mu \nonumber\\
& & \chi_\sharp = \sigma_\sharp \tau_3 \chi_\sharp \nonumber\\
& & \chi_{\dot{a}} = - \sigma_\sharp \tau_3 \chi_{\dot{a}} \nonumber \\
& & \chi_a = - \sigma_\sharp \tau_3 \chi_{a} \quad .\label{fermiontruncationD=8}
\end{eqnarray}
The supersymmetry parameter $\epsilon$ is truncated like the gravitino.
It can be checked that the constraints of eq. \eqref{constraintsonfermionsD=8}   and the Majorana condition of eq. \eqref{symplMajorana}
are consistent with the truncation and this implies that  one ends up with the correct number of fermions.\footnote{In particular the constraint \eqref{constraintsonfermionsD=8} implies that $\chi_\sharp = \sigma_{\dot{a}} \chi_{\dot{a}}$.} The chirality condition of eq. \eqref{sortofchirality} on the truncated fermions gives two spinors of opposite chirality that can be recast in a single Dirac spinor $\Psi$ satisfying the standard $D=8$ Majorana condition  $\Psi = \tilde{C}_8 \overline{\Psi}^T$, where $\tilde{C}_8= C_8 \gamma_9$ satisfies the condition
\begin{equation}
\tilde{C}_8^\dagger \gamma_\mu \tilde{C}_8 =  \gamma_\mu^T\quad ,
\end{equation}
which has opposite sign with respect to eq. \eqref{CdaggergammaCD=8}.

We can now figure  out how the truncation acts on the supersymmetry algebra. First of all, it is straightforward to check that the truncation on the scalars and the one on the fermions are consistent.  On the 1-forms, the fermionic truncation is consistent with keeping only the components $A_{\mu \, \dot{A}A}$, because the supersymmetry variation of $A_{\mu \, \sharp A}$
is identically zero. Similarly, for the 2-form only the singlet component survives because the variation of $A_{\mu\nu}^{\dot{A}}$ vanishes identically. Finally, the 3-form is fully projected out. The variation of the 1-forms and 2-form that survive the projection is
\begin{eqnarray}
& &  \delta A_{\mu \, \dot{A}A} = e^{-\Phi/2} V_{\dot{A} \dot{a}} V_{Aa} \left( i \bar{\epsilon} \sigma_{\dot{a}} \tau_a
\psi_\mu - \bar{\epsilon} \gamma_\mu \sigma_{\dot{a}} \chi_a - \bar{\epsilon} \gamma_\mu \tau_a \tau_3\chi_{\dot{a}} \right) \nonumber \\
& & \delta A_{\mu\nu}^\sharp = e^{-\Phi} \left(i\bar{\epsilon} \gamma_{[\mu} 
\psi_{\nu]} + \tfrac{1}{2} \bar{\epsilon} \gamma_{\mu\nu} \chi_\sharp \right)+ \tfrac{1}{4} \epsilon^{\dot{A}\dot{B}} \epsilon^{AB} A_{[\mu \, \dot{A}A} \delta A_{\nu] \, \dot{B}B} \quad .
\end{eqnarray}
To summarise, the gauge fields that survive are four vectors and one 2-form, which is precisely the content of the half-maximal theory. 

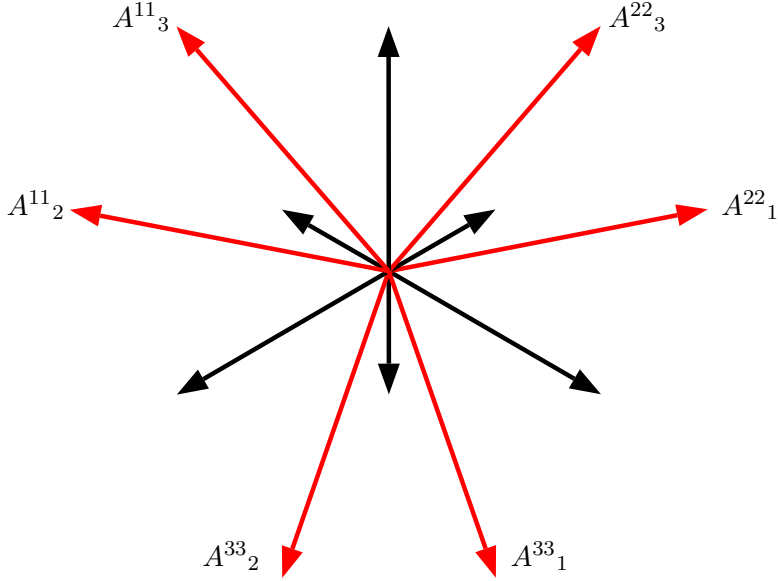
\begin{figure}[t!]
    \centering

    \scalebox{1} 
    {
        \begin{pspicture}(0,-3.9350965)(14.601648,4.2372804)
        \psline[linewidth=0.06cm,arrowsize=0.05291667cm
        4.0,arrowlength=1.4,arrowinset=0.0]{->}(7.00721,0.20367658)(7.008521,-1.4363229)
        \psdots[dotsize=0.12,dotangle=180.04582](7.00721,0.20367658)
       \psline[linewidth=0.06cm,arrowsize=0.05291667cm
        4.0,arrowlength=1.4,arrowinset=0.0]{->}(7.00721,0.20367658)(9.82852,-1.4340677)
        \psline[linewidth=0.06cm,arrowsize=0.05291667cm
        4.0,arrowlength=1.4,arrowinset=0.0,linecolor=red]{->}(7.00721,0.20367658)(5.5904727,-3.8774576)
        \psline[linewidth=0.06cm,arrowsize=0.05291667cm
        4.0,arrowlength=1.4,arrowinset=0.0,linecolor=red]{->}(7.00721,0.20367658)(11.246552,1.0270672)
        \psline[linewidth=0.06cm,arrowsize=0.05291667cm
        4.0,arrowlength=1.4,arrowinset=0.0]{->}(7.00721,0.20367658)(8.426554,1.0248119)
        \psline[linewidth=0.06cm,arrowsize=0.05291667cm
        4.0,arrowlength=1.4,arrowinset=0.0]{->}(7.00721,0.20367658)(4.188522,-1.4385781)
        \psline[linewidth=0.06cm,arrowsize=0.05291667cm
        4.0,arrowlength=1.4,arrowinset=0.0,linecolor=red]{->}(7.00721,0.20367658)(9.824601,3.4659307)
        \psline[linewidth=0.06cm,arrowsize=0.05291667cm
        4.0,arrowlength=1.4,arrowinset=0.0]{->}(7.00721,0.20367658)(5.586554,1.0225407)
        \psline[linewidth=0.06cm,arrowsize=0.05291667cm
        4.0,arrowlength=1.4,arrowinset=0.0]{->}(7.00721,0.20367658)(7.0046024,3.4636755)
        \psline[linewidth=0.06cm,arrowsize=0.05291667cm
        4.0,arrowlength=1.4,arrowinset=0.0,linecolor=red]{->}(7.00721,0.20367658)(2.766555,1.0202855)
        \psline[linewidth=0.06cm,arrowsize=0.05291667cm
        4.0,arrowlength=1.4,arrowinset=0.0,linecolor=red]{->}(7.00721,0.20367658)(4.184603,3.4614203)
       \psline[linewidth=0.06cm,arrowsize=0.05291667cm
        4.0,arrowlength=1.4,arrowinset=0.0,linecolor=red]{->}(7.00721,0.20367658)(8.430472,-3.8751864)
        \usefont{T1}{ppl}{m}{n}
        {\rput(9.01501,-3.5864835){$A^{33}{}_1$}}
                \usefont{T1}{ppl}{m}{n}
        {\rput(4.91501,-3.5864835){$A^{33}{}_2$}}
                        \usefont{T1}{ppl}{m}{n}
        {\rput(3.71501,3.5864835){$A^{11}{}_3$}}
                                \usefont{T1}{ppl}{m}{n}
        {\rput(10.31501,3.5864835){$A^{22}{}_3$}}
                                \usefont{T1}{ppl}{m}{n}
        {\rput(2.31501,1.0864835){$A^{11}{}_2$}}
                                       \usefont{T1}{ppl}{m}{n}
        {\rput(11.81501,1.0864835){$A^{22}{}_1$}}
        \end{pspicture}
    }

    \caption{\label{15withsl3components}  \sl The ${\bf 15}$ of ${\rm SL}(3,\mathbb{R})$ to which the 8-forms belong. 
     The longest weights are painted in red, and for each longest weight we have written the corresponding component of the potential. For simplicity the space-time indices are omitted. The shortest weights have multiplicity two.}
\end{figure}

By projecting the fermions  according to eq. \eqref{fermiontruncationD=8}  in the supersymmetry transformation of the 8-forms whose gravitino term is given in eq. \eqref{susytransf8forms}, one obtains that only the components $A_{\mu_1 ...\mu_8}{}^{MN}{}_\sharp$ survive. Out of these, only the 8-forms $A_{\mu_1 ...\mu_8}{}^{\dot{A}\dot{B}}{}_\sharp$ in the ${\bf 3}$ of ${\rm SL}(2,\mathbb{R})$ couple to 7-branes, and their supersymmetry transformations have the form
\begin{equation}
\delta A_{\mu_1 ...\mu_8}{}^{\dot{A}\dot{B}}{}_\sharp= e^{2\Phi} \tilde{V}^{\dot{A}}_{\dot{a}} \tilde{V}^{\dot{B}}_{\dot{a}}  \bar{\epsilon} \gamma_{[ \mu_1 ...\mu_7} \sigma_\sharp \psi_{\mu_8 ]} + ... \quad . \label{susytransf8formstruncated}
\end{equation}
In particular, we are interested in the brane components, which are the long weights of the  ${\bf 3}$, {\it i.e.} the two components $A_{\mu_1 ...\mu_8}{}^{\dot{A}\dot{A}}{}_\sharp$. We can understand better how the truncation acts by looking at the weight diagram of the ${\bf 15}$ in fig. \ref{15withsl3components}. We fix our conventions so that $\sharp =3$ corresponds to taking the  ${\rm SL}(2,\mathbb{R})$ subgroup as the one generated by the root $\alpha_1$ in fig. \ref{the8ofsl3}. This ${\rm SL}(2,\mathbb{R})$ acts on the indices 1 and 2, and the 8-form components that survive the projection are  $A_{\mu_1 ...\mu_8}{}^{11}{}_3$ and $A_{\mu_1 ...\mu_8}{}^{22}{}_3$. If $\sharp =2$,  the ${\rm SL}(2,\mathbb{R})$ subgroup is generated by the root $\alpha_2$ and acts on the indices 1 and 3. In this case the 8-form components that survive are   $A_{\mu_1 ...\mu_8}{}^{11}{}_2$ and $A_{\mu_1 ...\mu_8}{}^{33}{}_2$. Finally, if $\sharp =1$,  the ${\rm SL}(2,\mathbb{R})$ subgroup is generated by $\alpha_1 + \alpha_2$ and acts on the indices 2 and 3, and the 8-form components that survive are   $A_{\mu_1 ...\mu_8}{}^{22}{}_1$ and $A_{\mu_1 ...\mu_8}{}^{33}{}_1$.
To summarise, we find that for each truncation there are two space-filling branes that preserve the same supersymmetry of the truncation, precisely as expected from the analysis of the central charges \cite{Bergshoeff:2013sxa}.

We now want to understand this result from the perspective of the IIB theory. From IIB, one expects only four space-filling branes to arise by reducing to eight dimensions, which are the D9, the D7 and their S-duals. The remaining two 7-branes are exotic and couple to  8-forms that arise from  mixed-symmetry potentials in IIB. These potentials are derived from a suitable decomposition of the $E_{11}$ algebra \cite{West:2001as}, and can be found for instance in section 3.1 of ref. \cite{Bergshoeff:2012jb}. One can classify all the mixed-symmetry potentials that give rise to branes in lower dimensions in terms of the non-positive integer number $\alpha$ denoting how the tension of the corresponding  brane scales with respect to the string coupling $g_S$, and obviously T-duality relates different potentials with the same value of $\alpha$. Following \cite{Bergshoeff:2012ex}, we denote the potentials with $\alpha=-1,-2,-3...$ with the letters $C$, $D$, $E$ and so on. The 8-forms in eight dimensions then arise from the RR potentials $C_8$ and $C_{10}$ (with $\alpha=-1$), their  S-duals $E_8$ and $F_{10}$ (with $\alpha=-3$ and $-4$ respectively) and the mixed-symmetry potentials $E_{10,2,2}$ and $F_{10,2,2}$ (again with $\alpha=-3$ and $-4$ respectively).\footnote{In general we denote with  $A_{p,q,r,..}$  a mixed-symmetry potential in a representation such that $p,q,r, ...$ (with $p\ \geq q \geq r ...$) denote the length of each column of its Young Tableau.} Denoting with $x^i$, $i=1,2$ the internal directions in the reduction from ten to eight dimensions, these two  mixed-symmetry potentials give rise to the 8-form potentials $E_{\mu_1 ...\mu_8 \, x^1 x^2,x^1 x^2 ,x^1 x^2}$ and  $F_{\mu_1 ...\mu_8 \, x^1 x^2,x^1 x^2,x^1 x^2}$.

To derive which is the pair of 7-branes that is not projected out in each truncation, we move to the string frame, which corresponds to performing the redefinitions
\begin{equation}
e_\mu{}^\alpha = e^{-\tfrac{1}{3}\phi} e_\mu^{s}{}^\alpha \qquad \quad  \psi_\mu =  e^{-\tfrac{1}{6}\phi} \psi_\mu^s  \qquad \quad  \epsilon = e^{-\tfrac{1}{6}\phi} \epsilon^s \quad .
\end{equation}
As a convention, we associate to the case $\sharp =3$ the reduction to $D=8$ of the SO9 truncation. In this case the global symmetry of the truncated theory is perturbative and the dilaton $\Phi$ is proportional to the eight-dimensional string dilaton $\phi$. To get the right scaling in the string frame we impose
\begin{equation}
  {\rm SO9}: 
 \Phi = -\tfrac{2}{3} \phi  \quad .
\end{equation}
As a result, the supersymmetry transformations of the four vectors and the 1-form have no dilaton dependence in front of the gravitino term, as expected from the reduction of the SO9-truncated ten-dimensional theory. In particular, the 2-form is the NS-NS 2-form $B_{\mu\nu}$ and transforms exactly as the second equation  in \eqref{2formstransfintro}, while the vectors are $B_{\mu \, x^i}$ and  $g_{\mu \, x^i}$. 
Performing the same rescaling on eq. \eqref{susytransf8formstruncated}, we find that the transformation of both $A_{\mu_1 ...\mu_8}{}^{11}{}_3$ and $A_{\mu_1 ...\mu_8}{}^{22}{}_3$ has a factor $e^{-4 \phi}$ in front of the gravitino term, which implies that the corresponding branes have both $\alpha=-4$. These are the branes coming from the IIB potentials $F_{10}$ and $F_{10,2,2}$. 

We take the truncation identified by $\sharp =2$ to be  the O7 truncation. In this case the ${\rm SL}(2,\mathbb{R})$ symmetry is non-perturbative, and the components of the matrix $V_{\dot{A}\dot{a}}$ scale differently with respect to the string dilaton. In particular, we take the component with $\dot{A} = \dot{a} =1$ to scale like $e^{\phi/2}$, and the one with  $\dot{A} = \dot{a} =3$ to scale like $e^{-\phi/2}$. On top of this, the scalar $\Phi$ contains a term proportional to the string dilaton. The precise dependence on the string dilaton  of $\Phi$, $V_{11}$ and $V_{33}$ is
\begin{equation}
  {\rm O7}: \begin{cases}
 \Phi = \tfrac{1}{3} \phi + ...\\
 V_{11} = e^{\phi/2} ...\\
 V_{33} = e^{-\phi/2} ...\\
  \end{cases} \quad ,
\end{equation}
where we have ignored the contribution of  the additional scalars. One obtains that the transformation of the 2-form has an $e^{-\phi}$ factor, as expected because this is the RR 2-form $C_{\mu\nu \, x^1 x^2}$  and transforms as the third equation in \eqref{2formstransfintro}. Out of the four vectors, two have no dilaton factor (corresponding to $B_{\mu \, x^i}$) and two have a factor $e^{-\phi}$ (corresponding to $C_{\mu \, x^i}$).  By performing the rescaling on eq. \eqref{susytransf8formstruncated}, we find that the transformation of $A_{\mu_1 ...\mu_8}{}^{11}{}_2$ has a factor $e^{-3\phi}$, while the one of $A_{\mu_1 ...\mu_8}{}^{33}{}_2$ has a factor $e^{-\phi}$. We thus identify  the former with the potential $E_8$ and the latter  with the potential $C_8$.

Finally, the truncation identified by $\sharp =1$ is the reduction of the O9 truncation. Also in this case the ${\rm SL}(2,\mathbb{R})$ symmetry is non-perturbative, and we take the component of $V_{\dot{A}\dot{a}}$ with $\dot{A} = \dot{a} =2$ to scale like $e^{\phi/2}$, and the one with  $\dot{A} = \dot{a} =3$ to scale like $e^{-\phi/2}$. On top of this, the scalar $\Phi$ contains a term proportional to the string dilaton precisely as in the O7 case. We thus get
\begin{equation}
  {\rm O9}: \begin{cases}
 \Phi = \tfrac{1}{3}\phi+ ...  \\
  V_{22} = e^{\phi/2} ...\\
 V_{33} = e^{-\phi/2} ...\\
  \end{cases} \quad .
\end{equation}
One obtains that the transformation of the 2-form has an $e^{-\phi}$ factor, as expected because this is the RR 2-form $C_{\mu\nu}$  and transforms as the first equation in \eqref{2formstransfintro}. Out of the four vectors, two have no dilaton factor (corresponding to $g_{\mu \, x^i}$) and two have a factor $e^{-\phi}$ (corresponding to $C_{\mu \, x^i}$).  From eq. \eqref{susytransf8formstruncated} we read that the potential $A_{\mu_1 ...\mu_8}{}^{22}{}_1$ has a factor $e^{-3\phi}$ and thus corresponds to $E_{10,2,2}$, while the one of $A_{\mu_1 ...\mu_8}{}^{33}{}_1$ has a factor $e^{-\phi}$ and corresponds to $C_{10}$.

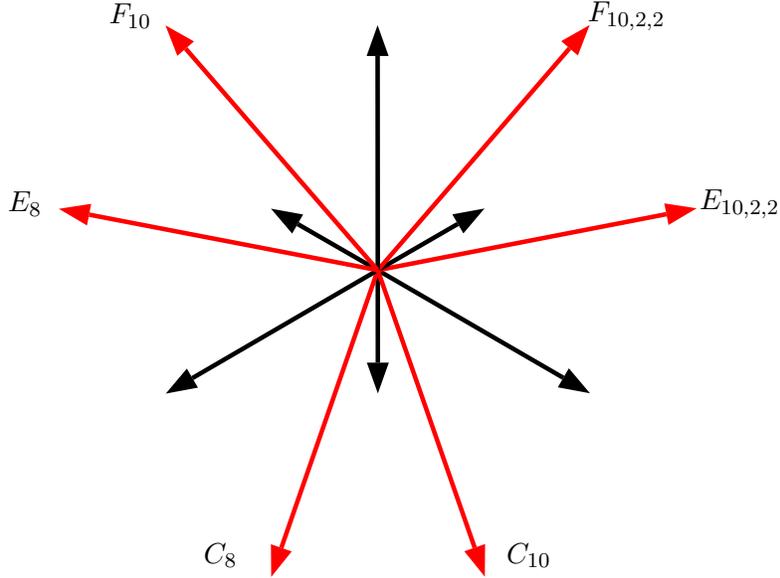
\begin{figure}[h]
    \centering

    \scalebox{1} 
    {
        \begin{pspicture}(0,-3.9350965)(14.601648,4.2372804)
        \psline[linewidth=0.06cm,arrowsize=0.05291667cm
        4.0,arrowlength=1.4,arrowinset=0.0]{->}(7.00721,0.20367658)(7.008521,-1.4363229)
        \psdots[dotsize=0.12,dotangle=180.04582](7.00721,0.20367658)
       \psline[linewidth=0.06cm,arrowsize=0.05291667cm
        4.0,arrowlength=1.4,arrowinset=0.0]{->}(7.00721,0.20367658)(9.82852,-1.4340677)
        \psline[linewidth=0.06cm,arrowsize=0.05291667cm
        4.0,arrowlength=1.4,arrowinset=0.0,linecolor=red]{->}(7.00721,0.20367658)(5.5904727,-3.8774576)
        \psline[linewidth=0.06cm,arrowsize=0.05291667cm
        4.0,arrowlength=1.4,arrowinset=0.0,linecolor=red]{->}(7.00721,0.20367658)(11.246552,1.0270672)
        \psline[linewidth=0.06cm,arrowsize=0.05291667cm
        4.0,arrowlength=1.4,arrowinset=0.0]{->}(7.00721,0.20367658)(8.426554,1.0248119)
        \psline[linewidth=0.06cm,arrowsize=0.05291667cm
        4.0,arrowlength=1.4,arrowinset=0.0]{->}(7.00721,0.20367658)(4.188522,-1.4385781)
        \psline[linewidth=0.06cm,arrowsize=0.05291667cm
        4.0,arrowlength=1.4,arrowinset=0.0,linecolor=red]{->}(7.00721,0.20367658)(9.824601,3.4659307)
        \psline[linewidth=0.06cm,arrowsize=0.05291667cm
        4.0,arrowlength=1.4,arrowinset=0.0]{->}(7.00721,0.20367658)(5.586554,1.0225407)
        \psline[linewidth=0.06cm,arrowsize=0.05291667cm
        4.0,arrowlength=1.4,arrowinset=0.0]{->}(7.00721,0.20367658)(7.0046024,3.4636755)
        \psline[linewidth=0.06cm,arrowsize=0.05291667cm
        4.0,arrowlength=1.4,arrowinset=0.0,linecolor=red]{->}(7.00721,0.20367658)(2.766555,1.0202855)
        \psline[linewidth=0.06cm,arrowsize=0.05291667cm
        4.0,arrowlength=1.4,arrowinset=0.0,linecolor=red]{->}(7.00721,0.20367658)(4.184603,3.4614203)
       \psline[linewidth=0.06cm,arrowsize=0.05291667cm
        4.0,arrowlength=1.4,arrowinset=0.0,linecolor=red]{->}(7.00721,0.20367658)(8.430472,-3.8751864)
        \usefont{T1}{ppl}{m}{n}
        {\rput(9.01501,-3.5864835){$C_{10}$}}
                \usefont{T1}{ppl}{m}{n}
        {\rput(4.91501,-3.5864835){$C_8$}}
                        \usefont{T1}{ppl}{m}{n}
        {\rput(3.71501,3.5864835){$F_{10}$}}
                                \usefont{T1}{ppl}{m}{n}
        {\rput(10.31501,3.5864835){$F_{10,2,2}$}}
                                \usefont{T1}{ppl}{m}{n}
        {\rput(2.31501,1.0864835){$E_8$}}
                                       \usefont{T1}{ppl}{m}{n}
        {\rput(11.81501,1.0864835){$E_{10,2,2}$}}
        \end{pspicture}
    }

    \caption{\label{15sl3withmixedsymmfig} \sl The identification of the long weights of the ${\bf 15}$ of ${\rm SL}(3,\mathbb{R})$ with the IIB potentials. The table should be compared with Table \ref{15withsl3components}.
         }
\end{figure}

We draw again in fig. \ref{15sl3withmixedsymmfig} the weight diagram of the ${\bf 15}$, where now the long weights are identified with the potentials of the IIB theory. We see from the diagram that the branes on the same horizontal line share the same value of $\alpha$. This is obvious from the fact that the ${\rm SL}(2,\mathbb{R})$ associated to the root $\alpha_1$ is part of the T-duality symmetry. In particular, the branes with $\alpha=-4$ belong to the ${\bf 3}$, the branes with $\alpha=-3$ belong to the ${\bf 4}$ and the branes with $\alpha=-1$ belong to the ${\bf 2}$ of this ${\rm SL}(2,\mathbb{R})$.
The table also shows that the ${\rm SL}(2,\mathbb{R})$ of the IIB theory is the one generated by $\alpha_2$. As we know, the 8-forms belong to the ${\bf 3}$, the 10-forms to the ${\bf 4}$ and the mixed-symmetry potentials to the ${\bf 2}$ of this other ${\rm SL}(2,\mathbb{R})$. Finally, the third ${\rm SL}(2,\mathbb{R})$, generated by $\alpha_1+ \alpha_2$, mixes $E_8$ and $F_{10}$, $C_8$ and $F_{10,2,2}$ and $C_{10}$ and $E_{10,2,2}$. For each truncation, it is the branes in the ${\bf 3}$ that survive, as we have already shown. It is known that in the case of the O7 truncation the potentials $C_8$ and $E_8$ both survive because the corresponding branes preserve the same supersymmetry, but what this analysis shows is that in the SO9 truncation one gets that both $F_{10}$ and $F_{10,2,2}$ are not projected out, while in the case of the O9 truncation both $C_{10}$ and $E_{10,2,2}$ survive the projection. For clarity, we summarise this result in Table \ref{summarytruncationsD=8max}.  In the next section we will show how this result can be generalised to identify in any dimension all truncations  and  the various space-filling branes that preserve the same supersymmetry of each truncation.

\begin{table}[h!]
\begin{center}
\scalebox{1}{
\begin{tabular}{|c||c|c|}
\hline {$D=8$ truncations} &  {potentials}  & { brane components}   \\
 \hline   \hline
 \rule[-2mm]{0mm}{2mm}  \multirow{2}{*}{$\textbf{O9}$} & $C_{10}$   &  $C_{8 \, x^1 x^2}$  \\
 & $E_{10,2,2}$ & $E_{8 \, x^1 x^2 , x^1x^2, x^1x^2}$  \\
\hline   \rule[-2mm]{0mm}{2mm} \multirow{2}{*}{$\textbf{SO9}$} & $F_{10}$  & $F_{8 \, x^1 x^2}$ \\
 & $F_{10,2,2}$  & $F_{8 \, x^1 x^2, x^1x^2, x^1x^2}$    \\  
 \hline   \rule[-2mm]{0mm}{2mm} \multirow{2}{*}{$\textbf{O7}$} & $C_{8}$  & $C_{8}$      \\
 & $E_{8}$  &  $E_{8}$   \\

\hline

\end{tabular}
}
\caption{\sl The $\mathbb{Z}_2$ truncations of the maximal theory in $D=8$ from the IIB perspective. The indices $x^i$, $i=1,2$, label the internal directions. \label{summarytruncationsD=8max}}
\end{center}
\end{table}

\section{$\mathbb{Z}_2$ truncations in any dimension}

In the previous section we have determined the three different $\mathbb{Z}_2$ truncations of $D=8$ maximal supergravity to the half-maximal theory, and for each truncation we have identified the two space-filling branes that preserve the same supersymmetry of the truncation. We have shown that in the case of the O9 and SO9 truncations, one of the two 7-branes is an exotic brane, which corresponds to the IIB mixed symmetry potentials $E_{10,2,2}$ in the  O9 case and $F_{10,2,2}$ in the SO9  case. In general,  exotic branes are associated to specific components of the ten-dimensional mixed-symmetry potentials $A_{p,q,r,..}$ (with $p\ \geq q \geq r ...$) determined as follows: first of all, only the $p$ set can contain space-time indices, while all the other sets of indices must be internal, because the space-time indices must be antisymmetric. On top of this, the $p$ indices must contain all the internal indices $q$, which must contain all the internal indices $r$ and so on \cite{Bergshoeff:2011zk,Bergshoeff:2011ee,Bergshoeff:2011se,Bergshoeff:2012ex}.  
In \cite{Lombardo:2016swq} a universal rule was derived that relates different brane components of mixed-symmetry potentials by a  T-duality transformation in a given direction. Specifically,
given  an $\alpha=-n$ brane associated to a mixed-symmetry potential such that the internal $x$ index occurs $N$ times (in $N$ different sets of antisymmetric indices),  this is mapped by T-duality along $x$ to the brane associated to the potential in which the $x$ index occurs $n-N$ times. Schematically, this can be written as
\begin{equation}
\alpha=-n \ : \qquad \quad \underbrace{x,x,...,x}_N \ \overset{{\rm T}_x}{\longleftrightarrow} \ \underbrace{x,x, ....,x }_{n-N} \quad . \label{allbranesallalphasrule}
\end{equation}
Using this T-duality rule, if one performs two T-dualities in the directions $x^1$ and $x^2$ not only $C_8$ is mapped to $C_{8 \, x^1 x^2 }$ as one naturally expects, but also $E_8$ is mapped to $E_{8 \, x^1 x^2 , x^1 x^2 , x^1 x^2 }$ and $F_{8 \, x^1 x^2 } $ is mapped to 
$F_{8 \, x^1 x^2 , x^1 x^2 , x^1 x^2 }$. We stress that performing two T-dualities maps states in IIB to other states in the same theory, and as far as representations of the perturbative ${\rm SL}(2,\mathbb{R})$ inside ${\rm SL}(3,\mathbb{R})$ are concerned, it  maps one long weight to the other. 
In other words, using the universal T-duality rule in eq. \eqref{allbranesallalphasrule} we could have immediately declared that the O7 truncation, in which $C_8$ and $E_8$ are not projected out, is mapped by two T-dualities to the O9 truncation, in which  $C_{8 \, x^1 x^2 }$  and $E_{8 \, x^1 x^2 , x^1 x^2 , x^1 x^2 }$ are not projected out, and the latter truncation is mapped to the SO9 truncation, in which the 8-form potentials that survive are 
$F_{8 \, x^1 x^2 } $ and $F_{8 \, x^1 x^2 , x^1 x^2 , x^1 x^2 }$.

The aim of this section is to show that using eq. \eqref{allbranesallalphasrule} and S-duality, one can characterise all truncations in any dimension, and for each truncation determine all the space-filling branes that are not projected out. We will first discuss the maximal case in any dimension from seven to three, and we will then move to  the $\mathbb{Z}_2$ truncations  of the half-maximal theories listed in Table \ref{Fromhalfmaximaltoquartermaximalsupergravity} and finally the $\mathbb{Z}_2$ truncations  of the quarter-maximal theories listed in Table \ref{Fromquartermaximaltominimalsupergravity}.

\subsection{From maximal to half-maximal supergravity}

${\bf D=7}$: \
We want to consider the truncation of maximal supergravity in $D=7$ to the half-maximal theory coupled to three vector multiplets, with symmetry $\mathbb{R}^+ \times {\rm SO}(3,3)$, which is isomorphic to ${\rm GL}(4,\mathbb{R})$. In the truncation, the vectors in the ${\bf \overline{10}}$ are truncated to the ${\bf 6}$ and the 2-forms in the ${\bf 5}$ are truncated to a singlet. 
There are 5 different ways of performing this truncation, corresponding to the five different ways in which ${\rm SL}(4,\mathbb{R})$ can be embedded in ${\rm SL}(5,\mathbb{R})$, and this agrees with the dimension of the vector central charge, which indeed belongs to the vector representation of the R-symmetry ${\rm SO}(5)$. We denote with $M=1,...,5$ the index of the fundamental of ${\rm SL}(5,\mathbb{R})$ and with $m=1,...,5$ the vector index of ${\rm SO}(5)$. As in the previous section, the scalars are encoded in the matrix $L_{Mm}$ satisfying identities analogous to those in eq. \eqref{scalaridentities}, with the Maurer-Cartan form defined as in eq. \eqref{MaurerCartanforms}.\footnote{We use here exactly the same index notation as for the ${\rm SL}(3,\mathbb{R})/{\rm SO}(3)$ coset of the previous section. We do not expect this to cause any confusion to the reader.} 

The 7-forms $A_{\mu_1 ...\mu_7}{}^{MN}{}_P$ belong to the ${\bf 70}$ of ${\rm SL}(5,\mathbb{R})$, which as in the eight-dimensional case is the irreducible  representation with two symmetric upper indices and one lower index. The gravitino-dependent part of its supersymmetry transformation is 
\begin{equation}
\delta A_{\mu_1 ...\mu_7}{}^{MN}{}_P = i \tilde{L}^M_m \tilde{L}^N_m L_{Pn} \bar{\epsilon} \gamma_{[ \mu_1 ...\mu_6} \Gamma_n \psi_{\mu_7 ]} + ... \quad , \label{susytransf7forms}
\end{equation}
where  we denote with $\Gamma_m$  the ${\rm SO}(5)$ gamma-matrices.  
The 1/2-BPS space-filling branes are the 20 components such that $M=N$ and $M\neq P$ \cite{Bergshoeff:2013sxa}.

We truncate the theory by splitting the $M$ index as $M = (\sharp, A )$, where $A=1,...,4$ is the index of the fundamental of ${\rm SL}(4,\mathbb{R})$. Similarly, $m$ splits as $m = (\sharp, a)$. 
The scalars are truncated to
\begin{equation}
L_{Mm} \rightarrow ( e^{\Phi} , e^{-\Phi/4} V_{{A}{a}} ) \quad ,
\end{equation}
where the dilaton $\Phi$ parametrises $\mathbb{R}^+$ and the matrix  $V_{{A}{a}} $ contains the scalars parametrising the coset ${\rm SL}(4,\mathbb{R})/{\rm SO}(4)$. On  $\psi_\mu$ and $\epsilon$ the truncation acts as 
\begin{equation}
\psi_\mu = \Gamma_\sharp \psi_\mu \qquad \qquad 
\epsilon = \Gamma_\sharp \epsilon \quad . \end{equation}
As a result, after the truncation only the 7-forms $A_{\mu_1 ...\mu_7}{}^{MN}{}_\sharp$ survive, and in particular only the components $A_{\mu_1 ...\mu_7}{}^{AB}{}_\sharp$ in the ${\bf 10}$ of ${\rm SL}(4,\mathbb{R})$ couple to 6-branes. Their supersymmetry transformations have the form 
\begin{equation}
\delta A_{\mu_1 ...\mu_7}{}^{{A}{B}}{}_\sharp= i e^{\tfrac{3}{2}\Phi} \tilde{V}^{{A}}_{{a}} \tilde{V}^{{B}}_{{a}}  \bar{\epsilon} \gamma_{[ \mu_1 ...\mu_6}  \psi_{\mu_7 ]} + ... \quad . \label{susytransf7formstruncated}
\end{equation}
In particular, there are four 6-branes in the ${\bf 10}$, that all preserve the same supersymmetry which is the supersymmetry preserved by the truncation. 

From the ten-dimensional IIB perspective, the five truncations are the SO9, preserving the 2-form $B_{\mu\nu}$, the O9, preserving the 2-form $C_{\mu\nu}$ and the three different O7$_{x^i}$ truncations, preserving the 2-form $C_{\mu\nu\, x^j x^k}$ (with $i,j,k$ all different). 
As in eight dimensions, we go to the string frame to get the tension of the 6-branes that are preserved in each truncation. In seven dimensions, this  corresponds to performing the redefinitions
\begin{equation}
e_\mu{}^\alpha = e^{-\tfrac{2}{5}\phi} e_\mu^{s}{}^\alpha \qquad \quad  \psi_\mu =  e^{-\tfrac{1}{5}\phi} \psi_\mu^s  \qquad \quad  \epsilon = e^{-\tfrac{1}{5}\phi} \epsilon^s \quad .
\end{equation}
In the case of the SO9 truncation, the scalar $\Phi$ is proportional to the string dilaton, and ${\rm SL}(4,\mathbb{R})$ is the perturbative T-duality symmetry. 
More precisely, one gets
\begin{equation}
  {\rm SO9}: 
 \Phi = -\tfrac{4}{5} \phi  \quad ,
\end{equation}
and from eq. \eqref{susytransf7formstruncated} one can then check that the four preserved 6-branes have $\alpha=-4$. For the other four truncations, the dilaton identification can be taken to be
\begin{equation}
  {\rm O9} \, , \, {\rm O7}_{x^i}: \begin{cases}
 \Phi = \tfrac{1}{5} \phi + ...\\
 {\rm diag} ( V_{Aa}) = ( e^{-\tfrac{3}{4} \phi} ..., e^{\tfrac{1}{4} \phi} ...,e^{\tfrac{1}{4} \phi} ...,e^{\tfrac{1}{4} \phi} ...)\\
  \end{cases} \quad ,
\end{equation}
where as in eight dimensions the dots stand for contributions of the other scalars. One can check that in this case from eq. \eqref{susytransf7formstruncated} one gets one $\alpha=-1$ 6-brane and three $\alpha=-3$ 6-branes. 

\begin{table}[t!]
\begin{center}
\scalebox{1.00}{
\begin{tabular}{|c|c||c|c|c|}
\hline $D=7$ truncations & \# &  potentials   & brane components & \#  branes \\
 \hline \hline   \rule[-2mm]{0mm}{2mm}   \multirow{2}{*}{\textbf{O9}} & \multirow{2}{*}{1} &  $C_{10}$   &  $C_{7 \, x^1 x^2 x^3}$ & 1  
 \\
 & & $E_{10,2,2}$ & $E_{7 \, x^1 x^2 x^3 , x^p x^q, x^p x^q}$ & 3 \\
\hline   \rule[-2mm]{0mm}{2mm}  \multirow{2}{*}{\textbf{SO9}} & \multirow{2}{*}{1} & $F_{10}$  & $F_{7 \, x^1 x^2 x^3}$ & 1   \\
&  & $F_{10,2,2}$  & $F_{7 \, x^1 x^2 x^3 , x^px^q, x^px^q}$ & 3\\
 \hline   \rule[-2mm]{0mm}{2mm}  \multirow{3}{*}{$\textbf{O7}_{x^i}$} & \multirow{3}{*}{3} & $C_{8}$  & $C_{7 \, x^i}$ & 1    \\ 
 & & $E_{8}$  &  $E_{7\,  x^i}$ & 1  
  \\
 &  & $E_{9,2,1}$  &  $ E_{7 \, x^ix^p,x^i x^p,x^i}$ & 2  \\

\hline
\end{tabular}
}
\caption{\sl The $\mathbb{Z}_2$ truncations of the maximal theory in $D=7$.  In the case of the O7 truncation, the $i=1,2,3$ index labels the truncation while the $p$ and $q$ indices are different from $i$. }
\label{truncationmaximal7}
\end{center}
\end{table}

These results could have been easily deduced by compactification from eight dimensions. Indeed, we know from the analysis of the previous section that $F_{10}$ and $F_{10,2,2}$ give rise to 7-branes that are both preserved under the SO9 truncation. This implies that in seven dimensions the 6-branes coupled to $F_{\mu_1 ...\mu_7 \, x^1 x^2 x^3 , x^1 x^2 , x^1 x^2}$, $F_{\mu_1 ...\mu_7 \, x^1 x^2 x^3 , x^1 x^3 , x^1 x^3}$ and $F_{\mu_1 ...\mu_7 \, x^1 x^2 x^3 , x^2 x^3 , x^2 x^3}$ must all preserve the same supersymmetry of the one coupled to $F_{\mu_1 ...\mu_7 \, x^1 x^2 x^3}$. The same applies to the O9 truncation, where $C_{10}$ and $E_{10,2,2}$ give rise to four 6-branes. We can then use the T-duality rules in eq. \eqref{allbranesallalphasrule} to determine the 6-branes that are not projected out in each of the O7 truncations. The truncation O7$_{x^i}$ is obtained by performing two T-dualities in the directions $x^p$ and $x^q$ different from $x^i$. One obtains, together with $C_{7 \, x^i}$ and $E_{7 \, x^i}$ coming from $C_8$ and $E_8$, also the two additional branes $E_{7 \, x^i x^p , x^i x^p , x^i}$ coming from the mixed-symmetry potential $E_{9,2,1}$. The full result of the different truncations in seven dimensions is summarised in Table \ref{truncationmaximal7}.

In any dimension $D=10-d$, the potentials $F_{10,2n,2n}$ are known to belong to a specific representation of ${\rm SO}(d,d)$, which is the self-dual representation with $d$ antisymmetric indices \cite{Bergshoeff:2012jb}. As we have seen, the corresponding branes are those preserved by the SO9 truncation. In the following we will show that one can continue the analysis starting from the SO9 truncation and mapping this to all the other possible truncations in any dimensions using S-duality and the T-duality transformations in eq. \eqref{allbranesallalphasrule}.

\begin{table}[t!]
\begin{center}
\scalebox{1}{
\begin{tabular}{|c|c||c|c|c|}
\hline $D=6{\rm A}$ truncations & \# &  potentials   & brane components & \#  branes \\
 \hline \hline 
 \cline{1-5}   \rule[-2mm]{0mm}{2mm}  \multirow{3}{*}{\textbf{O9}} & \multirow{3}{*}{1} & $C_{10}$  &  $C_{6 \, x^1 ...x^4}$ & 1   \\
 & &  $E_{10,2,2}$ & $E_{6 \, x^1 ...x^4 ,x^px^q,x^px^q}$  & 6    
  \\
& & $G_{10,4,4}$ & $G_{6 \, x^1 ...x^4 ,x^1 ...x^4,x^1 ...x^4}$ & 1 
\\

\hline   \rule[-2mm]{0mm}{2mm}  \multirow{3}{*}{\textbf{SO9}} & \multirow{3}{*}{1} & $F_{10}$  & $F_{6 \, x^1 ...x^4}$ & 1   
\\
 & & $F_{10,2,2}$  & $F_{6 \, x^1 ...x^4,x^px^q,x^px^q}$ & 6   
 \\
  & & $F_{10,4,4}$  &  $F_{6 \, x^1 ...x^4 ,x^1 ...x^4,x^1 ...x^4}$ & 1 
   \\
  \hline   \rule[-2mm]{0mm}{2mm}  \multirow{5}{*}{$\textbf{O7}_{x^i x^j}$}  & \multirow{5}{*}{6} & $C_8$  & $C_{6\, x^ix^j}$ & 1 
   \\
&  & $E_8$ & $E_{6\, x^ix^j}$ & 1  
 \\
&   & $E_{9,2,1}$ & $E_{6\, x^ix^jx^p, x^ix^p,x^i}$ & 4    
  \\
&  & $E_{10,4,2}$ & $E_{6 \, x^1 ...x^4 ,x^1 ...x^4,x^px^q}$ & 1  
 \\
&   & $G_{10,4,2}$ & $G_{6 \, x^1 ...x^4 , x^1 ...x^4,x^px^q}$ & 1  
  \\
   \hline   \rule[-2mm]{0mm}{2mm}  \multirow{3}{*}{$\textbf{O5}$} & \multirow{3}{*}{1} & $C_6$  & $C_{6}$ & 1   
   \\
&  & $E_{8,2}$ & $E_{6 \, x^px^q, x^px^q}$ & 6  
 \\
&   & $G_{10,4}$ & $G_{6 \, x^1 ...x^4,x^1 ...x^4}$ & 1  
    \\
     \hline   \rule[-2mm]{0mm}{2mm}  \multirow{3}{*}{$\textbf{SO5}$} & \multirow{3}{*}{1}& $D_6$  & $D_{6}$ & 1    
     \\
 & & $D_{8,2}$ & $D_{6\, x^px^q,x^px^q}$ & 6    
 \\
&   & $D_{10,4}$ & $D_{6 \, x^1 ...x^4,x^1 ...x^4}$ & 1    \\  \hline

\end{tabular}
}
\caption{\sl The $\mathbb{Z}_2$ truncations of the maximal theory in $D=6$ leading to the $\mathcal{N}=(1,1)$ theory. The $x$'s denote the four internal directions. As everywhere else in this section, for each truncation the $i,j,...$ indices label the truncation while the $p,q,...$ indices  are the remaining ones.}
\label{6A6Bbranes}
\end{center}
\end{table}

\vskip .5cm
\noindent
${\bf D=6}$: \ 
The maximal theory in six dimensions has global symmetry ${\rm SO}(5,5)$, and the 6-form potentials that couple to 5-branes can either support vector or tensor multiplets. The former belong to the ${\bf 320}$, which has 80 long weights, while the latter belong to the ${\bf \overline{126}}$, which has 16 long weights \cite{Bergshoeff:2011qk,Bergshoeff:2013sxa}.
The theory can be truncated to either $\mathcal{N}=(1,1)$ supergravity  coupled to four vector multiples, with global symmetry $\mathbb{R}^+\times {\rm SO}(4,4)$, or $\mathcal{N}=(2,0)$ supergravity coupled to five vector multiples, with global symmetry ${\rm SO}(5,5)$. As Table \ref{numberofbranesmax} shows, in the first case there are ten different truncations, which can be easily understood by observing that there are ten ways in which one can embed ${\rm SO}(4,4)$ in ${\rm SO}(5,5)$, while in the latter there is only one truncation because the truncated theory has the same symmetry of the maximal one. 

We start considering the truncations to the  $\mathcal{N}=(1,1)$  theory, which we also denote as 6A. The SO9 truncation is the one that leaves the perturbative T-duality symmetry ${\rm SO}(4,4)$ intact. The eight 5-branes charged under the potentials $F_{6 \, x^1 ...x^4}$, $F_{6 \, x^1 ...x^4,x^px^q,x^px^q}$ and  $F_{6 \, x^1 ...x^4 ,x^1 ...x^4,x^1 ...x^4}$ are all preserved by the truncation, and correspond to the eight long weights of the ${\bf 35}_{\rm V}$ of ${\rm SO}(4,4)$. This is mapped by S-duality to the O9 truncation, preserving the branes coupled to $C_{10}$ and $E_{10,2,2}$ together with the $\alpha=-5$ brane coupled to $G_{10,4,4}$. By T-dualising in the directions $x^p$ and $x^q$ one then gets the branes of the O7$_{x^i x^j}$ truncations, with $i,j$ different from $p,q$. There are six pairs of coordinates that one can choose, corresponding to six different O7 truncations. One can also perform four T-dualities on O9, which leads to the O5 truncation. Finally, the tenth truncation is obtained by performing an S-duality transformation of O5, and we dub this latter truncation SO5. We give in Table \ref{6A6Bbranes} the full result, which can be reproduced using eq. \eqref{allbranesallalphasrule} and the S-duality transformation rules of the various mixed-symmetry potentials.

\begin{table}[t!]
\begin{center}
\scalebox{1}{
\begin{tabular}{|c||c|c|c|}
\hline $D=6{\rm B}$ truncation  &  potentials   & brane components & \#  branes  \\   \hline \hline
   \rule[-2mm]{0mm}{2mm}  \multirow{4}{*}{\textbf{tensor}}  & $D_{7,1}$   &  $D_{6 \, x^p,x^p}$ & 4 
 \\
  & $D_{9,3}$ & $D_{6\, x^px^qx^r,x^px^qx^r}$ &4  
   \\
& $F_{9,3}$ & $F_{6\, x^px^qx^r,x^px^qx^r}$ &4 
   \\
 & $F_{10,4,1,1}$ & $F_{6\, x^1 ...x^4 ,x^1 ...x^4,x^p,x^p}$ &4 
 \\ \hline
\end{tabular}
}
\caption{\sl The $\mathbb{Z}_2$ truncation of the maximal theory in $D=6$ leading to the $\mathcal{N}=(2,0)$ theory. 
The branes that survive the projection are 16 and in the maximal theory support tensor multiplets in their world-volume.}
\label{6A6BbranesN20}
\end{center}
\end{table}

The truncation to the  $\mathcal{N}=(2,0)$  or 6B theory is unique. All the 5-branes that support tensor multiplets on their world-volume preserve the same supersymmetry and all survive the projection, that we simply label ``tensor'' truncation. These branes are charged under the mixed-symmetry potentials $D_{7,1}$, $D_{9,3}$ with $\alpha=-2$ and $F_{9,3}$ and $F_{10,4,1,1}$ with $\alpha=-4$, and they correspond to the 16 long weights of the ${\bf \overline{126}}$ of ${\rm SO}(5,5)$. We report the result in Table \ref{6A6BbranesN20}. 

Apart from the space-filling branes that preserve the same supersymmetry of the truncation, starting from six dimensions that are additional space-filling branes that survive the truncation and are 1/2-BPS states of the truncated theory. In the case of the 6A truncation, these branes are in the ${\bf 35}_{\rm V} \oplus {\bf 35}_{\rm S}\oplus {\bf 35}_{\rm C}$, as shown in Table \ref{numberofbraneshalfmax}. This result can be understood in the case of the SO9 truncation, in which case the ${\rm SO}(4,4)$ symmetry is perturbative and the branes that are left are those of the heterotic theory \cite{Bergshoeff:2012jb}. In particular, the eight branes in the ${\bf 35}_{\rm V}$ are the SO5 branes in Table \ref{6A6Bbranes} and the 16 branes in the  ${\bf 35}_{\rm S}\oplus {\bf 35}_{\rm C}$ are  all those in Table \ref{6A6BbranesN20} (more precisely the first representation contains the $\alpha=-2$ branes and the second the $\alpha=-4$ branes \cite{Bergshoeff:2011qk}). In \cite{Bergshoeff:2012jb,Bergshoeff:2017gpw} the world-volume multiplets for each of these branes in the truncated theory were determined, and in particular it was shown that the SO5 branes and the $\alpha=-2$ branes in Table \ref{6A6BbranesN20} support a hypermultiplet, while the $\alpha=-4$ branes in Table \ref{6A6BbranesN20} support a tensor multiplet.

For any other truncation to the 6A theory, the space-filling branes that remain as 1/2-BPS states of the truncated theory can be determined using the properties of the various truncations under S and T dualities. In particular, the branes in Table \ref{6A6BbranesN20} are present in every truncation. On top of this, in the O9 case one gets the O5 branes and vice-versa, in the SO5 case one gets the SO9 branes and in the O7$_{x^i x^j}$ case one gets the O7$_{x^k x^l}$ branes, where $i,j \neq k,l$. 

\begin{table}[t!]
\begin{center}
\scalebox{1.0}{
\begin{tabular}{|c|c||c|c|c|}
\hline $D=5$ truncations & \# &  potentials   & brane components & \#  branes \\
 \hline \hline   \rule[-2mm]{0mm}{2mm}    \multirow{3}{*}{\textbf{O9}} &\multirow{3}{*}{1} &  $C_{10}$   &  $C_{5 \, x^1 ...x^5}$  & 1 
\\
& & $E_{10,2,2}$ & $E_{5 \, x^1 ...x^5, x^px^q, x^px^q}$ & 10 
\\
& & $G_{10,4,4}$ & $G_{5 \, x^1 ...x^5, x^px^qx^rx^s, x^px^qx^rx^s}$ & 5 
\\
\hline   \rule[-2mm]{0mm}{2mm}   \multirow{3}{*}{\textbf{SO9}} &\multirow{3}{*}{1} & $F_{10}$  & $F_{5 \, x^1 ...x^5}$ & 1  
\\
& & $F_{10,2,2}$  &  $F_{5 \, x^1 ...x^5, x^px^q,x^px^q}$ & 10  
   \\
& & $F_{10,4,4}$  & $F_{5\, x^1 ...x^5, x^px^qx^rx^s, x^px^qx^rx^s}$ & 5  
 \\
\hline   \rule[-2mm]{0mm}{2mm}   \multirow{6}{*}{$\textbf{O7}_{x^i x^j x^k }$} &\multirow{6}{*}{10} & $C_{8}$  & $C_{5\, x^i x^j x^k}$ & 1   \\
  & & $E_{8}$  & $E_{5\, x^ix^jx^k}$ & 1   
   \\
&  & $E_{9,2,1}$  & $E_{5 \, x^ix^jx^k x^p, x^ix^p, x^i}$ & 6  
   \\
 & & $E_{10,4,2}$  &  $E_{5\, x^1 ...x^5 ,x^ix^jx^p x^q,x^ix^j}$ & 3 
  \\
&  & $G_{10,4,2}$  &  $G_{5 \, x^1 ...x^5 ,x^ix^jx^px^q,x^ix^j}$ & 3  
 \\
&    & $G_{10,5,4,1}$  & $G_{5 \, x^1 ...x^5 ,x^1 ...x^5,x^ix^jx^kx^p,x^p}$ & 2  
   \\
\hline   \rule[-2mm]{0mm}{2mm}   \multirow{5}{*}{$\textbf{O5}_{x^i}$} &\multirow{5}{*}{5} & $C_{6}$  & $C_{5 \, x^i}$ & 1    
\\
& & $E_{8,2}$  &  $E_{5\, x^i x^p x^q, x^p x^q}$ & 6 
  \\
&  & $E_{9,4,1}$  & $E_{5\, x^i x^px^qx^r,x^ix^px^qx^r,x^i}$ & 4 
 \\
&  & $G_{10,4}$  &$G_{5 \, x^1 ...x^5,x^px^qx^rx^s}$ & 1  
 \\
&  & $G_{10,5,2,1}$  & $G_{5 \, x^1 ...x^5,x^1 ...x^5,x^ix^p,x^p}$ & 4   
 \\
 \hline   \rule[-2mm]{0mm}{2mm}   \multirow{5}{*}{$\textbf{SO5}_{x^i}$} &\multirow{5}{*}{5}& $D_{6}$  & $D_{5\, x^i}$ & 1 
  \\
& & $D_{8,2}$  &  $D_{5 \, x^i x^px^q,x^px^q}$ & 6 
\\
& & $D_{10,4}$  &  $D_{5 \, x^1 ...x^5 ,x^px^qx^rx^s}$ & 1    
\\
& & $F_{9,4,1}$  &$F_{5\, x^i x^px^qx^r,x^ix^p x^q x^r,x^i}$ & 4 
\\

& &   $F_{10,5,2,1}$  & $F_{5\, x^1 ...x^5 ,x^1 ...x^5 ,x^ix^p,x^p}$ & 4   
\\
   
\hline   \rule[-2mm]{0mm}{2mm}   \multirow{4}{*}{$\textbf{T}^2\textbf{SO5}_{x^i}$} & \multirow{4}{*}{5} & $D_{7,1}$  & $D_{5 \, x^ix^p,x^p}$ & 4 
\\
& & $D_{9,3}$  &  $D_{5\, x^ix^px^qx^r,x^px^qx^r}$ & 4    
 \\
&  & $F_{9,3}$  & $F_{5\, x^i x^px^qx^r,x^px^qx^r}$ & 4 
 \\
&   &  $F_{10,4,1,1}$ &  $F_{5\, x^1 ...x^5 ,x^px^qx^rx^s,x^p,x^p}$ &4  
  \\
\hline

\end{tabular}
}
\caption{\sl The $\mathbb{Z}_2$ truncations of the maximal theory in $D=5$. }
\label{maximalbranesD5}
\end{center}
\end{table}

The 6A truncation has also a natural interpretation as the untwisted sector of IIA reduced on the orbifold $T^4/\mathbb{Z}_2$. The space-filling branes of this theory were discussed in \cite{Bergshoeff:2012jb,Bergshoeff:2013spa,Bergshoeff:2017gpw}, and performing a single T-duality transformation to translate the results of those papers in the IIB language, one finds that the $\mathbb{Z}_2$ truncation in this  case in the SO5 one, while the remaining 1/2-BPS space-filling branes in the truncated theory are the SO9 branes and the branes in Table \ref{6A6BbranesN20}. Indeed, one can for instance show that the mixed-symmetry potentials that are coupled to 5-branes in table 9 of \cite{Bergshoeff:2017gpw} are mapped to the SO9 potentials in Table \ref{6A6Bbranes} and to the potentials in Table \ref{6A6BbranesN20} by performing a single T-duality using eq. \eqref{allbranesallalphasrule}.

In the case of the unique 6B truncation, as Table \ref{numberofbraneshalfmax} shows, all the 80 5-branes in the ${\bf 320}$, {\it i.e.} all the branes in Table \ref{6A6Bbranes}, are 1/2-BPS states of the truncated theory. The truncation has a natural geometric interpretation as the 
untwisted sector of IIB reduced on  $T^4/\mathbb{Z}_2$, and indeed it is easy to show that the 5-branes listed in table 10 of \cite{Bergshoeff:2017gpw}, where the orbifold analysis was performed in detail,  are exactly the branes in Table \ref{6A6Bbranes}.

\vskip .5cm
\noindent
${\bf D=5}$: \ As Table \ref{numberofbranesmax} shows,
there are 432 1/2-BPS 4-branes in five dimensions, which are the long weights of the $\overline{\textbf{1728}}$ of E$_{6(6)}$. There are 27 different $\mathbb{Z}_2$ truncations to the half-maximal theory coupled to five vector multiplets, with symmetry $\mathbb{R}^+ \times {\rm SO}(5,5)$, and in each of these we expect 16 space-filling branes  to preserve the same supersymmetry of the truncation \cite{Bergshoeff:2013sxa}.

As in the 6A case, we can start from the SO9 truncation and then obtain all the others using dualities. The 16 4-branes that preserve the same supersymmetry of the SO9 truncation couple to the potentials $F_{5 \, x^1 ...x^5}$, $F_{5 \, x^1 ...x^5, x^px^q,x^px^q}$ and  $F_{5\, x^1 ...x^5, x^px^qx^rx^s, x^px^qx^rx^s}$, that correspond to the long weights of the ${\bf \overline{126}}$ of the ${\rm SO}(5,5)$ symmetry of the truncated theory. By S-duality, these are mapped to the O9 truncation, and then by T-dualities the latter is mapped to 10 O7$_{x^ix^jx^k}$ truncations and 5 O5$_{x^i}$ truncations. The O5 truncations can then be mapped by S-duality to 5 SO5$_{x^i}$ truncations. Finally, we want to derive what happens if one performs two T-dualities on the SO5$_{x^i}$ branes. Using eq. \eqref{allbranesallalphasrule}, one can show that if both T-dualities are along directions different from $x^i$, then the SO5$_{x^i}$ branes are mapped into themselves. On the other hand, if one T-duality is along $x^i$, then the set of branes that one ends up with is always the same regardless of which direction one chooses for the other T-duality transformation. We call the corresponding truncation T$^2$SO5$_{x^i}$. The full result of this analysis is summarised in Table \ref{maximalbranesD5}. 

As in six dimensions, for each truncation there are additional space-filling branes that survive the projection and are 1/2-BPS states of the truncated theory. As Table \ref{numberofbraneshalfmax} shows, 
these branes are in the ${\bf 320} \oplus {\bf 210}$ of ${\rm SO}(5,5)$, and they are 80 for each representation. To determine what these branes are, we can again consider the SO9 truncation, in which case the ${\rm SO}(5,5)$ symmetry is perturbative. From table 5 of ref. \cite{Bergshoeff:2012ex}
we find that for this truncation the branes in the ${\bf 210}$ have $\alpha=-2$ and those in the ${\bf 320}$ have $\alpha=-4$. Therefore these branes are all the SO5$_{x^i}$ and T$^2$SO5$_{x^i}$ branes. Similarly, for different truncations one can determine the branes that are 1/2-BPS states of the truncated theory using dualities. As a particular case, the T$^2$SO5$_{x^i}$ truncation is the geometric truncation corresponding to IIB on $T^4/\mathbb{Z}_2 \times S^1$, where $x^i$ is the circle direction.

\begin{table}[t!]
\begin{center}
\scalebox{0.85}{
\begin{tabular}{|c|c||c|c|c|}
\hline $D=4$ truncations & \# &  potentials   & brane components & \#  branes \\
 \hline \hline   \rule[-2mm]{0mm}{2mm}    \multirow{4}{*}{\textbf{O9}} &\multirow{4}{*}{1}  & $C_{10}$   &  $C_{4 \, x^1 ...x^6}$  & 1 
 \\
& & $E_{10,2,2}$ & $E_{4 \, x^1 ...x^6, x^px^q, x^px^q} $ & 15  
 \\
& & $G_{10,4,4}$ & $G_{4\, x^1 ...x^6,x^px^qx^rx^s,x^px^qx^rx^s}$ & 15  
 \\
& & $I_{10,6,6}$ & $I_{4 \, x^1 ...x^6 ,x^1 ...x^6,x^1 ...x^6}$ & 1  
  \\
\hline   \rule[-2mm]{0mm}{2mm} \multirow{4}{*}{\textbf{SO9}} & \multirow{4}{*}{1} & $F_{10}$  & $F_{4 \, x^1...x^6}$ & 1  \\
 & & $F_{10,2,2}$  & $F_{4 \, x^1 ...x^6,x^px^q,x^px^q}$ & 15    
   \\
&  & $F_{10,4,4}$  & $F_{4 \, x^1 ...x^6,x^px^qx^rx^s,x^px^qx^rx^s}$ & 15 
 \\
&   & $F_{10,6,6}$  & $F_{4 \, x^1 ...x^6,x^1 ...x^6,x^1 ...x^6}$ & 1 
   \\

\hline   \rule[-2mm]{0mm}{2mm} \multirow{8}{*}{$\textbf{O7}_{x^ix^jx^kx^l}$} &  \multirow{8}{*}{15} & $C_{8}$  & $C_{4 \, x^ix^jx^kx^l}$  & 1   
\\
& & $E_{8}$  & $E_{4\, x^ix^jx^kx^l}$ & 1   
 \\
 &  & $E_{9,2,1}$  & $E_{4\, x^i x^jx^k x^l x^p,x^ix^p,x^i}$ & 8   
  \\
& & $E_{10,4,2}$  & $E_{4 \, x^1 ...x^6,x^ix^jx^px^q,x^ix^j}$ & 6    
 \\
& & $G_{10,4,2}$  & $G_{4 \, x^1 ...x^6,x^ix^jx^px^q,x^ix^j}$ & 6  
 \\
&
 & $G_{10,5,4,1}$  & $G_{4 \, x^1 ...x^6,x^ix^jx^kx^px^q,x^ix^jx^kx^p,x^p}$ & 8    
 \\
   & & $G_{10,6,6,2}$  & $G_{4 \, x^1 ...x^6,x^1 ...x^6,x^1 ...x^6,x^px^q}$ & 1    
 \\
&   & $I_{10,6,6,2}$  & $I_{4 \, x^1 ...x^6,x^1 ...x^6,x^1...x^6,x^px^q}$ & 1    
 \\
\hline   \rule[-2mm]{0mm}{2mm} \multirow{8}{*}{$\textbf{O5}_{x^ix^j}$} &  \multirow{8}{*}{15}  & $C_{6}$  & $C_{4\, x^ix^j}$ & 1 
\\
& & $E_{8,2}$ & $E_{4\, x^ix^jx^px^q,x^px^q}$ & 6  
\\
& &  $E_{9,4,1}$  & $E_{4 \, x^ix^jx^px^qx^r,x^ix^px^qx^r,x^i}$ & 8  
 \\
& & $E_{10,6,2}$  & $E_{4\, x^1 ...x^6,x^1 ...x^6,x^ix^j}$ & 1  
 \\
& & $G_{10,4}$  &  $G_{4\, x^1 ...x^6,x^px^qx^rx^s}$ & 1  
\\
& & $G_{10,5,2,1}$  & $G_{4\, x^1 ...x^6,x^ix^px^qx^rx^s,x^ix^p,x^p}$ & 8    
\\
& & $G_{10,6,4,2}$  & $G_{4 \, x^1 ...x^6,x^1 ...x^6,x^ix^jx^px^q,x^px^q}$ & 6   
\\
& & $I_{10,6,6,4}$  & $I_{4 \, x^1 ...x^6,x^1 ...x^6,x^1 ...x^6,x^px^qx^rx^s}$ & 1   
\\

\hline   \rule[-2mm]{0mm}{2mm} \multirow{4}{*}{$\textbf{O3}$} &  \multirow{4}{*}{1} & $C_{4}$  & $C_{4}$ & 1 \\
& & $E_{8,4}$ & $E_{4 \, x^px^qx^rx^s,x^px^qx^rx^s}$ & 15 
\\
& &  $G_{10,6,2,2}$ & $G_{4 \, x^1 ...x^6, x^1 ...x^6,x^px^q,x^px^q}$ & 15     
 \\
& & $I_{10,6,6,6}$ & $I_{4 \, x^1 ...x^6,x^1 ...x^6,x^1...x^6, x^1 ...x^6}$ & 1  
\\
\hline   \rule[-2mm]{0mm}{2mm} \multirow{8}{*}{$\textbf{SO5}_{x^ix^j}$}  &  \multirow{8}{*}{15}& $D_{6}$  & $D_{4 \, x^ix^j}$ & 1   
\\
& & $D_{8,2}$ & $D_{4\, x^ix^jx^px^q,x^px^q}$ & 6   
\\
& & $D_{10,4}$  &  $D_{4 \, x^1 ...x^6,x^p x^q x^r x^s}$ & 1  
 \\
& & $F_{9,4,1}$  & $F_{4\, x^ix^jx^px^qx^r,x^ix^px^qx^r,x^i}$ &  8   
 \\
 & & $F_{10,5,2,1}$  & $F_{4 \, x^1 ...x^6,x^i x^p x^q x^r x^s ,x^ix^p,x^p}$ & 8 
\\
& & $H_{10,6,2}$  & $H_{4 \, x^1 ...x^6,x^1 ...x^6,x^i x^j}$ & 1 
\\
& & $H_{10,6,4,2}$  & $H_{4 \, x^1 ...x^6,x^1 ...x^6,x^ix^jx^px^q,x^px^q}$ & 6 
\\
& &  $H_{10,6,6,4}$  & $H_{4 \, x^1 ...x^6,x^1 ...x^6,x^1 ...x^6,x^p x^q x^r x^s}$ & 1  
\\
\hline   \rule[-2mm]{0mm}{2mm} \multirow{8}{*}{$\textbf{T}^2\textbf{SO5}_{x^ix^j}$} & \multirow{8}{*}{15} & $D_{7,1}$  & $D_{4 \, x^i x^j x^p,x^p}$ & 4    
\\
& & $D_{9,3}$ &  $D_{4\, x^i x^j x^px^qx^r,x^px^qx^r}$ & 4  
 \\
&  & $F_{9,3}$   & $F_{4\, x^ix^jx^px^qx^r, x^px^q x^r}$ & 4   
 \\
& & $F_{9,5,2}$ & $F_{4 \, x^ix^jx^px^qx^r,x^ix^jx^px^qx^r, x^ix^j}$ & 4  
\\
& & $F_{10,4,1,1}$ & $F_{4 \, x^1 ...x^6,x^p x^q x^r x^s,x^p,x^p}$ & 4   
\\
& & $F_{10,6,3,1}$ & $F_{4 \, x^1 ...x^6,x^1 ...x^6,x^ix^jx^p,x^p}$ & 4   
\\
& & $H_{10,6,3,1}$ & $H_{4 \, x^1 ... x^6 ,x^1 ...x^6,x^i x^jx^p,x^p}$ & 4   
\\
& & $H_{10,6,5,3}$ & $H_{4 \, x^1 ...x^6,x^1 ...x^6,x^ix^jx^px^qx^r,x^px^qx^r}$ & 4  
\\
 \hline

\end{tabular}
}
\caption{\sl The $\mathbb{Z}_2$ truncations of the maximal theory in $D=4$.}
\label{maximaltruncationsbranesD4}
\end{center}
\end{table}

\vskip .5cm
\noindent
${\bf D=4}$: \
From Table \ref{numberofbranesmax} we read that the maximal theory in $D=4$ possesses 63 different $\mathbb{Z}_2$ truncations to the half-maximal theory, whose symmetry is ${\rm SL}(2,\mathbb{R}) \times {\rm SO}(6,6)$. There are 2016 space-filling branes belonging to the ${\bf 8645}$ of ${\rm E}_{7(7)}$, and the degeneracy for each truncation is 32. The mixed-symmetry potentials that couple to the 3-branes that preserve the same supersymmetry of the SO9 truncation are $F_{10}$, $F_{10,2,2}$, $F_{10,4,4}$ and $F_{10,6,6}$, and the 32 brane components are the long weights of the ${\bf (1,\overline{462})}$ representation of the symmetry of the truncated theory. By S and T dualities, one can determine all the branes that preserve the same supersymmetry of the truncation for each of the 63 truncations. The result is summarised in Table \ref{maximaltruncationsbranesD4}. 

We can also determine the  3-branes that in each truncation are not projected out and become 1/2-BPS states of the truncated theory. These branes are in the ${\bf (3,495)} \oplus {\bf (1,2079)}$ of the  ${\rm SL}(2,\mathbb{R}) \times {\rm SO}(6,6)$ symmetry of the truncated theory.  In particular, in the case of the SO9 truncation one finds (see {\it e.g.} table 6 of ref. \cite{Bergshoeff:2012ex})  that there are 240 $\alpha=-2$ and 240 $\alpha=-6$ branes in the ${\bf 495}$ and 480 $\alpha=-4$ branes in the ${\bf 2079}$ of ${\rm SO}(6,6)$. These are all the SO5$_{x^i x^j}$ and T$^2$SO5$_{x^i x^j}$ branes. Using S and T dualities one can determine the 960 3-branes that are 1/2-BPS states of the truncated theory for all the other truncations. The truncations T$^2$SO5$_{x^i x^j}$ are  identified with the geometric compactifications of IIB on $T^4/\mathbb{Z}_2 \times T^2$, where $x^ix^j$ are the two torus directions.

\vskip .5cm
\noindent
${\bf D=3}$: \
Finally we consider the three-dimensional case. In the maximal theory there are 17280 space-filling 1/2-BPS 2-branes belonging to the ${\bf 147250}$ of ${\rm E}_{8(8)}$. The $\mathbb{Z}_2$ truncation leads to the half-maximal theory with symmetry ${\rm SO}(8,8)$. There are 135 different truncations, and for each truncation there are 128 1/2-BPS 2-branes preserving the same supersymmetry of the truncation. To determine these branes in each truncation, one first has to further decompose ${\rm SO}(8,8)$ as $\mathbb{R}^+ \times {\rm SO}(7,7)$. In the case of the SO9 truncation, as usual the ${\rm SO}(7,7)$ symmetry is perturbative. The $\alpha=-4$ potentials $F_{10}$, $F_{10,2,2}$, $F_{10,4,4}$ and $F_{10,6,6}$ give rise to 64 2-branes that are the long weights of  the ${\bf \overline{1716}}$ of ${\rm SO}(7,7)$. This representation embeds in the ${\bf 6435}$ of ${\rm SO}(8,8)$, which  also contains the ${\bf {1716}}$ of ${\rm SO}(7,7)$, whose 64 long weights correspond to the $\alpha=-8$ potentials $J_{10,7,7,1,1}$, $J_{10,7,7,3,3}$, $J_{10,7,7,5,5}$ and $J_{10,7,7,7,7}$.These $\alpha=-4 $ and $\alpha=-8$ branes together give the whole set of 128 branes that preserve the supersymmetry of the SO9 truncations. 
By performing all possible S and T duality transformations, one then determines all the other truncations. The result is summarised in Table \ref{maximaltruncationsbranesD3}. 

As in any other dimension below seven, there are additional space-filling branes that are preserved by the truncation and become 1/2-BPS states of the truncated theory. These branes are 8960 and correspond to the long weights of the ${\bf 60060}$ of ${\rm SO}(8,8)$. In the case of the SO9 truncation, this representation contains 560  $\alpha=-2$  and 560 $\alpha=-10$ branes in the ${\bf 1001}$ of  
${\rm SO}(7,7)$, 2240 $\alpha=-4$ and 2240 $\alpha=-8$ in the ${\bf 11648}$ of ${\rm SO}(7,7)$, and finally 3360 $\alpha=-6$ branes in the ${\bf 24024}$ of ${\rm SO}(7,7)$ (see table 7 of ref. \cite{Bergshoeff:2012ex}). These are all the  SO5$_{x^i x^j x^k}$ and T$^2$SO5$_{x^i x^j x^k}$ branes in Table \ref{maximaltruncationsbranesD3}. We end the analysis of the maximal theories by observing that the truncations T$^2$SO5$_{x^i x^j x^k}$ are  identified with the geometric compactifications of IIB on $T^4/\mathbb{Z}_2 \times T^3$, where $x^ix^j x^k $ are the three torus directions.

\begin{table}[t!]
\begin{center}
\scalebox{0.732}{
\begin{tabular}{|c|c||c|c|c|}
\hline $D=3$ truncations & \# &  potentials   & brane components & \#  branes  \\ 
\hline \hline   \rule[-2mm]{0mm}{2mm}    \multirow{8}{*}{\textbf{O9}} &\multirow{8}{*}{1} & $C_{10}$   &  $C_{3 \, x^1 ...x^7}$  & 1 
 \\
& & $E_{10,2,2}$ & $E_{3 \, x^1 ...x^7, x^px^q, x^px^q} $ & 21 
 \\
&  & $G_{10,4,4}$ & $G_{3 \, x^1 ...x^7,x^px^qx^rx^s,x^px^qx^rx^s}$ & 35 
 \\
&  & $G_{10,7,7,1,1}$  & $G_{3 \, x^1 ...x^7 ,x^1 ...x^7,x^1 ...x^7,x^p,x^p}$ & 7 
   \\
&   & $I_{10,6,6}$ & $I_{3 \, x^1 ...x^7,x^p x^q x^r x^s x^t x^u,x^p x^q x^r x^s x^t x^u}$ & 7  
  \\
&     & $I_{10,7,7,3,3}$  & $I_{3\, x^1 ...x^7,x^1 ...x^7,x^1 ...x^7,x^px^qx^r,x^p x^q x^r}$ & 35 
   \\
&       & $K_{10,7,7,5,5}$  & $K_{3 \, x^1 ...x^7,x^1 ...x^7,x^1 ...x^7,x^p x^q x^r x^s x^t,x^p x^q x^r x^s x^t}$ & 21 
   \\
&      & $M_{10,7,7,7,7}$  & $M_{3 \, x^1 ...x^7,x^1 ...x^7,x^1 ...x^7,x^1 ...x^7,x^1 ...x^7}$ & 1 
   \\
\hline   \rule[-2mm]{0mm}{2mm} \multirow{8}{*}{\textbf{SO9}} & \multirow{8}{*}{1}& $F_{10}$  & $F_{3 \, x^1 ...x^7}$ & 1  \\
 & &  $F_{10,2,2}$  & $F_{3 \, x^1 ...x^7,x^px^q,x^px^q}$ & 21    
   \\
 & & $F_{10,4,4}$  & $F_{3 \, x^1 ...x^7,x^px^qx^rx^s,x^px^qx^rx^s}$ & 35  
 \\
 &  & $F_{10,6,6}$  & $F_{3\, x^1 ...x^7,x^p x^q x^r x^s x^t x^u, x^p x^q x^r x^s x^t x^u}$ & 7 
   \\
&      & $J_{10,7,7,1,1}$  & $J_{3 \, x^1 ...x^7,x^1 ...x^7,x^1 ...x^7,x^p,x^p}$ & 7 
   \\
&      & $J_{10,7,7,3,3}$  & $J_{3 \, x^1 ...x^7,x^1 ...x^7,x^1 ...x^7,x^px^qx^r,x^p x^q x^r}$ & 35 
   \\
&        & $J_{10,7,7,5,5}$  & $J_{3\, x^1 ...x^7,x^1 ...x^7,x^1 ...x^7,x^p x^q x^r x^s x^t,x^p x^q x^r x^s x^t}$ & 21 
   \\
&       & $J_{10,7,7,7,7}$  & $J_{3 \, x^1 ...x^7,x^1 ...x^7,x^1 ...x^7,x^1 ...x^7,x^1 ...x^7}$ & 1 
   \\
\hline   \rule[-2mm]{0mm}{2mm} \multirow{18}{*}{$\textbf{O7}_{x^ix^j x^k x^l x^m}$} &  \multirow{18}{*}{21}  & $C_{8}$  & $C_{3\, x^ix^jx^kx^l x^m}$  
& 1 \\
& & $E_{8}$  & $E_{3\, x^ix^jx^kx^lx^m}$ & 1 
 \\
&   & $E_{9,2,1}$  & $E_{3\, x^i x^jx^kx^l x^m x^p,x^ix^p,x^i}$ & 10 
  \\
& & $E_{10,4,2}$  & $E_{3\, x^1 ...x^7,x^ix^jx^px^q,x^ix^j}$ & 10  
 \\ 
&   & $G_{9,6,5}$  & $G_{3\, x^ix^jx^kx^l xm5 x^p,x^ix^j x^k x^l  x^m x^p, x^i x^jx^kx^l x^m}$ & 2   
 \\
& & $G_{10,4,2}$  & $G_{3 \, x^1 ...x^7,x^ix^jx^p x^q,x^ix^j}$ & 10   
 \\
& & $G_{10,5,4,1}$  & $G_{3\, x^1 ...x^7,x^ix^jx^kx^p x^q,x^ix^jx^kx^p,x^p}$ & 20    
 \\
&  & $G_{10,6,6,2}$  & $G_{3\, x^1 ...x^7,x^ix^jx^kx^lx^p x^q,x^ix^jx^kx^lx^px^q,x^px^q}$ & 5   
 \\
&   & $G_{10,7,5,1,1}$  & $G_{3\, x^1 ...x^7,7,x^ix^j x^k x^l x^m,x^i, x^i}$ & 5      
 \\
  &   & $I_{10,6,6,2}$  & $I_{3\, x^1 ...x^7,x^ix^jx^kx^lx^p x^q,x^ix^jx^kx^lx^p x^q,x^p x^q}$ & 5   
 \\
 &  & $I_{10,7,7,2,1,1}$  & $I_{3 \, x^1 ...x^7,x^1 ...x^7,x^1 ...x^7,x^px^q,x^p,x^p}$ & 2    
 \\ 
&  & $I_{10,7,7,5,3}$  & $I_{3 \, x^1 ...x^7,x^1 ...x^7, x^1 ...x^7,x^ix^jx^kx^p x^q,x^i x^j x^k}$ & 10  
 \\
   &   & $I_{10,7,5,1,1}$  & $I_{10,7,x^1x^2x^3x^4x^5,x^i, x^i}$ & 5 
 \\
&    & $I_{10,7,6,3,2}$  & $I_{3\, x^1 ...x^7,x^1 ...x^7,x^ix^jx^kx^lx^m x^p,x^ix^j x^p, x^i x^j}$ & 20      
 \\
&  & $K_{10,7,7,6,5,1}$  & $K_{3\, x^1 ...x^7,x^1 ...x^7,x^1 ...x^7,x^i x^j x^k x^l x^p x^q, x^ix^j x^k x^l x^p, x^p}$ & 10   
 \\
 &   & $K_{10,7,7,5,3}$  & $K_{3\, x^1 ...x^7,x^1 ...x^7,x^1 ...x^7,x^ix^jx^kx^px^q,x^i x^j x^k}$ & 10    
 \\
 &  & $K_{10,7,7,7,7,2}$  & $K_{3\, x^1 ...x^7,x^1 ...x^7,x^1 ...x^7,x^1 ...x^7,x^1 ...x^7, x^px^q}$ & 1 
 \\ 
&   & $M_{10,7,7,7,7,2}$  & $I_{3\, x^1 ...x^7,x^1 ...x^7,x^1 ...x^7,x^1 ...x^7,x^1 ...x^7, x^px^q}$ & 1  
 \\
\hline   \rule[-2mm]{0mm}{2mm} \multirow{20}{*}{$\textbf{O5}_{x^i x^jx^k}$}  &  \multirow{20}{*}{35} & $C_{6}$  & $C_{3\, x^ix^jx^k}$ & 1 
\\
& & $E_{8,2}$ & $E_{3 \, x^ix^jx^kx^px^q,x^px^q}$ & 6  
\\
& & $E_{9,4,1}$  & $E_{3\, x^i x^j x^k x^px^qx^r,x^ix^px^qx^r,x^i}$ & 12  
 \\
& & $E_{10,6,2}$  & $E_{3 \, x^1 ...x^7,x^ix^j x^p x^q x^r x^s,x^ix^j}$ & 3 
 \\
& & $G_{9,6,3}$  & $G_{3 \, x^1 ...x^7, x^i x^j x^k x^p x^q x^r, x^i x^j x^k}$ & 4   
\\
& & $G_{10,4}$  &  $G_{3\, x^1 ...x^7,x^px^qx^rx^s}$ & 1  
\\
& & $G_{10,5,2,1}$  & $G_{3\, x^1 ...x^7,x^ix^px^qx^rx^s,x^ix^p,x^p}$ & 12  
\\
& & $G_{10,6,4,2}$  & $G_{3 \, x^1 ...x^7,x^i x^j x^p x^q x^r x^s,x^ix^jx^px^q,x^px^q}$ & 18 
\\
& & $G_{10,7,6,3}$  & $G_{3 \, x^1 ...x^7,x^1 ...x^7,x^i x^j x^k x^p x^q x^r,x^px^q x^r}$ & 4   
\\
& & $G_{10,7,3,1,1}$  & $G_{3 \, x^1 ...x^7,x^1 ...x^7,x^ix^jx^k,x^i,x^i}$ & 3   
\\
& & $I_{10,6,6,4}$  & $I_{3\, x^1 ...x^7,x^i x^j x^p x^q x^r x^s, x^i x^j x^p x^q x^r x^s,x^p x^q x^r x^s}$ & 3  
\\
& & $I_{10,7,7,4,1,1}$  & $I_{3\, x^1 ...x^7,x^1 ...x^7,x^1 ...x^7, x^p x^q x^r x^s, x^p, x^p}$ & 4   
\\
& & $I_{10,7,7,7,3}$  & $I_{3\, x^1 ...x^7,x^1 ...x^7,x^1 ...x^7,x^1 ...x^7,x^i x^j x^k}$ & 1  
\\
& & $I_{10,7,6,5,2}$  & $I_{3\, x^1 ...x^7,x^1 ...x^7,x^i x^j x^k x^p x^q x^r, x^i x^j x^p x^q x^r,x^i x^j}$ & 12  
\\
& & $I_{10,7,5,3,1}$  & $I_{3\, x^1 ...x^7,x^1 ...x^7,x^i x^j x^k x^p x^q, x^i x^p x^q, x^i}$ & 18 
\\
& & $I_{10,7,4,1}$  & $I_{3\, x^1 ...x^7,x^1 ...x^7,x^i x^j x^k x^p, x^p}$ & 4   
\\
& & $K_{10,7,7,7,5,2}$  & $K_{3\, x^1 ...x^7,x^1 ...x^7,x^1 ...x^7,x^1 ...x^7, x^i x^j x^k x^p x^q, x^p x^q}$ & 6 
\\
& & $K_{10,7,7,5,1}$  & $K_{3\, x^1 ...x^7,x^1 ...x^7,x^1 ...x^7, x^i x^p x^q x^r x^s, x^i}$ & 3  
\\
& & $K_{10,7,7,6,3,1}$  & $K_{3\, x^1 ...x^7,x^1 ...x^7,x^1...x^7,x^i x^j x^p x^q x^r x^s, x^i x^j x^p, x^p}$ & 12   
\\
& & $M_{10,7,7,7,7,4}$  & $M_{3\, x^1 ...x^7,x^1 ...x^7,x^1 ...x^7,x^1 ...x^7,x^1 ...x^7, x^p x^q x^r x^s}$ & 1   
\\
\hline  
 \end{tabular}
}

\end{center}
\end{table}

\begin{table}[t!]
\begin{center}
\scalebox{0.736}{
\begin{tabular}{|c|c||c|c|c|}
\hline $D=3$ truncations & \# &  potentials   & brane components & \#  branes  \\ \hline \hline  
 \rule[-2mm]{0mm}{2mm} \multirow{14}{*}{$\textbf{O3}_{x^i}$}  &  \multirow{14}{*}{7} & $C_{4}$  & $C_{3\, x^i}$ & 1 \\
& & $E_{8,4}$ & $E_{3 \, x^i x^p x^q x^r x^s,x^p x^q x^r x^s}$ & 15 
\\
& & $E_{9,6,1}$ & $E_{3\, x^ix^p x^q x^r x^s x^t,x^i x^p x^q x^r x^s x^t,x^i }$ & 6  
\\
& & $G_{10,6,2,2}$ & $G_{3\, x^1 ...x^7,x^p x^q x^r x^s x^t x^u,x^px^q,x^px^q}$ & 15     
 \\
&  & $G_{10,7,4,3}$ & $G_{3\, x^1 ...x^7,x^1 ...x^7,x^i x^p x^q x^r, x^p x^q x^r}$ & 20     
 \\ 
&  & $G_{9,6,1}$ & $G_{3\, x^i x^p x^q x^r x^s x^t,x^i x^p x^q x^r x^s x^t,x^i}$ & 6     
 \\
 &   & $G_{10,7,1,1,1}$ & $G_{3\, x^1 ...x^7,x^1 ...x^7,x^i, x^i, x^i}$ & 1     
 \\
& & $I_{10,6,6,6}$ & $I_{3\, x^1 ...x^7 ,x^p x^q x^r x^s x^t x^u, x^p x^q x^r x^s x^t x^u, x^p x^q x^r x^s x^t x^u}$ & 1  
\\
& & $I_{10,7,7,6,1,1}$ & $I_{3\, x^1 ...x^7,x^1 ...x^7,x^1 ...x^7,x^p x^q x^r x^s x^t x^u, x^p, x^p}$ & 6  
\\
& & $I_{10,7,5,5,1}$ & $I_{3 \, x^1 ...x^7,x^1 ...x^7, x^i x^p x^q x^r x^s, x^i x^p x^q x^r x^s, x^i}$ & 15  
\\
& & $I_{10,7,4,3}$ & $I_{3\, x^1 ...x^7,x^1 ...x^7, x^i x^p x^q x^r, x^p x^q x^r }$ & 20  
\\
& & $K_{10,7,7,6,1,1}$ & $K_{3\, x^1 ...x^7,x^1 ...x^7,x^1 ...x^7, x^p x^q x^r x^s x^t x^u ,x^p, x^p}$ & 6  
\\ 
&  & $K_{10,7,7,7,3,2}$ & $K_{3\, x^1 ...x^7,x^1 ...x^7,x^1 ...x^7,x^1 ...x^7,x^i x^p x^q, x^p x^q}$ & 15 
\\
&    & $M_{10,7,7,7,7,6}$  & $M_{3\, x^1 ...x^7,x^1 ...x^7,x^1 ...x^7,x^1 ...x^7,x^1 ...x^7,x^p x^q x^r x^s x^t x^u}$ & 1 
   \\
\hline   \rule[-2mm]{0mm}{2mm} \multirow{20}{*}{$\textbf{SO5}_{x^i x^jx^k}$}&  \multirow{20}{*}{35}  &  $D_{6}$  & $D_{3\, x^ix^jx^k}$ & 1 
\\
& & $D_{8,2}$ & $D_{3\, x^i x^jx^kx^px^q,x^px^q}$ & 6  
\\
& & $D_{10,4}$  &  $D_{3\, x^1 ...x^7,x^px^qx^rx^s}$ & 1  
\\
& & $F_{9,4,1}$  & $F_{3\, x^ix^jx^kx^px^qx^r,x^ix^px^qx^r,x^i}$ & 12  
 \\
&  & $F_{9,6,3}$  & $F_{3 \, x^i x^j x^k x^p x^q x^r, x^i x^j x^k x^p x^q x^r, x^i x^j x^k}$ & 4   
\\
&  & $F_{10,5,2,1}$  & $F_{3 \, x^1 ...x^7,x^ix^p x^qx^rx^s,x^ix^p,x^p}$ & 12    
\\
& 
& $F_{10,7,4,1}$  & $F_{3 \, x^1 ...x^7,x^1 ...x^7,x^i x^j x^k x^p, x^p}$ & 4   
\\
& 
& $H_{10,6,2}$  & $H_{3\, x^1 ...x^7,x^ix^j x^p x^q x^r x^s,x^ix^j}$ & 3  
 \\
& & $H_{10,6,4,2}$  & $H_{3\, x^1 ...x^7,x^i x^j x^p x^q x^r x^s,x^ix^jx^px^q,x^px^q}$ & 18   
\\
& 
& $H_{10,6,6,4}$  & $H_{3\, x^1 ...x^7,x^i x^j x^p x^q x^r x^s, x^i x^j x^p x^q x^r x^s,x^p x^q x^r x^s}$ & 3   
\\
& 
& $H_{10,7,3,1,1}$  & $H_{3\, x^1 ...x^7,x^1 ...x^7,x^ix^jx^k,x^i,x^i}$ & 3   
\\
& & $H_{10,7,5,3,1}$  & $H_{3\, x^1 ...x^7,x^1 ...x^7,x^i x^j x^k x^p x^q, x^i x^p x^q, x^i}$ & 18   
\\
& & $H_{10,7,7,5,1}$  & $H_{3\, x^1 ...x^7,x^1 ...x^7,x^1 ...x^7, x^i x^p x^q x^r x^s, x^i}$ & 3   
\\
& 
& $J_{10,7,6,3}$  & $J_{3\, x^1 ...x^7,x^1 ...x^7,x^i x^j x^k x^p x^q x^r,x^px^q x^r}$ & 4  
\\
&
& $J_{10,7,7,4,1,1}$  & $J_{3\, x^1 ...x^7,x^1 ...x^7,x^1 ...x^7, x^p x^q x^r x^s, x^p, x^p}$ & 4   
\\
& & $J_{10,7,6,5,2}$  & $J_{3\, x^1 ...x^7,x^1 ...x^7,x^i x^jx^k x^p x^q x^r, x^i x^j x^p x^q x^r,x^i x^j}$ & 12   
\\
&
& $J_{10,7,7,6,3,1}$  & $J_{3\, x^1 ...x^7,x^1 ...x^7,x^1 ...x^7,x^i x^j x^p x^q x^r x^s, x^i x^j x^p, x^p}$ & 12   
\\
& 
& $L_{10,7,7,7,3}$  & $L_{3\, x^1 ...x^7,x^1 ...x^7,x^1 ...x^7,x^1 ...x^7,x^i x^j x^k}$ & 1   
\\
& & $L_{10,7,7,7,5,2}$  & $L_{3\, x^1 ...x^7,x^1 ...x^7,x^1 ...x^7,x^1 ...x^7, x^i x^j x^k x^p x^q, x^p x^q}$ & 6   
\\
& & $L_{10,7,7,7,7,4}$  & $L_{3\, x^1 ...x^7,x^1 ...x^7,x^1 ...x^7,x^1 ...x^7,x^1 ...x^7, x^p x^q x^r x^s}$ & 1   
\\
\hline   \rule[-2mm]{0mm}{2mm} \multirow{16}{*}{$\textbf{T}^2\textbf{SO5}_{x^i x^jx^k}$} &  \multirow{16}{*}{35}  &  $D_{7,1}$  & $D_{3\, x^i x^jx^kx^p,x^p}$ & 4 
\\
&&  $D_{9,3}$ & $D_{3\, x^ix^jx^kx^px^q x^r,x^px^q x^r}$ & 4  
\\
& &  $F_{9,3}$  & $F_{3\, x^i x^j x^k x^p x^q x^r,x^p x^q x^r}$ & 4  
 \\
&  
 & $F_{9,5,2}$  & $F_{3\, x^i x^j x^k x^p x^q x^r, x^i x^j x^p x^q x^r,x^i x^j}$ & 12  
 \\
&  
  & $F_{10,4,1,1}$  & $F_{3\, x^1 ...x^7,x^p x^q x^r x^s, x^p, x^p}$ & 4  
 \\
&  
 & $F_{10,6,3,1}$  & $F_{3\, x^1 ...x^7,x^i x^j x^p x^q x^r x^s, x^i x^j x^p, x^p}$ & 12  
 \\
 & 
& $H_{10,6,3,1}$  & $H_{3\, x^1 ...x^7,x^i x^j x^p x^q x^r x^s, x^i x^j x^p, x^p}$ & 12  
 \\
&  
 & $H_{10,7,4,2,1}$  & $H_{3\, x^1 ...x^7,x^1 ...x^7,x^ix^jx^kx^p,x^i x^p,x^i}$ & 12  
 \\
& 
& $H_{10,6,5,3}$  & $H_{3\, x^1 ...x^7,x^i x^j x^p x^q x^r x^s,x^i x^j x^p x^q x^r, x^p x^q x^r}$ & 12   
\\
& 
& $H_{10,7,6,4,1}$  & $H_{3\, x^1 ...x^7,x^1 ...x^7,x^i x^j x^k x^p x^q x^r,x^i x^p x^q x^r,x^i}$ & 12   
\\
& 
& $J_{10,7,7,5,2,1}$  & $J_{3\, x^1 ...x^7,x^1 ...x^7,x^1 ...x^7,x^i x^p x^q x^r x^s,x^i x^p,x^p}$ & 12   
\\
& 
& $J_{10,7,7,7,4,1}$  & $J_{3\, x^1 ...x^7,x^1 ...x^7,x^1 ...x^7,x^1 ...x^7,x^i x^j x^k x^p,x^p}$ & 4   
\\
& 
& $J_{10,7,6,4,1}$  & $J_{3\, x^1 ...x^7,x^1 ...x^7,x^i x^j x^k x^p x^q x^r, x^i x^p x^q x^r, x^i}$ & 12   
\\
& 
& $J_{10,7,6,6,3}$  & $J_{3\, x^1 ...x^7,x^1 ...x^7,x^i x^j x^k x^p x^q x^r,x^i x^j x^k x^p x^q x^r, x^i x^j x^k }$ & 4   
\\
& 
& $L_{10,7,7,7,4,1}$  & $L_{3\, x^1 ...x^7,x^1 ...x^7,x^1 ...x^7,x^1 ...x^7, x^i x^j x^k x^p, x^p}$ & 4   
\\
& 
& $L_{10,7,7,7,6,3}$  & $L_{3\, x^1 ...x^7,x^1 ...x^7,x^1 ...x^7,x^1 ...x^7,x^ix^jx^kx^p x^q x^r, x^p x^q x^r}$ & 4   
\\
 \hline
 \end{tabular}
}
\caption{\sl The $\mathbb{Z}_2$ truncations of the maximal theory in $D=3$.}
\label{maximaltruncationsbranesD3}

\end{center}
\end{table}

\subsection{From half-maximal to quarter-maximal supergravity}
\label{Fromhalfmaximaltoquartermaximalsupergravity}

${\bf D=6}$: \ Both the $\mathcal{N}=(1,1)$ (or 6A) and the $\mathcal{N}=(2,0)$ (or 6B) theories admit $\mathbb{Z}_2$ truncations to the $\mathcal{N}=(1,0)$ supergravity theory coupled to one tensor  multiplet and four hypermultiplets, with  the hyper-scalars parametrising the coset manifold ${\rm SO}(4,4)/({\rm SO}(4) \times {\rm SO(4)}$. The truncation is unique in the case of the $\mathcal{N}=(1,1)$ theory because the global symmetry stays the same, while in the case of the $\mathcal{N}=(2,0)$ there are five different truncations. We now want to determine for any of these truncations  what are the branes  that preserve the same supersymmetry of the truncation.

We start considering the 6A theory. 
As Table \ref{numberofbraneshalfmax} shows, the 1/2-BPS space-filling branes of the 6A theory belong to the ${\bf 35}_{\rm V} \oplus {\bf 35}_{\rm S} \oplus {\bf 35}_{\rm C}$, and all preserve the same supersymmetry of the truncated theory. As already discussed in the previous subsection, in the case of the SO9 truncation to the half-maximal theory, the branes in the ${\bf 35}_{\rm V}$ are the SO5 branes 
while the branes in the ${\bf 35}_{\rm S} \oplus {\bf 35}_{\rm C}$ are the ``tensor'' branes in Table \ref{6A6BbranesN20}. As we mentioned already, the SO5 truncation of the maximal theory is the T-dual of the untwisted sector of IIA on $T^4/\mathbb{Z}_2$, and in this case the 1/2-BPS space-filling branes of the 6A theory are the SO9 and the tensor branes.  
In general, 
from the point of view of the maximal theory, the ${\bf 35}_{\rm V}$ comes from the ${\bf 320}$, while the other two representations come from the ${\bf \overline{126}}$.

In the case of the 6B theory, there are five different truncations. As mentioned already, the truncated theory can be viewed as the untwisted sector of IIB on $T^4/\mathbb{Z}_2$, and we know that on this theory the O9 and O5 orientifolds preserve the same supersymmetry.\footnote{This model was originally constructed in \cite{Pradisi:1988xd}. For the systematics of orientifold model building, see \cite{Bianchi:1990yu}.} We therefore label the corresponding truncation as $\text{O}9/\text{O}5$. By dualities one gets the other four truncations, and the final result is 
\begin{align}
\text{O}9/\text{O}5 \ , \hspace{0.2 cm} \text{SO}9/\text{SO}5 \ , \hspace{0.2 cm} 3 \times \text{O}7_{x^i x^j}/\text{O}7_{x^k x^l} \quad .
\end{align}
In each case, these branes are the long weights of the ${\bf 35}_{\rm V} \oplus {\bf 35}_{\rm V}$ representation of the ${\rm SO}(4,4)$ symmetry of the truncated theory.

\vskip .5cm
\noindent
${\bf D=5}$: \
The five-dimensional half-maximal theory admits five different $\mathbb{Z}_2$ truncations to the quarter-maximal theory with symmetry $(\mathbb{R}^+ )^2 \times {\rm SO}(4,4)$, containing gravity coupled to two vector multiplets and four hypermultiplets. As Table \ref{numberofbraneshalfmax} shows, the space-filling branes belong to the ${\bf 320} \oplus {\bf 210}$ of ${\rm SO}(5,5)$. A particular way of realising the half-maximal theory is by compactifying IIB on SO9, which corresponds to the heterotic truncation, so that the ${\rm SO}(5,5)$ symmetry is perturbative. In particular, the branes in the ${\bf 210}$ have $\alpha=-2$ while the branes in the ${\bf 320}$ have $\alpha=-4$ \cite{Bergshoeff:2012jb}. 
These branes  are the SO5$_{x^i}$ and T$^2$SO5$_{x^i}$ ones in Table \ref{maximalbranesD5}, where $i=1,...,5$ labels each of the five truncations. In other words, the ${\rm SO}(4,4)$ of the truncated theory is the one that fixes the particular $x^i$ that identifies the truncation. By looking at Table \ref{maximalbranesD5}, one can show that out of the ${\bf 320}$ one selects the ${\bf 35}_{\rm V} \oplus {\bf 35}_{\rm V}$, while out of the ${\bf 210}$ one selects the 
${\bf 35}_{\rm S} \oplus {\bf 35}_{\rm C}$.

One can also obtain the half-maximal theory by compactifying IIB on $(T^4/\mathbb{Z}_2) \times S^1$, corresponding to the truncation T$^2$SO5$_{y}$, where $y$ is the $S^1$ coordinate. It is instructive to show how using duality symmetries one can find the five truncations of the T$^2$SO5$_{y}$ theory starting from the truncations of the SO9 theory. Denoting the five internal coordinates as $x^i, y$, with $i=1,...,4$, we first perform an S-duality transformation to go to O9, then four T-dualities along the $x^i$  directions to go to O5$_{y}$, then again S-duality to go to SO5$_{y}$, and finally two T-dualities along $y$ and any other $x$ direction to go to T$^2$SO5$_{y}$. If one performs this chain of transformations on the truncations of the SO9 theory,
\begin{align}
\text{SO}9 : \qquad 4 \times \text{SO}5_{x^i}/\text{T}^2\text{SO}5_{x^i} \ , \hspace{0.2 cm} \text{SO}5_{y}/\text{T}^2\text{SO}5_{y} \quad , 
\end{align}
one finds the truncations of the $\text{T}^2\text{SO}5_{y}$ theory, which are
\begin{align}
\text{T}^2\text{SO}5_{y} : \qquad
\text{O}9/\text{O}5_{y} \ , \hspace{0.2 cm} 3 \times \text{O}7_{x^i x^j y}/\text{O}7_{x^k x^l y} \ , \hspace{0.2 cm} \text{SO}9/\text{SO}5_{y} \quad . 
\end{align}

\vskip .5cm
\noindent
${\bf D=4}$: \
The ${\rm SL}(2,\mathbb{R}) \times {\rm SO}(6,6)$ theory in four dimensions can be truncated to the ${\cal N}=2$ theory with symmetry ${\rm SL}(2,\mathbb{R})^3 \times {\rm SO}(4,4)$, describing gravity coupled to three vector multiplets and four hypermultiplets. There are 15 different truncations, corresponding to the different ways in which one can embed ${\rm SO}(4,4)$ into ${\rm SO}(6,6)$, and correspondingly the vector central charge belongs to the ${\bf 15}$ of the R-symmetry group U(4). As reported in Table \ref{numberofbraneshalfmax}, the 960 space-filling branes are the long weights of the ${\bf (3,495) \oplus (1, 2079)}$ \cite{Bergshoeff:2012jb}, and for each truncation one expects 64 branes to preserve the same supersymmetry of the truncation. 
To figure out what these branes are in each truncation, we focus on the particular half-maximal theory that results from the SO9 truncation of the maximal one. In this case the ${\rm SO}(6,6)$ symmetry is perturbative, and we expect only the branes with even $\alpha$ to survive as 1/2-BPS states. Therefore we get the 15 truncations
\begin{align}
\text{SO}9 : \qquad 15 \times \text{SO}5_{x^ix^j}/\text{T}^2\text{SO}5_{x^ix^j}   \quad , \label{truncationsSO9truncD=4}
\end{align}
where for any $i,j=1,...,6$  the SO5$_{x^ix^j}$ and T$^2$SO5$_{x^i x^j}$  branes preserve the same supersymmetry.
Fixing $i$ and $j$ to identify a specific truncation to the ${\cal N}=2$ theory, the symmetry ${\rm SO}(4,4)$ acts on the remaining coordinates, and by looking at Table \ref{maximaltruncationsbranesD4} one finds that the 64 branes that preserve the same supersymmetry of the truncated theory are
\begin{eqnarray} 
& & {\bf (3,495)} \rightarrow ({\bf 3,1,1,35}_{\rm S}) \oplus ({\bf 3,1,1,35}_{\rm C}) \nonumber \\
& & {\bf (1,2079)} \rightarrow ({\bf 1,3,1,35}_{\rm V}) \oplus ({\bf 1,1,3,35}_{\rm V}) \quad . \label{branesofN=4truncationD=4}
\end{eqnarray}

Apart from the branes in eq. \eqref{branesofN=4truncationD=4}, that preserve the same supersymmetry of the truncated theory, in four dimensions there are additional space-filling branes that survive the projection and are 1/2-BPS states of the ${\cal N}=2$ theory. These branes are the 384 long weights of the following representations \cite{Bergshoeff:2014lxa}:
\begin{equation}
{\bf (3,3,1,28) \oplus (3,1,3,28) \oplus (1,3,3,28)\oplus (1,1,1,350)} \quad . \label{1/2BPSbranesN=2}
\end{equation} 
In the particular case of the $\text{SO}5_{x^ix^j}/\text{T}^2\text{SO}5_{x^ix^j}$ truncation of the SO9-truncated theory, these all the SO5$_{x^kx^l}$ and T$^2$SO5$_{x^k x^l}$  branes with $k,l \neq i,j$.

Similarly to the five-dimensional case, one can obtain the half-maximal theory by compactifying IIB on $(T^4/\mathbb{Z}_2) \times T^2$, corresponding to the truncation T$^2$SO5$_{y^1 y^2}$, where $y^1$ and $y^2$ are the $T^2$ coordinates. As in $D=5$, one can show how using duality symmetries one can find the 15 truncations of the T$^2$SO5$_{y^1 y^2}$ half-maximal theory starting from the truncations of the SO9 half-maximal theory given in eq \eqref{truncationsSO9truncD=4}. Denoting the six internal coordinates as $x^i, y^m$, with $i=1,...,4$ and $m=1,2$, we first perform an S-duality transformation to go to O9, then four T-dualities along the $x^i$  directions to go to O5$_{y^1 y^2}$, then again S-duality to go to SO5$_{y^1 y^2}$, and finally two T-dualities along one $y$ and any other $x$ direction to go to T$^2$SO5$_{y^1 y^2}$. One finds
 \begin{align}
\text{T}^2\text{SO}5_{y^1 y^2} : \quad   &\text{O}9/\text{O}5_{y^1y^2} \ , \hspace{0.2 cm} \text{O7}_{x^1...x^4}/\text{O3}\ , \hspace{0.2 cm} 3 \times \text{O7}_{x^ix^jy^1y^2}/\text{O7}_{x^kx^l y^1 y^2} \ , \nonumber \\
 &  3 \times \text{O5}_{x^ix^j}/\text{O5}_{x^kx^l} \ , \hspace{0.2 cm} \text{SO9}/\text{SO5}_{y^1y^2} \ , \hspace{0.2 cm} 3 
 \times \text{SO5}_{x^ix^j}/\text{SO5}_{x^kx^l} \ , \nonumber \\
 &  3 \times \text{T}^2\text{SO5}_{x^ix^j}/\text{T}^2\text{SO5}_{x^kx^l} \quad .  
 \end{align}

\vskip .5cm
\noindent
${\bf D=3}$: \ In three dimensions the ${\rm SO}(8,8)$ theory is truncated to the ${\rm SO}(4,4) \times {\rm SO}(4,4)$ theory, describing supergravity coupled to eight hyper-multiplets. The vector central charge belongs to the ${\bf 35}$ of the R-symmetry ${\rm SO}(8)$, and correspondingly there are 35 different truncations, which can be seen as the $\tfrac{1}{2} \cdot {8 \choose 4}$ ways in which one can embed ${\rm SO}(4,4) \times {\rm SO}(4,4)$ into ${\rm SO}(8,8)$. As Table \ref{numberofbraneshalfmax} shows, there are 8960 1/2-BPS 2-branes which are the long weights of the ${\bf 600060}$, and for each truncation one expects 256 branes preserving the same supersymmetry of the truncation \cite{Bergshoeff:2012jb}. As in the higher-dimensional cases, to identify each truncation and the corresponding branes we consider the half-maximal theory resulting from the SO9 truncation of the maximal one. We get
\begin{align}
\text{SO}9 : \qquad 35 \times \text{SO}5_{x^ix^j x^k}/\text{T}^2\text{SO}5_{x^ix^j x^k}   \quad , \label{truncationsSO9truncD=3}
\end{align} 
where the indices $i,j,k$ run from 1 to 7. Fixing $i,j,k$ identifies a specific truncation to the quarter-maximal theory, and by carefully looking at Table \ref{maximaltruncationsbranesD3} and identifying one ${\rm SO}(4,4)$ as acting on the four unfixed indices and the other as grouping different potentials together,  one can show that the $\text{SO}5_{x^ix^j x^k}/\text{T}^2\text{SO}5_{x^ix^j x^k}$ branes belong to the representation
\begin{equation}
({\bf 35}_{\rm V} ,  {\bf 35}_{\rm S}) \oplus ({\bf 35}_{\rm V} ,  {\bf 35}_{\rm C}) \oplus ({\bf 35}_{\rm S} ,  {\bf 35}_{\rm V}) \oplus ({\bf 35}_{\rm S} ,  {\bf 35}_{\rm C}) \quad .\label{branesofN=4truncationD=3}
\end{equation}

Apart from the branes in eq. \eqref{branesofN=4truncationD=3}, there are additional space-filling branes that survive the truncation and are 1/2-BPS of the truncated theory. The representation of such branes is \cite{Bergshoeff:2014lxa}
\begin{equation}
({\bf 350} ,  {\bf 28}) \oplus ({\bf 28} ,  {\bf 350})
 \label{1/2BPSN=2D=3}
 \end{equation}
 corresponding to 4608 branes. In the particular case of the $\text{SO}5_{x^ix^jx^k}/\text{T}^2\text{SO}5_{x^ix^jx^k}$ truncation of the SO9-truncated theory, these all the SO5$_{x^i x^lx^m}$ and T$^2$SO5$_{x^i x^l x^m}$  branes with $l,m \neq j,k$. This indeed gives in total $3 \cdot {4 \choose 2} \cdot 2 \cdot 128 = 4608$ branes. 

Analogously to the  five and four-dimensions cases, one can also obtain the half-maximal theory as a geometric orbifold $(T^4/\mathbb{Z}_2) \times T^3$ truncation of IIB. Denoting with $x^i$, $i=1,...,4$ the orbifold coordinates and with $y^m$, $m=1,2,3$,  the torus coordinates, this corresponds to the truncation
T$^2$SO5$_{y^1 y^2 y^3}$. One can obtain the 35 truncations of the T$^2$SO5$_{y^1 y^2 y^3}$ half-maximal theory starting from the truncations of the SO9 half-maximal theory given in eq \eqref{truncationsSO9truncD=3}. We first perform an S-duality transformation to go to O9, then four T-dualities along the $x^i$  directions to go to O5$_{y^1 y^2 y^3}$, then again S-duality to go to SO5$_{y^1 y^2 y^3}$, and finally two T-dualities along one $y$ and one $x$ direction to go to T$^2$SO5$_{y^1 y^2 y^3}$. The result is
 \begin{align}
\text{T}^2\text{SO}5_{y^1 y^2 y^3} : \quad   &\text{O}9/\text{O}5_{y^1y^2 y^3} \ , \hspace{0.2 cm} 3 \times \text{O7}_{x^1...x^4 y^m}/\text{O3}_{y^m}\ , \hspace{0.2 cm} 3 \times \text{O7}_{x^ix^jy^1y^2 y^3}/\text{O7}_{x^kx^l y^1 y^2 y^3} \ , \nonumber \\
 &  9 \times \text{O5}_{x^ix^j y^m}/\text{O5}_{x^kx^l y^m} \ , \hspace{0.2 cm} \text{SO9}/\text{SO5}_{y^1y^2 y ^3} \ , \hspace{0.2 cm} 9 
 \times \text{SO5}_{x^ix^j y^m}/\text{SO5}_{x^kx^l y^m}\ , \nonumber \\
 &  9 \times \text{T}^2\text{SO5}_{x^ix^j y ^m}/\text{T}^2\text{SO5}_{x^kx^l y^m} \quad .  
 \end{align}

\subsection{From quarter-maximal to $1/8$-maximal supergravity}
\label{Fromquartermaximaltominimalsupergravity}

${\bf D=4}$: \ The four-dimensional $\mathcal{N}=2$ theory with symmetry $({\rm SL}(2,\mathbb{R}))^3 \times {\rm SO}(4,4)$ can be finally truncated to minimal supergravity coupled to seven chiral multiplets and global symmetry $({\rm SL}(2,\mathbb{R}))^7$. The vector central charge belongs to the ${\bf 3}$ of the R-symmetry U(2), and correspondingly there are three different  $\mathbb{Z}_2$ truncations, as expected from the fact that there are three different ways of embedding  ${\rm SO}(2,2) \times {\rm SO}(2,2)$ inside ${\rm SO}(4,4)$, with ${\rm SO}(2,2)$ being isomorphic to  $({\rm SL}(2,\mathbb{R}))^4$. The 384 1/2-BPS space-filling branes are the long weights of the representations given in eq. \eqref{1/2BPSbranesN=2}, and one expects that for each of the three truncations 128 branes are not projected out and preserve the same supersymmetry of the truncation. We can for instance identify these truncations as follows: first truncate from the maximal theory to the half-maximal one via SO9, and then from the half-maximal to the $\mathcal{N}=2$ theory  via $\text{SO}5_{x^5 x^6}/\text{T}^2\text{SO}5_{x^5 x^6}$.  One is then left with the $\text{SO}5_{x^ix^j}$ and $\text{T}^2\text{SO}5_{x^ix^j}$ space-filling branes in Table \ref{maximaltruncationsbranesD4}, with $i,j=1,...,4$, and in each of the three truncations one identifies the branes as follows: 
\begin{align}
\text{SO}9 \rightarrow \text{SO}5_{x^5 x^6}/\text{T}^2\text{SO}5_{x^5 x^6}  : \quad 3 \times \text{SO}5_{x^ix^j}/\text{SO}5_{x^kx^l} / \text{T}^2\text{SO}5_{x^ix^j} / \text{T}^2\text{SO}5_{x^kx^l}  \quad , \label{truncationsSO9SO556T2SO556truncD=4}
\end{align}
where the indices $i,j,k,l$ are all different. By analysing Table \ref{maximaltruncationsbranesD4} one can determine the representations of the $\text{SO}5_{x^ix^j}/\text{SO}5_{x^kx^l} / \text{T}^2\text{SO}5_{x^ix^j} / \text{T}^2\text{SO}5_{x^kx^l}$ branes for fixed $i,j,k,l$ that preserve the supersymmetry of a given truncation. Starting from the representations of the 1/2-BPS branes in the $\mathcal{N}=2$ theory given in eq. \eqref{1/2BPSbranesN=2} one gets
\begin{align}
{\bf (3,3,1,28)}  \rightarrow   &  {\bf (3,3,1,3,1,1,1) \oplus (3,3,1,1,3,1,1) } \nonumber \\ 
  \oplus & {\bf  (3,3,1,1,1,3,1) \oplus (3,3,1,1,1,1,3) } \nonumber \\
 {\bf (3,1,3,28)} \rightarrow &  {\bf (3,1,3,3,1,1,1) \oplus (3,1,3,1,3,1,1) }\nonumber \\
 \oplus  & {\bf(3,1,3,1,1,3,1) \oplus (3,1,3,1,1,1,3) } \nonumber \\
 {\bf (1,3,3,28) } \rightarrow  & {\bf (1,3,3,3,1,1,1) \oplus (1,3,3,1,3,1,1) }\nonumber \\
  \oplus & {\bf(1,3,3,1,1,3,1) \oplus (1,3,3,1,1,1,3) } \nonumber \\
   {\bf (1,1,1,350)} \rightarrow & {\bf (1,1,1,3,3,3,1) \oplus (1,1,1,3,3,1,3) }\nonumber \\
  \oplus  & {\bf(1,1,1,3,1,3,3) \oplus (1,1,1,1,3,3,3) } \quad . \label{branestruncationD=4minimal}
\end{align}

The $\mathcal{N}=2$ theory has also a natural geometric origin as IIB compactified on the orbifold  $T^6/(\mathbb{Z}_2 \times \mathbb{Z}_2 )$. 
In this case the torus $T^6$ is factorised as $T^6= \otimes^3_{i=1} T_{i}^2$, where $T_{i}^2$ indicates the two-dimensional torus with coordinates $x^i$ and $y^i$, and the two  $\mathbb{Z}_2$'s act as $(+,-,-)$ and $(-,+,-)$ on the three pairs of coordinates. We can think of the orbifold action as the sequence of two truncations, which are the $\text{T}^2\text{SO}5_{x^1 y^1}$ from the maximal to the half-maximal theory, and then  $\text{T}^2\text{SO}5_{x^2 y^2}/\text{T}^2\text{SO}5_{x^3 y^3}$ from the half-maximal to the $\mathcal{N}=2$ theory. The three different truncations of the $\mathcal{N}=2$ theory can then be derived by a suitable series of dualities starting from eq. \eqref{truncationsSO9SO556T2SO556truncD=4}, and the result is 
\begin{align}
&\text{O}9/\text{O}5_{x^1y^1}/\text{O}5_{x^2y^2}/\text{O}5_{x^3 y^3} \hspace{0.2 cm}, \quad  \text{O}3/\text{O}7_{x^1y^1x^2 y^2}/\text{O}7_{x^1y^1x^3y^3}/ \text{O}7_{x^2y^2x^3y^3}\hspace{0.2 cm}, \nonumber \\
&\text{SO}9/\text{SO}5_{x^1y^1}/\text{SO}5_{x^2y^2}/\text{SO}5_{x^3y^3} \quad .
\end{align}
One can recognise the first two as the two T-dual orientifold projections of the IIB theory on  $T^6/(\mathbb{Z}_2 \times \mathbb{Z}_2 )$ \cite{Antoniadis:1999ux,Angelantonj:1999ms}, and the last as the projection on the heterotic sector.
As we will show in detail in section 5, the 128 branes in each of these truncations are precisely those that are responsible for the cancellation of the tadpole conditions when all possible geometric and non-geometric fluxes are turned on, and for the O3/O7$_{x^1y^1x^2 y^2}$/O7$_{x^1y^1x^3y^3}$/ O7$_{x^2y^2x^3y^3}$ orientifold case, they have been determined in \cite{Lombardo:2016swq, Lombardo:2017yme}.

\vskip .5cm
\noindent
${\bf D=3}$: \ The three-dimensional theory with symmetry ${\rm SO}(4,4) \times {\rm SO}(4,4)$ can be truncated to the supergravity theory with four supersymmetry coupled to eight scalar multiplets and global symmetry $({\rm SL}(2,\mathbb{R}))^8$. As Table \ref{numberofbranesquartermax} shows, there are nine different truncations, corresponding to the fact that each of the two ${\rm SO}(4,4)$ can be decomposed in $({\rm SL}(2,\mathbb{R}))^4$ in three different ways. We expect that for each of the nine truncations 512  out of the 4608 2-branes in the representations of eq. \eqref{1/2BPSN=2D=3} survive the are not projected out. These are the branes that preserve the same supersymmetry of the truncation. To identify these branes and the corresponding truncations, we proceed as in four dimensions, considering the particular case of the SO9 truncation of the maximal theory, and the further $\text{SO}5_{x^5 x^6 x^7}/\text{T}^2\text{SO}5_{x^5 x^6 x^7}$ truncation of the half-maximal theory to the theory with eight supersymmetries. One gets
\begin{align}
\text{SO}9 \rightarrow \text{SO}5_{x^5 x^6 x^7}/\text{T}^2\text{SO}5_{x^5 x^6 x^7}  : \ 9 \times \text{SO}5_{x^ix^j x^k}/\text{SO}5_{x^i x^l x^m} / \text{T}^2\text{SO}5_{x^ix^j x^k } / \text{T}^2\text{SO}5_{x^i x^l x^m}  \quad , \label{truncationsSO9SO556T2SO556truncD=3}
\end{align}
where $i=5,6,7$ and $j,k,l,m = 1,...,4$ and all different. Fixing $i,j,k,l,m$ to identify a particular truncation, one reads off the corresponding branes from Table \ref{maximaltruncationsbranesD3}. These branes belong to representations made of four triplets and four singlets of $({\rm SL}(2,\mathbb{R}))^8$
which arise from decomposing all the ${\bf 350}$ and ${\bf 28}$ representations in eq. \eqref{1/2BPSN=2D=3} as
\begin{align}
{\bf 350}  \rightarrow   &  {\bf (1,3,3,3) \oplus (3,1,3,3) \oplus (3,3,1,3) \oplus (3,3,3,1) } \nonumber \\ 
 {\bf 28} \rightarrow &  {\bf (3,1,1,1) \oplus (1,3,1,1) \oplus (1,1,3,1) \oplus (1,1,1,3) } \quad . \label{branestruncationD=3minimal}
\end{align} 

As in four dimensions, this theory can also be considered as IIB compactified on the orbifold  $T^6/(\mathbb{Z}_2 \times \mathbb{Z}_2 )\times S^1$. 
Denoting the orbifold coordinates $x^i y^i$, $i=1,2,3$ as in four dimensions, and the circle coordinate $z$, we can think of the orbifold action as the sequence of two truncations, which are the $\text{T}^2\text{SO}5_{x^1 y^1 z}$ from the maximal to the half-maximal theory, and then  $\text{T}^2\text{SO}5_{x^2 y^2 z }/\text{T}^2\text{SO}5_{x^3 y^3 z }$ from the half-maximal to the quarter-maximal theory. The nine different truncations of the $\mathcal{N}=2$ theory can then be derived by a suitable series of dualities starting from eq. \eqref{truncationsSO9SO556T2SO556truncD=3}, and the result is 
\begin{align}
&\text{O}9/\text{O}5_{x^1y^1z}/\text{O}5_{x^2y^2z}/\text{O}5_{x^3 y^3z} \nonumber \\
&  \text{SO}9/\text{SO}5_{x^1y^1z}/\text{SO}5_{x^2y^2z}/\text{SO}5_{x^3y^3z}\nonumber \\
& \text{O}5_{y^1y^2y^3}/\text{O}5_{y^1x^2x^3}/\text{O}5_{x^1y^2x^3}/\text{O}5_{x^1 x^2 y^3}   \nonumber \\
&  \text{O}5_{x^1x^2x^3}/\text{O}5_{x^1y^2y^3}/\text{O}5_{y^1x^2y^3}/\text{O}5_{y^1 y^2 x^3} \nonumber \\
& \text{SO}5_{y^1y^2y^3}/\text{SO}5_{y^1x^2x^3}/\text{SO}5_{x^1y^2x^3}/\text{SO}5_{x^1 x^2 y^3}  \nonumber \\
& 
\text{SO}5_{x^1x^2x^3}/\text{SO}5_{x^1y^2y^3}/\text{SO}5_{y^1x^2y^3}/\text{SO}5_{y^1 y^2 x^3}  \nonumber \\
&  \text{O}3_z/\text{O}7_{x^1y^1x^2 y^2z}/\text{O}7_{x^1y^1x^3y^3z}/ \text{O}7_{x^2y^2x^3y^3z} \nonumber \\
& 
\text{T}^2\text{SO}5_{x^1x^2x^3}/\text{T}^2\text{SO}5_{x^1y^2y^3}/\text{T}^2\text{SO}5_{y^1x^2y^3}/\text{T}^2\text{SO}5_{y^1 y^2 x^3}  \nonumber \\
&\text{T}^2\text{SO}5_{y^1y^2y^3}/\text{T}^2\text{SO}5_{y^1x^2x^3}/\text{T}^2\text{SO}5_{x^1y^2x^3}/\text{T}^2\text{SO}5_{x^1 x^2 y^3} \quad ,
\end{align}
where the corresponding branes can be read in Table \ref{maximaltruncationsbranesD3}. 
\vskip .5cm
To conclude this section, we give in Table \ref{summaryalltruncations} the summary of all the results, in which in any dimension and for any supersymmetry we denote in red the representations of the branes that preserve the same supersymmetry of the truncation, and in black the ones that are 1/2-BPS states.

\begin{table}[h!]
\begin{center}
\scalebox{.9}{
\begin{tabular}{|c||c|c|c|c|}
\hline  \backslashbox{$D$}{\# susy} &  {32}  & {16}  & 8 & 4  \\
 \hline   \hline
  \rule[1mm]{0mm}{3mm}  
\multirow{2}{*}{8} & ${\rm SL}(3,\mathbb{R})\times {\rm SL}(2,\mathbb{R})$   &  $\mathbb{R}^+\times ({\rm SL}(2,\mathbb{R}))^2$ & &  \\
 \cline{2-3}
  \rule[1mm]{0mm}{3mm}
 & ${\bf (15,1)}$ & ${\red {\bf (3,1)}}$  & & \\
\hline \hline
  \rule[1mm]{0mm}{3mm}  
\multirow{2}{*}{7} & ${\rm SL}(5,\mathbb{R})$   &  $\mathbb{R}^+\times {\rm SL}(4,\mathbb{R})$ & &  \\
 \cline{2-3}
  \rule[1mm]{0mm}{3mm}
 & ${\bf 70}$ & ${\red {\bf 10}}$  & & \\
 \hline \hline
  \rule[1mm]{0mm}{3mm}  
\multirow{6}{*}{6} & \multirow{2}{*}{${\rm SO}(5,5)$}   &  $\mathbb{R}^+\times {\rm SO}(4,4)$  &  \multirow{2}{*}{$\mathbb{R}^+\times {\rm SO}(4,4)$} & \\
  \cline{3-3}
  \rule[1mm]{0mm}{3mm}
& & ${\rm SO}(5,5)$ & & \\
 \cline{2-4}
  \rule[1mm]{0mm}{3mm}
 & \multirow{2}{*}{${\bf 320}$} & ${\red {\bf 35}_{\rm V}} \ \  {\bf 35}_{\rm V}$ & ${\red {\bf 35}_{\rm V}}$ &  \\
   \cline{3-4}
  \rule[1mm]{0mm}{3mm}
 & & ${\bf 320}$ & ${\red {\bf 35}_{\rm V} \oplus {\bf 35}_{\rm V}}$ & \\
  \cline{2-4}
  \rule[1mm]{0mm}{3mm}
 & \multirow{2}{*}{${\bf \overline{126}}$} & ${\bf 35}_{\rm S} \oplus {\bf 35}_{\rm C}$ & ${\red {\bf 35}_{\rm S} \oplus {\bf 35}_{\rm C}}$ & \\
  \cline{3-4}
  \rule[1mm]{0mm}{3mm}
 & & ${\red \bf \overline{126}}$ &  & \\
 \hline \hline
  \rule[1mm]{0mm}{3mm}  
\multirow{3}{*}{5} & ${\rm E}_{6(6)}$   &  $\mathbb{R}^+\times {\rm SO}(5,5)$ &  $(\mathbb{R}^+)^2 \times {\rm SO}(4,4)$ & \\
 \cline{2-4}
  \rule[1mm]{0mm}{3mm}
 & \multirow{2}{*}{${\bf \overline{1728}}$} & \multirow{2}{*}{${\red {\bf \overline{126}}}$} \ \ \ \ \  ${\bf 320}$ & ${\red {\bf 35}_{\rm V} \oplus {\bf 35}_{\rm V}}$&  \\
 & & \ \ \ \ \ \ \ \  \ \ \ ${\bf 210}$ &${\red {\bf 35}_{\rm S} \oplus {\bf 35}_{\rm C}}$ & \\
 \hline \hline 
  \rule[1mm]{0mm}{3mm}  
\multirow{3}{*}{4} & ${\rm E}_{7(7)}$   &  ${\rm SL}(2,\mathbb{R})\times {\rm SO}(6,6)$ & $({\rm SL}(2,\mathbb{R}))^3\times {\rm SO}(4,4)$ &  $({\rm SL}(2,\mathbb{R}))^7$ \\
 \cline{2-5}
  \rule[1mm]{0mm}{3mm}
 & \multirow{2}{*}{${\bf {8645}}$} & \multirow{2}{*}{${\red {\bf (1,\overline{462})}}$} \  \ ${\bf (3,495)}$  & \multirow{2}{*}{${\red {\bf (3, 35)}}$} \ \ ${\bf (3^2, 28)}$ & \multirow{2}{*}{${\red {\bf 3^3}}$}\\
 & & \ \ \ \ \ \ \ \ \ \ \ \  \ ${\bf (1, 2079)}$& \ \ \ \ \ \ \ \ \ \  \ \ ${\bf (1, 350)}$ & \\
 \hline \hline
  \rule[1mm]{0mm}{3mm}  
\multirow{3}{*}{3} & ${\rm E}_{8(8)}$   &  $ {\rm SO}(8,8)$ & ${\rm SO}(4,4) \times {\rm SO}(4,4)$ &  $({\rm SL}(2,\mathbb{R}))^8$ \\
 \cline{2-5}
  \rule[1mm]{0mm}{3mm}
 & \multirow{2}{*}{${\bf {147250}}$} & \multirow{2}{*}{${\red {\bf {6435}}} \ \ \ {\bf 60060}$} & \multirow{2}{*}{${\red {\bf (35, 35)}}$} \ \ ${\bf (350, 28)}$& \multirow{2}{*}{${\red {\bf 3^4}}$}\\
 & & & \ \ \ \ \ \ \ \ \ \ \ \  \ ${\bf (28, 350)}$ & \\
\hline

\end{tabular}
}
\caption{\sl The representations of the space-filling branes that preserve the same supersymmetry of the truncation (in red) and those that are 1/2-BPS states (in black). In six dimensions the case with 16 supersymmetry is divided in two rows, with the upper row corresponding to the 6A and the lower row to the 6B truncation. In four dimensions, the representations of the cases with eight and four supersymmetry are written in a short-hand notation which stands for eqs. \eqref{branesofN=4truncationD=4} and \eqref{1/2BPSbranesN=2} (eight supersymmetries) and \eqref{branestruncationD=4minimal}. Similarly, in three dimensions the red representations of the theory with eight supersymmetries are given in eqs. \eqref{branesofN=4truncationD=3}, while those of the theory with four supersymmetries can be derived from \eqref{branestruncationD=3minimal}.
\label{summaryalltruncations}}
\end{center}
\end{table}

\section{Embedding tensor, quadratic constraints and space-filling branes}

The previous analysis yields a concrete prescription for finding the quadratic constraint (QC) irrep's containing all space-filling branes preserving the same supersymmetry as the given truncation. The argument we used is inspired by the properties of mixed symmetry potentials and how they are used to source exotic branes through Wess-Zumino coupling.
The aim of this section is to show that the very same conclusion can be drawn independently by purely studying $\mathbb{Z}_{2}$ truncations within gauged supergravities in various dimensions to obtain theories with less supersymmetries. The concrete prescription is to compare all the QC's required by the theory with more supersymmetry when restricted to its 
$\mathbb{Z}_{2}$-even sector with the ones imposed by the consistency of the theory with halved amount of supersymmetry.  
The mismatch between the two exactly identifies the irrep's containing space-filling branes allowed by the less supersymmetric theory. From the technical viewpoint of gauged  supergravity, such a mismatch represents the set of all closure conditions specified within the $[\textrm{odd},\,\textrm{odd}]$ sector.

We will now carry out the aforementioned analysis by treating each number of dimensions separately. This will involve the study of the allowed embedding tensor irrep's and the corresponding set of QC's it is subject to throughout a chain of $\mathbb{Z}_{2}$ truncations.

\subsection{Gauged supergravities in $D=9$}

In nine spacetime dimensions $32$ and $16$ are the only amounts of real supercharges which are compatible with Lorentz symmetry. We shall refer to these cases as the maximal and half-maximal theory, respectively. Hence there is only one $\mathbb{Z}_{2}$ truncation to be considered here, \emph{i.e.} the one relating them (see \cite{Dibitetto:2012rk} for details).
The most general gauged maximal theory was studied in \cite{FernandezMelgarejo:2011wx}, it enjoys global symmetry $\mathbb{R}^{+}\times \textrm{SL}(2,\mathbb{R})$ and it has embedding tensor 
\begin{equation}
\Theta \, \in \, \underbrace{\textbf{2}_{(+3)}}_{\theta^{i}}\,\oplus\,\underbrace{\textbf{3}_{(-4)}}_{\kappa^{(ij)}}\ .
\end{equation}
The discrete truncation yielding the half-maximal theory reads
\begin{equation}
\begin{array}{lclc}
\mathbb{R}^{+}\times \textrm{SL}(2,\mathbb{R}) & \longrightarrow & \left(\mathbb{R}^{+}\right)^{2} & ,
\end{array}
\end{equation}
however this case turns out to be rather trivial since the half-maximal theory has no embedding tensor irrep's with long weights. This means that even the only space-filling brane that
one expects cannot be sourced by flux tadpoles.

\subsection{Gauged supergravities in $D=8$}

Moving to the $D=8$ case, the only two possibilities available are still just the maximal and half-maximal theories, which are then related again by a $\mathbb{Z}_{2}$ truncation.
The most general gauged maximal theory was studied in \cite{AlonsoAlberca:2000gh,deRoo:2011fa,Andino:2016bwk}, and it has $\textrm{SL}(3,\mathbb{R})\times \textrm{SL}(2,\mathbb{R})$ global symmetry. The truncation relating the two theories is given by \cite{Dibitetto:2012rk}\footnote{We are using here the notation  commonly used in the gauged supergravity literature, which differs from the one adopted in section 2. This should not cause any confusion to the reader.}
\begin{equation}
\begin{array}{lclc}
\textrm{SL}(3,\mathbb{R})\times \textrm{SL}(2,\mathbb{R}) & \longrightarrow & \mathbb{R}^{+}\times\left(\textrm{SL}(2,\mathbb{R}))^{2}\right) & , \\[2mm]
\Theta \, \in \, \underbrace{\left(\textbf{3},\textbf{2}\right)}_{\xi_{\alpha m}}\,\oplus\,\underbrace{\left(\overline{\textbf{6}},\textbf{2}\right)}_{{f_{\alpha}}^{(mn)}} & 
& \underbrace{\left(\textbf{2},\textbf{2}\right)_{(-1)}}_{a_{\alpha i}}\,\oplus\,\underbrace{\left(\textbf{2},\textbf{2}\right)_{(-1)}}_{b_{\alpha i}} & .
\end{array} 
\end{equation}
The QC of the maximal theory
\begin{equation}\label{QC_Max_8D}
\begin{array}{rcccc}
\epsilon^{\alpha\beta}\,\xi_{\alpha m}\,\xi_{\beta n} & = & 0 & , & \left(\overline{\textbf{3}},\textbf{1}\right) \\[2mm]
{f_{(\alpha}}^{np}\,\xi_{\beta) p} & = & 0 & , & \left(\overline{\textbf{3}},\textbf{3}\right) \\[2mm]
\epsilon^{\alpha\beta}\,\left(\epsilon_{mqr}\,{f_{\alpha}}^{qn}\,{f_{\beta}}^{rp}\,+\,{f_{\alpha}}^{np}\,\xi_{\beta m}\right)  & = & 0 & , & 
\left(\overline{\textbf{15}},\textbf{1}\right)\,\oplus\,\left(\overline{\textbf{3}},\textbf{1}\right) \\[2mm]
\end{array}
\end{equation}
reduce upon truncating to the following set of QC 
\begin{equation}
\begin{array}{rcccc}
\epsilon^{\alpha\beta}\,\epsilon^{ij}\,a_{\alpha i}\,a_{\beta j} & = & 0 & , & \left(\textbf{1},\textbf{1}\right)_{(-2)} \\[2mm]
\epsilon^{\alpha\beta}\,\epsilon^{ij}\,b_{\alpha i}\,b_{\beta j} & = & 0 & , & \left(\textbf{1},\textbf{1}\right)_{(-2)} \\[2mm]
\epsilon^{\alpha\beta}\,\epsilon^{ij}\,a_{\alpha i}\,b_{\beta j} & = & 0 & , & \left(\textbf{1},\textbf{1}\right)_{(-2)} \\[2mm]
\epsilon^{ij}\,a_{(\alpha i}\,b_{\beta) j} & = & 0 & , & \left(\textbf{3},\textbf{1}\right)_{(-2)} \\[2mm]
\epsilon^{\alpha\beta}\,a_{\alpha (i}\,b_{\beta j)} & = & 0 & , & \left(\textbf{1},\textbf{3}\right)_{(-2)} 
\end{array}
\end{equation}
while the consistency of the half-maximal theory requires the last two constraints, while only two combinations of the three singlets have to vanish. However, that does not affect our counting of space-filling objects. As far as these are concerned, the peculiarity here is that the decomposition of the QC in \eqref{QC_Max_8D} would in principle yield an extra 
$\left(\textbf{3},\textbf{1}\right)$ containing one space-filling brane which cannot be sourced by any fluxes. This is due to the fact that the three-form representation of 
$\textrm{SO(2,2)}\,\sim\,\left(\textrm{SL}(2,\mathbb{R})\right)^{2}$ does not have any long weights.

\subsection{Gauged supergravities in $D=7$}

As far as theories in seven dimensions are concerned, the situation remains unchanged, \emph{i.e.} only maximal and half-maximal supersymmetry are allowed and they are related by a 
$\mathbb{Z}_{2}$ truncation. The embedding tensor formulation of gauged maximal $D=7$ supergravities was developed in \cite{Samtleben:2005bp}. The global symmetry is given by 
$\textrm{SL}(5,\mathbb{R})$, which can be subsequently broken as follows to yield a half-maximal theory
\begin{equation}
\begin{array}{lclc}
\textrm{SL}(5,\mathbb{R}) & \longrightarrow & \mathbb{R}^{+}\times\textrm{SL}(4,\mathbb{R}) & , \\[2mm]
\Theta \, \in \, \underbrace{\overline{\textbf{15}}}_{Y_{(MN)}}\,\oplus\,\underbrace{\textbf{40}}_{{Z^{[MN],P}}} & 
& \underbrace{\textbf{1}_{(-8)}}_{\theta}\,\oplus\,
\underbrace{\textbf{6}_{(+2)}}_{\xi_{[mn]}}\,\oplus\,\underbrace{\overline{\textbf{10}}_{(+2)}}_{Q_{(mn)}}\,\oplus\,\underbrace{\textbf{10}_{(+2)}}_{\tilde{Q}^{(mn)}} & .
\end{array} 
\end{equation}
The QC constraints of the maximal theory 
\begin{equation}
\begin{array}{rcccc}
Y_{MQ}\,Z^{QN,P}\,+\,2\epsilon_{MRSTU}\,Z^{RS,N}\,Z^{TU,P} & = & 0 & , & \left(\textbf{5}\,\oplus\,\textbf{45}\,\oplus\,\textbf{70}\right)
\end{array}
\end{equation}
reduce to 
\begin{equation}
\begin{array}{rcccc}
\theta\,\xi_{mn} & = & 0 & , & \left(\textbf{6}_{(-6)}\right) \\[2mm]
\left(\tilde{Q}^{mp}\,+\,\xi^{mp}\right)\,Q_{pn} & = & 0 & , & \left(\textbf{1}_{(+4)}\,\oplus\,\textbf{15}_{(+4)}\right) \\[2mm]
Q_{mp}\,\xi^{pn}\,+\,\xi_{mp}\,\left(\tilde{Q}^{pn}\,+\,\xi^{pn}\right)& = & 0 & , & \left(\textbf{1}_{(+4)}\,\oplus\,\textbf{15}_{(+4)}\right) \\[2mm]
\theta\,\tilde{Q}^{mn} & = & 0 & , & \left(\textbf{10}_{(-6)}\right) 
\end{array}
\end{equation}
where all of the above constraints are demanded for consistency of the half-maximal theory, except for the singlet part of the second constraint and the last constraint transforming in
the $\textbf{10}$. As we are interested in the space-filling branes, we immediately see that the four objects that we are looking for exactly coincide with the long weights in the 
$\textbf{10}$. Note that this is perfect agreement with what presented in table~\ref{summaryalltruncations}.

\subsection{Gauged supergravities in $D=6$}

Moving down to six dimensions, one encounters for the first time the possibility of constructing quarter-maximal theories, \emph{i.e.} preserving only eight real supercharges arranged within a symplectic Majorana-Weyl (SMW) doublet. 
Starting from the maximal theory \cite{Bergshoeff:2007ef} enjoying $\textrm{SO}(5,5)$ global symmetry, one can perform the following $\mathbb{Z}_2$ truncation 
\begin{equation}
\begin{array}{lclc}
\textrm{SO}(5,5) & \longrightarrow & \mathbb{R}^{+}\times\textrm{SO}(4,4) & , \\[2mm]
\Theta \, \in \, \underbrace{\textbf{144}_{\rm C}}_{\theta^{\alpha A}} & 
& \underbrace{\textbf{8}_{\rm C}^{(+3)}}_{\zeta_{M}}\,\oplus\,
\underbrace{\textbf{8}_{\rm C}^{(-1)}}_{\xi_{M}}\,\oplus\,\underbrace{\textbf{56}_{\rm C}^{(-1)}}_{f_{[MNP]}} & ,
\end{array} 
\end{equation}
to obtain a theory with sixteen supercharges realising $\mathcal{N}=(1,1)$ supersymmetry and where the gravity multiplet is coupled to four vector multiplets.

The QC of the $\mathcal{N}=(2,2)$ theory read
\begin{equation}
\begin{array}{rcccc}
\theta^{\alpha A}\,\theta^{\beta B}\,\eta_{AB} & = & 0 & , & (\textbf{10}\,\oplus\,\textbf{126}_{\rm C}) \\[2mm]
\theta^{\alpha A}\,\theta^{\beta [B}\,\left(\gamma^{C]}\right)_{\alpha\beta} & = & 0 & , & (\textbf{320})
\end{array}
\end{equation}
where $\eta$ is the $\textrm{SO}(5,5)$-invariant metric, while $\left\{\gamma^{A}\right\}$ represent the $\textrm{SO}(5,5)$ Dirac matrices. When restricting oneself to the even sector, the above set of constraints takes the following form (we furthermore set $\xi=0$)
\begin{equation}
\begin{array}{rcccc}
f_{R[MN}\,{f_{PQ]}}^{R} & = & 0 & , & \left(\textbf{35}_{\rm S}^{(-2)}\,\oplus\,\textbf{35}_{\rm C}^{(-2)}\right) \\[2mm]
f_{MNP}\,\zeta^{P} & = & 0 & , & \left(\textbf{28}^{(+2)}\right) \\[2mm]
f_{MNP}\,f^{MNP} & = & 0 & , & \left(\textbf{1}^{(-2)}\right) \\[2mm]
f_{[MNP}\,\zeta_{Q]}|_{\textrm{SD}} & = & 0 & , & \left(\textbf{35}_{\rm S}^{(+2)}\right) 
\end{array}
\end{equation}
the last two constraints not being required for the consistency of the $(1,1)$ theory. Hence, the $8$ space-filling branes found in table~\ref{summaryalltruncations} arise here as the long weights of the $\textbf{35}_{\rm S}^{(+2)}$ of $\mathbb{R}^{+}\times\textrm{SO}(4,4)$.

Furthermore, a second \emph{inequivalent} $\mathbb{Z}_2$ truncation is the one mentioned in the previous section giving rise to the chiral half-maximal theory with $\mathcal{N}=(2,0)$ supersymmetry and tensorial matter. This truncation leaves the whole $\textrm{SO}(5,5)$ global symmetry unbroken but it has no embedding tensor deformations. This means that, even if it has space-filling brane states available in the spectrum, we have no possibility of cancelling their charge by means of flux tadpoles.

The last step we still need to discuss within the $D=6$ case is the one taking us from the $\mathcal{N}=(1,1)$ to the $\mathcal{N}=(1,0)$ theory by means of the following truncation
\begin{equation}
\begin{array}{lclc}
\mathbb{R}^{+}\times\textrm{SO}(4,4) & \longrightarrow & \mathbb{R}^{+}\times\textrm{SO}(4,4) & , \\[2mm]
\Theta \, \in \, \underbrace{\textbf{8}_{\rm C}^{(+3)}}_{\zeta_{M}}\,\oplus\,
\underbrace{\textbf{8}_{\rm C}^{(-1)}}_{\xi_{M}}\,\oplus\,\underbrace{\textbf{56}_{\rm C}^{(-1)}}_{f_{[MNP]}} & 
& \textrm{None} & ,
\end{array} 
\end{equation}
where the resulting $\mathcal{N}=(1,0)$ theory contains tensor as well as hypermultiplets. However, no deformation parameters survive the truncation and hence no space-filling branes may be added by consistently cancelling their charge by means of flux tadpoles.

\subsection{Gauged supergravities in $D=5$}

Now let us move to the five-dimensional case. Here again one can have theories with $32$, $16$ or $8$ real supercharges. Starting from the maximal theory \cite{deWit:2004nw} with 
$\textrm{E}_{6(6)}$ global symmetry, the following $\mathbb{Z}_{2}$ truncation
\begin{equation}
\begin{array}{lclc}
\textrm{E}_{6(6)} & \longrightarrow & \mathbb{R}^{+}\times\textrm{SO}(5,5) & , \\[2mm]
\Theta \, \in \, \underbrace{\overline{\textbf{351}}}_{Z^{[AB]}} & 
& \underbrace{\textbf{10}_{(+2)}}_{\xi_{M}}\,\oplus\,
\underbrace{\textbf{45}_{(-4)}}_{\zeta_{[MN]}}\,\oplus\,\underbrace{\textbf{120}_{(+2)}}_{f_{[MNP]}} & ,
\end{array} 
\end{equation}
produces a half-maximal gauged supergravity \cite{Schon:2006kz} with five extra vector multiplets. 
Decomposing the QC's of the maximal theory and restricting ourselves to the even part, we find the following set of QC's
\begin{equation}\label{QC_Half_5D}
\begin{array}{rcccc}
\xi_{M}\,\xi^{M} & = & 0 &, & \left(\textbf{1}_{(+4)}\right) \\[2mm]
\zeta_{MN}\,\xi^{N} & = & 0 &, & \left(\textbf{10}_{(-2)}\right) \\[2mm]
f_{MNP}\,\xi^{P} & = & 0 &, & \left(\textbf{45}_{(+4)}\right) \\[2mm]
f_{R[MN}\,{f_{PQ]}}^{R} \,+\, f_{[MNP} \, \xi_{Q]} & = & 0 & , & \left(\textbf{210}_{(+4)}\right)  \\[2mm]
{\zeta_{M}}^{Q}\,f_{NPQ}\,+\,\xi_{M}\,\zeta_{NP} & = & 0 & , & \left(\textbf{10}_{(-2)}\,\oplus\,\textbf{120}_{(-2)}\,\oplus\,\textbf{320}_{(-2)}\right) \\[2mm]
f_{MNP}\,f^{MNP} & = & 0 & , & \left(\textbf{1}_{(+4)}\right) \\[2mm]
f_{[MNP}\,\zeta_{QR]}|_{\textrm{SD}} & = & 0 & , & \left(\overline{\textbf{126}}_{(-2)}\right) 
\end{array}
\end{equation}
which exactly correspond to the QC of the half-maximal theory, plus the last two lines transforming in the $\textbf{1}_{(+4)}$ and $\overline{\textbf{126}}_{(-2)}$ as two additional ones. Therefore, the spacefilling branes of this theory are given by the $16$ long weights inside the latter extra QC irrep. This is once again in agreement with what found in the previous section.

We now want to further truncate the half-maximal theory to obtain a quarter-maximal theory coupled to four hypermultiplets. This is done through
\begin{equation}
\begin{array}{lclc}
\mathbb{R}^{+}\times\textrm{SO}(5,5) & \longrightarrow & \left(\mathbb{R}^{+}\right)^{2}\times\textrm{SO}(4,4) & , \\[2mm]
\Theta \, \in \, \underbrace{\textbf{10}}_{\xi_{M}}\,\oplus\,\underbrace{\textbf{45}}_{\zeta_{[MN]}}\,\oplus\,\underbrace{\textbf{120}}_{f_{[MNP]}} & 
& 3\,\times\,
\left(\textbf{1}\,\oplus\,\textbf{28}\right) & ,
\end{array} 
\end{equation}
where the resulting embedding tensor can be rearranged into the \emph{reducible} object denoted by ${\Theta_{\Lambda}}^{\alpha}$, where in turn, the index $\Lambda$ labels the vectors, while $\alpha$ runs over $\textrm{adj}\left(\mathbb{R}^{+}\right)\oplus\textrm{adj}\left(\mathbb{R}^{+}\right)\oplus\textrm{adj}\left(\textrm{SO}(4,4)\right)$.
The QC's \eqref{QC_Half_5D} then reduce to the following irrep's
\begin{equation}\label{QC_Quart_5D}
\begin{array}{lclc}
\textbf{1} & \longrightarrow & \textbf{1} & , \\[2mm]
\textbf{10} & \longrightarrow & \textbf{1}\,\oplus\,\textbf{1}\,\oplus\,\textrm{odd} & , \\[2mm]
\textbf{45} & \longrightarrow & \textbf{1}\,\oplus\,\textbf{28}\,\oplus\,\textrm{odd} & , \\[2mm]
\textbf{210} & \longrightarrow & \textbf{28}\,\oplus\,\textbf{28}\,\oplus\,\textbf{35}_{\rm S}\,\oplus\,\textbf{35}_{\rm C}\,\oplus\,\textrm{odd} & , \\[2mm]
\textbf{120} & \longrightarrow & \textbf{28}\,\oplus\,\textbf{28}\,\oplus\,\textrm{odd} & , \\[2mm]
\textbf{320} & \longrightarrow & \textbf{1}\,\oplus\,\textbf{1}\,\oplus\,\textbf{28}\,\oplus\,\textbf{28}
\,\oplus\,\textbf{35}_{\rm V}\,\oplus\,\textbf{35}_{\rm V}\,\oplus\,\textrm{odd} & , 
\end{array}
\end{equation}
to be compared with those ones appearing in the QC's of the $\mathcal{N}=1$ \cite{deWit:2005ub}
\begin{equation}
{\Theta_{\Lambda}}^{\alpha}\,{\Theta_{\Sigma}}^{\beta}\,{f_{\alpha\beta}}^{\gamma}\,+\,{\left[t_{\alpha}\right]_{\Lambda}}^{\Gamma}\,
{\Theta_{\Sigma}}^{\alpha}\,{\Theta_{\Gamma}}^{\gamma} \,=\, 0 \ ,
\end{equation}
where $\left\{{f_{\alpha\beta}}^{\gamma}\right\}$ and $\left\{{\left[t_{\alpha}\right]_{\Lambda}}^{\Gamma}\right\}$ are the global symmetry generators in the adjoint and vector representation, respectively. The unneeded QC irrep's containing the longest weights turn out to be 
$\textbf{35}_{\rm S}\,\oplus\,\textbf{35}_{\rm C}\,\oplus\,\textbf{35}_{\rm V}\,\oplus\,\textbf{35}_{\rm V}$, once again in perfect agreement with the analysis of section~3.

\subsection{Gauged supergravities in $D=4$}

In $D=4$, the minimal amount of supersymmetry which is consistent with Lorentz symmetry is given by four real supercharges rearranged into a single Majorana spinor. This implies the extra possibility in this case to further truncate a quarter-maximal theory down to a minimal one.
The most general maximal gauged theory was studied in \cite{deWit:2007kvg} and it turns out to enjoy $\textrm{E}_{7(7)}$ global symmetry. Furthermore, in \cite{Dibitetto:2011eu}, the 
truncation defined by 
\begin{equation}
\begin{array}{lclc}
\textrm{E}_{7(7)} & \longrightarrow & \textrm{SL}(2,\mathbb{R})\times\textrm{SO}(6,6) & , \\[2mm]
\Theta \, \in \, \underbrace{\textbf{912}}_{X_{\mathcal{M}\mathcal{N}\mathcal{P}}} & 
& \underbrace{(\textbf{2},\textbf{12})}_{\xi_{\alpha M}}\,\oplus\,
\underbrace{(\textbf{2},\textbf{220})}_{f_{\alpha [MNP]}} & ,
\end{array} 
\end{equation}
was found to yield a half-maximal theory \cite{Schon:2006kz} coupled to six extra vector multiplets.

The QC of the maximal theory
\begin{equation}
\Omega^{\mathcal{M}\mathcal{Q}}\,X_{\mathcal{M}\mathcal{N}\mathcal{P}}\,X_{\mathcal{Q}\mathcal{R}\mathcal{S}} \ = \ 0 \ , \ (\textbf{133}\,\oplus\,\textbf{8645})
\end{equation}
give rise to the following QC's for $\mathbb{Z}_{2}$ even objects
\begin{equation}\label{QC_Half_4D}
\begin{array}{rcccc}
\xi_{\alpha M}\,{\xi_{\beta}}^{M} & = & 0 &, & \left(\textbf{3},\textbf{1}\right) \\[2mm]
{\xi_{(\alpha}}^{P}\,f_{\beta)MNP} & = & 0 &, & \left(\textbf{3},\textbf{66}\right) \\[2mm]
3\,f_{\alpha R[MN}\,{f_{\beta PQ]}}^{R} \,-\, 2\,f_{(\alpha[MNP} \, \xi_{\beta)Q]} & = & 0 & , & \left(\textbf{3},\textbf{495}\right) \\[2mm]
\epsilon^{\alpha\beta}\,\left({\xi_{\alpha}}^{P}\,f_{\beta PMN}\,+\,\xi_{\alpha M}\,\xi_{\beta N}\right) & = & 0 & , & \left(\textbf{1},\textbf{66}\right)\\[2mm]
\epsilon^{\alpha\beta}\,f_{\alpha MNR}\,{f_{\beta PQ}}^{R}\,+\, \left(f\,\xi \ \textrm{ terms}\right) & = & 0 & , & 
\left(\textbf{1},\textbf{66}\right)\,\oplus\,\left(\textbf{1},\textbf{2079}\right) \\[2mm]
f_{\alpha MNP}\,{f_{\beta}}^{MNP} & = & 0 & , & \left(\textbf{3},\textbf{1}\right) \\[2mm]
\epsilon^{\alpha\beta}\,\,f_{\alpha[MNP}\,f_{\beta QRS]}|_{\textrm{SD}} & = & 0 & , & \left(\textbf{1},\overline{\textbf{462}}\right)
\end{array}
\end{equation}
where one can recognise all the QC's of $\mathcal{N}=4$ supergravity, plus the last two lines which therefore contain space-filling branes. In particular, the 
$\left(\textbf{1},\overline{\textbf{462}}\right)$ contains exactly the $32$ long weights that we expect from the results of the previous section.

To further truncate to the quarter-maximal theories, we perform the following truncation \cite{Dibitetto:2012ia}
\begin{equation}
\begin{array}{lclc}
\textrm{SL}(2,\mathbb{R})\times\textrm{SO}(6,6) & \longrightarrow & \left(\textrm{SL}(2,\mathbb{R})\right)^{3}\times\textrm{SO}(4,4) & , \\[2mm]
\Theta \, \in \, \underbrace{(\textbf{2},\textbf{12})}_{\xi_{\alpha M}}\,\oplus\,
\underbrace{(\textbf{2},\textbf{220})}_{f_{\alpha [MNP]}} & 
& (\textbf{2},\textbf{2},\textbf{2},\textbf{1})\,\oplus\,
(\textbf{2},\textbf{2},\textbf{2},\textbf{28}) & ,
\end{array} 
\end{equation}
where once more we may regroup the embedding tensor of the $\mathcal{N}=2$ theory into the object ${\Theta_{\Lambda}}^{A}$, where $\Lambda$ takes values in the 
$(\textbf{2},\textbf{2},\textbf{2},\textbf{1})$, while the index $A$ spans the whole adjoint representation of the global symmetry group.

The $\mathcal{N}=4$ QC irrep's break into
\begin{equation}\label{QC_Quart_4D}
\begin{array}{lclc}
(\textbf{3},\textbf{1}) & \longrightarrow & (\textbf{3},\textbf{1},\textbf{1},\textbf{1}) & , \\[2mm]
(\textbf{3},\textbf{66}) & \longrightarrow & (\textbf{3},\textbf{3},\textbf{1},\textbf{1})\,\oplus\,(\textbf{3},\textbf{1},\textbf{3},\textbf{1})\,\oplus\,
(\textbf{3},\textbf{1},\textbf{1},\textbf{28})\,\oplus\,\textrm{odd} & , \\[2mm]
(\textbf{3},\textbf{495}) & \longrightarrow & (\textbf{3},\textbf{1},\textbf{1},\textbf{1})\,\oplus\,(\textbf{3},\textbf{1},\textbf{1},\textbf{35}_{\rm S})
\,\oplus\,(\textbf{3},\textbf{1},\textbf{1},\textbf{35}_{\rm C}) \\[1mm]
 & & \,\oplus\,(\textbf{3},\textbf{3},\textbf{1},\textbf{28})\,\oplus\,(\textbf{3},\textbf{1},\textbf{3},\textbf{28})\,\oplus\,\textrm{odd} & , \\[2mm]
(\textbf{1},\textbf{66}) & \longrightarrow & (\textbf{1},\textbf{3},\textbf{1},\textbf{1})\,\oplus\,(\textbf{1},\textbf{1},\textbf{3},\textbf{1})\,\oplus\,
(\textbf{1},\textbf{1},\textbf{1},\textbf{28})\,\oplus\,\textrm{odd} & ,\\[2mm]
(\textbf{1},\textbf{2079}) & \longrightarrow & (\textbf{1},\textbf{3},\textbf{1},\textbf{1})\,\oplus\,(\textbf{1},\textbf{1},\textbf{3},\textbf{1})\,\oplus\,
(\textbf{1},\textbf{3},\textbf{3},\textbf{1}) \\[1mm]
 & & \,\oplus\,(\textbf{1},\textbf{1},\textbf{1},\textbf{28})\,\oplus\,(\textbf{1},\textbf{3},\textbf{1},\textbf{28})\,\oplus\,
(\textbf{1},\textbf{1},\textbf{3},\textbf{28}) \\[1mm]
& & \,\oplus\,(\textbf{1},\textbf{3},\textbf{3},\textbf{28})\,\oplus\,(\textbf{1},\textbf{1},\textbf{1},\textbf{350}) \\[1mm]
 & & \,\oplus\,(\textbf{1},\textbf{3},\textbf{1},\textbf{35}_{\rm V})\,\oplus\,(\textbf{1},\textbf{1},\textbf{3},\textbf{35}_{\rm V})
\oplus\,\textrm{odd} & .
\end{array}
\end{equation}
The QC's demanded for consistency of the $\mathcal{N}=2$ theory read \cite{deWit:2005ub}
\begin{equation}
\begin{array}{rclc}
{\Theta_{\Lambda}}^{\alpha}\,{\Theta_{\Sigma}}^{\beta}\,{f_{\alpha\beta}}^{\gamma}\,+\,{\left[t_{\alpha}\right]_{\Lambda}}^{\Gamma}\,
{\Theta_{\Sigma}}^{\alpha}\,{\Theta_{\Gamma}}^{\gamma} & = & 0 & , \\[2mm]
{\Theta}^{\Lambda[\alpha}\,{\Theta_{\Lambda}}^{\beta]} & = & 0 & ,
\end{array}
\end{equation}
where $\left\{{f_{\alpha\beta}}^{\gamma}\right\}$ and $\left\{{\left[t_{\alpha}\right]_{\Lambda}}^{\Gamma}\right\}$ are the global symmetry generators in the adjoint and vector representation, respectively. 
The unneeded QC irrep's containing the longest weights turn out to be 
$(\textbf{3},\textbf{1},\textbf{1},\textbf{35}_{\rm S})\,\oplus\,(\textbf{3},\textbf{1},\textbf{1},\textbf{35}_{\rm C})\,\oplus\,
(\textbf{1},\textbf{3},\textbf{1},\textbf{35}_{\rm V})\,\oplus\,(\textbf{1},\textbf{1},\textbf{3},\textbf{35}_{\rm V})$, once again in perfect agreement with the analysis performed in the previous section.

A last step that can be discussed here in four dimensions is the one further breaking supersymmetry to $\mathcal{N}=1$. This truncation is concretely realised as follows
\begin{equation}
\begin{array}{lclc}
\left(\textrm{SL}(2,\mathbb{R})\right)^{3}\times\textrm{SO}(4,4) & \longrightarrow & \left(\textrm{SL}(2,\mathbb{R})\right)^{7} & , \\[2mm]
\Theta \, \in \, (\textbf{2},\textbf{2},\textbf{2},\textbf{1})\,\oplus\,(\textbf{2},\textbf{2},\textbf{2},\textbf{28}) & 
&  (\textbf{2},\textbf{2},\textbf{2},\textbf{2},\textbf{2},\textbf{2},\textbf{2}) & .
\end{array} 
\end{equation}
However, the minimal theory one ends up with is purely coupled to chiral multiplets and hence it possesses \emph{no vector fields}. As a consequence, the obtained supergravity model will reorganise the embedding tensor deformations surviving the above truncation into massive deformations not associated with any gauging. In particular, in this model all of them may be interpreted as parameters inducing a holomorphic superpotential in the seven complex scalar fields. Due to this, we do not have any QC's required for consistency and we do expect all of them to be sourced by space-filling branes.
This class of theories will be studied in detail in the next section, where these parameters will be interpreted as the $2^7=128$ generalised fluxes coming from an orbifold compactification of type IIB string theory.

\vskip .6cm

To conclude this section, we quickly comment on the three-dimensional case. The embedding tensor of the maximal theory belongs to the ${\bf 1 \oplus 3875}$ of ${\rm E}_{8(8)}$ \cite{Nicolai:2000sc}, while the quadratic constraint belongs to the ${\bf 3875 \oplus 147250}$ \cite{deWit:2008ta}. By truncating  the theory to the half-maximal one with symmetry ${\rm SO}(8,8)$, one can show that the embedding tensor is truncated to the one of the half-maximal theory \cite{Nicolai:2001ac}, while the quadratic constraints are truncated to the quadratic constraints of the half-maximal theory plus extra constraints, and the highest-dimensional representation of such remaining constraints is the ${\bf 6435}$, in agreement with the results of the previous section (see Table \ref{summaryalltruncations}). Similarly, one can study how the further truncations to the theories with eight and four supersymmetries give patterns for the quadratic constraints in agreement with Table \ref{summaryalltruncations}. Moreover, the theory with four supersymmetries in three dimensions can be further truncated to the minimal theory. 
Just as in the 4D minimally supersymmetric case, we expect no quadratic constraint to be present here and hence all highest weights surviving the branching of the QC's will correspond to exotic space-filling brane states. Effective descriptions of this type can be \emph{e.g.} obtained by compactifying M-theory on Joyce 8-manifolds of $\textrm{Spin}(7)$ holonomy. Internal manifolds of this type admit an orbifold limit where they are described as $T^{8}/\Gamma$, where the discrete symmetry $\Gamma$ can be \emph{e.g.} $\mathbb{Z}_{2}^{4}$. These M-theory backgrounds are also known to have perturbative corners given by type IIA orientifolds of Joyce 7-manifolds of $\textrm{G}_{2}$ holonomy \cite{Majumder:2001dx}, or heterotic strings on such $\textrm{G}_{2}$-manifolds \cite{Acharya:1996ef}.
However, we leave a careful analysis of all these features of the three-dimensional case as a future project.

\section{IIB on $T^6/(\mathbb{Z}_2 \times \mathbb{Z}_2)$, fluxes and Bianchi identities}

As explained at the end of the previous section, performing three $\mathbb{Z}_{2}$ truncations on a maximal gauged supergravity theory in four dimensions yields an $\mathcal{N}=1$ supergravity model where the supergravity multiplet is coupled to seven chiral multiplets. 
The scalar sector of the theory contains seven complex fields spanning the coset space $\left(\textrm{SL}(2,\mathbb{R})/\textrm{SO}(2)\right)^{7}$ which are usually denoted by 
$\Phi^{\alpha}\,\equiv\,\left(S,T_{i},U_{i}\right)$ with $i=1,2,3$. 
The kinetic Lagrangian follows from the K\"ahler potential
\begin{equation}
\label{Kaehler_STU}
K\,=\,-\log\left(-i\,(S-\overline{S})\right)\,-\,\sum_{i=1}^{3}{\log\left(-i\,(T_{i}-\overline{T}_{i})\right)}\,-\,\sum_{i=1}^{3}{\log\left(-i\,(U_{i}-\overline{U}_{i})\right)}\ ,
\end{equation}
yielding
\begin{equation}
\mathcal{L}_{\textrm{kin}} = \frac{\partial
S\partial \overline{S}}{\left(-i(S-\overline{S})\right)^2} + \, \sum_{i=1}^{3}\left(\frac{\partial
T_{i}\partial \overline{T}_{i}}{\left(-i(T_{i}-\overline{T}_{i})\right)^2} + \frac{\partial
U_{i}\partial \overline{U}_{i}}{\left(-i(U_{i}-\overline{U}_{i})\right)^2}\right) \ .
\end{equation}

The presence of fluxes induces a scalar potential $V$ for the scalar fields which is given in terms of the above K\"ahler potential and a holomorphic superpotential $W$ by
\begin{equation}
\label{V_N=1}
V\,=\,e^{K}\left(-3\,|W|^{2}\,+\,K^{\alpha\bar{\beta}}\,D_{\alpha}W\,D_{\bar{\beta}}\overline{W}\right)\ ,
\end{equation}
where $K^{\alpha\bar{\beta}}$ is the inverse of the aformentioned K\"ahler metric and $D_{\alpha}$ denotes the K\"ahler-covariant derivative.

As already mentioned earlier, the superpotential $W$ is induced by the $128$ deformation parameters surviving the truncation of the $\textbf{912}$ of $\textrm{E}_{7(7)}$ w.r.t.~$\mathbb{Z}_{2}^{3}$. The form of such superpotential is given by the most general polynomial in $\left(S,T_{i},U_{i}\right)$ without any mixed terms, \emph{i.e.}
\begin{equation}
\label{W_Form}
W\,\sim\,1\ + \ \dots \ + \ S\,T_{1}\,T_{2}\,T_{3}\,U_{1}\,U_{2}\,U_{3}\ ,
\end{equation}
which precisely includes $128$ terms, each of which is induced by its own deformation parameter.

In what follows we will review how the above class of minimal supergravities arise from orbifold compactifications of type IIB string theory down to four dimensions. Before moving to that analysis though, let us note that the consistency of the $\mathcal{N}=1$ theory \emph{only} requires all massive deformations of this type to be arranged into a holomorphic superpotential. In our specific case this requirement is automatically fulfilled by any polynomial function of the type given in \eqref{W_Form}, while no further QC's on its coefficients are needed. This means that any QC in terms of the superpotential couplings surviving the $\mathbb{Z}_{2}^{3}$ truncation is expected to be relaxed in our compactification yielding minimal supersymmetry by means of space-filling branes.

As previously anticipated, the above minimal supergravity models with seven chiral multiplets arise from dimensional reductions of type IIB string theory on $T^{6}/\left(\mathbb{Z}_{2}\times\mathbb{Z}_{2}\right)$ with O3/O7-planes.\footnote{See for instance \cite{Blumenhagen:2006ci} for a review on orientifold models with fluxes.} The $\mathbb{Z}_{2}\times\mathbb{Z}_{2}$ orbifold acts on the six internal coordinates precisely as described below equation~\eqref{branestruncationD=4minimal}. What further realises a supersymmetry breaking down to a minimal amount is a $\mathbb{Z}_{2}$ flipping the sign of all the coordinates on the $T^{6}$. There need to be O3-planes located at each fixed point of this involution, while a triplet of O7-planes are placed at fixed points of those involutions obtained by combining this last 
$\mathbb{Z}_{2}$ with the three non-trivial generators of the orbifold group. 

Adopting the type IIB language, the seven complex scalars of the $\mathcal{N}=1$ model have the following physical interpretation
\begin{equation}
\begin{array}{lclc}
S & \leftrightarrow & \textrm{Axiodilaton} & , \\[2mm]
T_{i} & \leftrightarrow & \textrm{K\"ahler moduli} & , \\[2mm]
U_{i} & \leftrightarrow & \textrm{Complex structure moduli} & .
\end{array}
\end{equation}
The first understanding of the mechanism that perturbatively induces a superpotential from fluxes in this context is due to \cite{Gukov:1999ya}, where a superpotential of the form
\begin{equation}
\label{W_GVW}
W_{\rm GVW}\,=\underbrace{\,P_{F}(U_{i})\,}_{F \textrm{ flux}}\,+\,S\,\underbrace{\,P_{H}(U_{i})\,}_{H \textrm{ flux}}\ ,
\end{equation}
where $P_{F}$ \& $P_{H}$ are cubic polynomials in the complex structure moduli controlled by R-R and NS-NS three-form fluxes with different legs in internal space.
Superpotentials of this type were found to describe special type IIB Minkowski backgrounds \cite{Giddings:2001yu} with no-scale symmetry due to the absence of $T$-dependence in $W$.

In \cite{Shelton:2005cf} it was argued, on the basis of string dualities, that the superpotential in \eqref{W_GVW} should be generalised to contain new fluxes which are named \emph{non-geometric}. Subsequently, in \cite{Aldazabal:2006up} the set of generalised fluxes was further enlarged to include the complete set of objects closed under perturbative and non-perturbative string dualities. The corresponding superpotential reads
\begin{equation}
\label{W_Complete}
\begin{array}{lclc}
W &= & \underbrace{\,P_{F}(U_{i})\,}_{F \textrm{ flux}}\,+\,S\,\underbrace{\,P_{H}(U_{i})\,}_{H \textrm{ flux}}\,+\,\sum\limits_{k}{T_{k}\,\underbrace{\,P_{Q}^{(k)}(U_{i})\,}_{Q \textrm{ flux}}}\,+\,S\,\sum\limits_{k}{T_{k}\,\underbrace{\,P_{P}^{(k)}(U_{i})\,}_{P \textrm{ flux}}} & \\[2mm]
&+& T_{1}\,T_{2}\,T_{3}\,\big(\underbrace{\,P_{F'}(U_{i})\,}_{F' \textrm{ flux}}\,+\,S\,\underbrace{\,P_{H'}(U_{i})\,}_{H' \textrm{ flux}}\big)\,+\,
\sum\limits_{k}{T_{i}\,T_{j}\,\big(\underbrace{\,P_{Q'}^{(k)}(U_{i})\,}_{Q' \textrm{ flux}}}\,+\,S\,\underbrace{\,P_{P'}^{(k)}(U_{i})\,}_{P' \textrm{ flux}}\big) & ,
\end{array}
\end{equation}
where $P_{F}$, $P_{H}$, $P_{Q}^{(k)}$ and $P_{P}^{(k)}$ are cubic polynomials in the complex structure moduli given by
\begin{equation}
\begin{array}{cclc}
P_{F}(U_{i}) & = & a_{0}\,-\,\sum\limits_{i}{a_{1}^{(i)}\,U_{i}}\,+\,\sum\limits_{i}{a_{2}^{(i)}\,\dfrac{U_{1}\,U_{2}\,U_{3}}{U_{i}}}\,-\,a_{3}\,U_{1}\,U_{2}\,U_{3} & , \\[3mm]
P_{H}(U_{i}) & = & -b_{0}\,+\,\sum\limits_{i}{b_{1}^{(i)}\,U_{i}}\,-\,\sum\limits_{i}{b_{2}^{(i)}\,\dfrac{U_{1}\,U_{2}\,U_{3}}{U_{i}}}\,+\,b_{3}\,U_{1}\,U_{2}\,U_{3} & , \\[3mm]
P_{Q}^{(k)}(U_{i}) & = & c_{0}^{(k)}\,+\,\sum\limits_{i}{c_{1}^{(ik)}\,U_{i}}\,-\,\sum\limits_{i}{c_{2}^{(ik)}\,\dfrac{U_{1}\,U_{2}\,U_{3}}{U_{i}}}\,-\,c_{3}^{(k)}\,U_{1}\,U_{2}\,U_{3} & , \\[3mm]
P_{P}^{(k)}(U_{i}) & = & -d_{0}^{(k)}\,-\,\sum\limits_{i}{d_{1}^{(ik)}\,U_{i}}\,+\,\sum\limits_{i}{d_{2}^{(ik)}\,\dfrac{U_{1}\,U_{2}\,U_{3}}{U_{i}}}\,+\,d_{3}^{(k)}\,U_{1}\,U_{2}\,U_{3} & ,
\end{array}
\end{equation}
while $P_{F'}$, $P_{H'}$, $P_{Q'}^{(k)}$ and $P_{P'}^{(k)}$ are cubic polynomials in the complex structure moduli given by
\begin{equation}
\begin{array}{cclc}
 P_{F'}(U_{i}) & = & a'_{3}\,-\,\sum\limits_{i} {{a'}_{2}^{(i)}\,U_{i}} \,+\, \sum\limits_{i}{{a'}_{1}^{(i)}\,\dfrac{U_{1}\,U_{2}\,U_{3}}{U_{i}}}\,-\,{a'}_{0}
\,U_{1}\,U_{2}\,U_{3} & , \\[3mm]
P_{H'}(U_{i}) & = & -{b'}_{3}\,+\,\sum\limits_{i}{{b'}_{2}^{(i)}\,U_{i}}\,-\,\sum\limits_{i}{{b'}_{1}^{(i)}\,\dfrac{U_{1}\,U_{2}\,U_{3}}{U_{i}}}\,+\,{b'}_{0}
 \,U_{1}\,U_{2}\,U_{3} & , \\[3mm]
P_{Q'}^{(k)}(U_{i}) & = & {c'}_{3}^{(k)}\,+\,\sum\limits_{i}{{c'}_{2}^{(ik)}\,U_{i}}\,-\,\sum\limits_{i}{{c'}_{1}^{(ik)}\,\dfrac{U_{1}\,U_{2}\,U_{3}}{U_{i}}}
\,-\,{c'}_{0}^{(k)}\,U_{1}\,U_{2}\,U_{3} & , \\[3mm]
P_{P'}^{(k)}(U_{i}) & = & -{d'}_{3}^{(k)}\,-\,\sum\limits_{i}{{d'}_{2}^{(ik)}\,U_{i}}\,+\,\sum\limits_{i}{{d'}_{1}^{(ik)}\,\dfrac{U_{1}\,U_{2}\,U_{3}}{U_{i}}}
 \,+\,{d'}_{0}^{(k)}\,U_{1}\,U_{2}\,U_{3} & .
\end{array}
\end{equation}
Note that the superpotential in \eqref{W_Complete} exactly comprises the aforementioned $128$ terms coming from the $\mathbb{Z}_{2}^{3}$ truncation of the embedding tensor of maximal gauged supergravity in four dimensions.

Once we understood these orbifold compactifications of type IIB with fluxes as $\mathbb{Z}_{2}^{3}$ truncations of maximal gauged supergravities in $D=4$, we would now like to interpret all the QC irrep's accordingly as quadratic conditions for the fluxes which may be sourced by space-filling objects in string theory. 
By applying our prescription, we can identify all the space-filling branes as the longest weights of the QC irrep's of the $\mathcal{N}=2$ theory decomposed by performing the last  
$\mathbb{Z}_{2}$ truncation. All of the other long weights contained in the rest of the $\mathbb{Z}_{2}$ even QC irrep's which were unneeded in the two previous steps of the truncation are interpreted as \emph{Bianchi Identities} (BI), \emph{i.e.} consistency constraints of the background itself. To be more explicit, the BI's can be seen as conditions for the absence of extra space-filling objects whose negative-tension counterparts realise the background itself. Mathematically, those quadratic constraints can be interpreted as conditions enforcing the closure of the flux-twisted exterior derivative operator.

By following the above prescription, let us now proceed to identify the space-filling brane states and the BI's for type IIB compactifications on 
$T^{6}/(\mathbb{Z}_{2}\times\mathbb{Z}_{2})$. The truncation net reads\footnote{The consistent identification for the different $\textrm{SL}(2,\mathbb{R})$ labels is 
$\textrm{SL}(2,\mathbb{R})^{7}\,\equiv\,\textrm{SL}(2,\mathbb{R})_{U_{1}}\times\textrm{SL}(2,\mathbb{R})_{U_{2}}\times\textrm{SL}(2,\mathbb{R})_{U_{3}}\times
\textrm{SL}(2,\mathbb{R})_{T_{1}}\times\textrm{SL}(2,\mathbb{R})_{T_{2}}\times\textrm{SL}(2,\mathbb{R})_{T_{3}}\times\textrm{SL}(2,\mathbb{R})_{S}$.}
\begin{equation}
\begin{array}{ccc}
\textrm{E}_{7(7)} & & \\[2mm]
\downarrow \, \mathbb{Z}_{2} & & \\[2mm]
\textrm{SL}(2,\mathbb{R})\times\textrm{SO}(6,6) & \rightarrow & \left(\textbf{1},\overline{\textbf{462}}\right)\\[2mm]
\downarrow \, \mathbb{Z}_{2} & & \\[2mm]
\textrm{SL}(2,\mathbb{R})^{3}_{U}\times\textrm{SO}(4,4) & \rightarrow & 
\begin{array}{l}
\left(\textbf{3},\textbf{1},\textbf{1},\textbf{35}_{\rm S}\right) \oplus 
\left(\textbf{3},\textbf{1},\textbf{1},\textbf{35}_{\rm C}\right) \oplus \\[1mm]
\oplus  \left(\textbf{1},\textbf{3},\textbf{1},\textbf{35}_{\rm V}\right) \oplus 
\left(\textbf{1},\textbf{1},\textbf{3},\textbf{35}_{\rm V}\right)
\end{array}
\\[2mm]
\downarrow \, \mathbb{Z}_{2} & & \\[2mm]
\textrm{SL}(2,\mathbb{R})^{7} & &
\end{array}
\end{equation}
where the irrep's pulled out on the right of the above diagram will precisely give rise to the BI's of our theory, upon truncation down to $\textrm{SL}(2,\mathbb{R})^{7}$, while the
space-filling branes will be captured by those QC irrep's surviving the triple discrete truncation, \emph{i.e.}
\begin{equation}\nonumber
\left(\textbf{1},\textbf{1},\textbf{1},\textbf{350}\right) \oplus \left(\textbf{1},\textbf{3},\textbf{3},\textbf{28}\right) \oplus 
\left(\textbf{3},\textbf{1},\textbf{3},\textbf{28}\right) \oplus \left(\textbf{3},\textbf{3},\textbf{1},\textbf{28}\right) \oplus \ \textrm{shorter weights}\ ,
\end{equation}
of $\textrm{SL}(2,\mathbb{R})^{3}_{U}\times\textrm{SO}(4,4)$. Further decomposition down to $\textrm{SL}(2,\mathbb{R})^{7}$ yields the following space-filling branes
\begin{equation}\label{T6Z2Z2branes}
\begin{array}{ccccc}
\left(\textbf{3}_{T_{i}},\textbf{3}_{T_{j}},\textbf{3}_{T_{k}}\right) & & & (1 \ \textrm{irrep}) & , \\[2mm]
\left(\textbf{3}_{S},\textbf{3}_{T_{i}},\textbf{3}_{T_{j}}\right) & & & (3 \ \textrm{irrep's}) & , \\[2mm]
\left(\textbf{3}_{S},\textbf{3}_{U_{i}},\textbf{3}_{U_{j}}\right) & & & (3 \ \textrm{irrep's}) & , \\[2mm]
\left(\textbf{3}_{T_{i}},\textbf{3}_{U_{j}},\textbf{3}_{U_{k}}\right) & & & (3 \ \textrm{irrep's}) & , \\[2mm]
\left(\textbf{3}_{T_{i}},\textbf{3}_{U_{i}},\textbf{3}_{U_{j}}\right) & & & (6 \ \textrm{irrep's}) & , 
\end{array}
\end{equation}
which precisely contain the $16\times 2^{3}\,=\, 2^{7}\,=\,128$ space-filling branes we needed from the previous analysis, 
while for the BI's one finds
\begin{equation}\label{T6Z2Z2BI}
\begin{array}{ccccc}
\left(\textbf{3}_{T_{i}},\textbf{3}_{T_{j}},\textbf{3}_{U_{j}}\right) & & & (6 \ \textrm{irrep's}) & , \\[2mm]
\left(\textbf{3}_{S},\textbf{3}_{T_{i}},\textbf{3}_{U_{j}}\right) & & & (6 \ \textrm{irrep's}) & , 
\end{array}
\end{equation}
containing $12\times 2^{3}\,=\,96$ BI's in total.
Note that this perfectly matches the results of \cite{Lombardo:2017yme}, to which the present analysis can be regarded as an independent check. 

A further physical comment concerning the possibility of relaxing the constraints. As already stated above, each and every quadratic condition for type IIB fluxes appearing in 
\eqref{T6Z2Z2branes} can be relaxed by adding the correspondent space-filling (exotic) brane sourcing the associated flux tadpole. All those branes are consistently preserving the same four real supercharges as the very background. 
On the other hand, it becomes very natural to ask about a similar possibility for the BI's in \eqref{T6Z2Z2BI}. Indeed, just as any other algebraic constraint, these are not needed for consistency. Their physical interpretation is that of enforcing the condition for closure of the flux-twisted exterior derivative operator defined on our CY background. 
Our intuition seems to suggest that the original CY will be deformed into an $G$-structure manifold as an effect of the backreaction of fluxes to the background geometry. 
In this context, the twisted exterior derivative operator will now receive torsion-induced contributions. Following the philosophy of \cite{Danielsson:2014ria}, part of the contributions to the internal curvature can be interpreted as the presence of space-filling KK-monopoles and T-duals thereof.

\section{Conclusions}

In this paper we have considered the supergravity theories that arise as sequences of $\mathbb{Z}_2$ truncations of the maximal theories. We have determined in all cases the  1/2-BPS space-filling branes that preserve the supersymmetry of the truncated theory and the representations of the symmetry of such theory to which they belong.  We have then discussed all the possible gaugings of these theories as described in terms of the embedding tensor. We have shown that for any theory, among the representations of the quadratic constraint on the embedding tensor that survive the truncation but are not needed for supersymmetry, the highest-dimensional ones are precisely those of the 1/2-BPS space-filling branes that preserve the same supersymmetry of the truncated theory. This can be naturally interpreted as the fact that these quadratic constraints after the truncation become tadpole conditions for such branes.

We point out that the number of different $\mathbb{Z}_2$  truncations of a given supergravity theory, that from a group-theory point of view is given by the number of different ways in which the symmetry of the truncated theory can be embedded in the symmetry of the original one, is also given by the number of vector central charges of the supersymmetry algebra. This rather intriguing result shows once again the deep relation between supersymmetry and group theory.

Although the analysis in this paper was performed uniquely in terms of the branes of the IIB theory, it would be interesting  to reinterpret this from the point of view of IIA and also from the point of view of M-theory. In particular, the analysis of the tadpole conditions for all the possible fluxes that can be included in the  $T^{6}/\left(\mathbb{Z}_{2}\times\mathbb{Z}_{2}\right)$ orientifold was performed in \cite{Lombardo:2017yme} also in the case of the IIA O6 orientifold, and this could then be compared with the techniques developed in this paper.  

Moreover,
 one can extend the three-dimensional case to include a more detailed analysis. In particular, the minimally supersymmetric gauged theories turn out to have fewer constraints coming from consistency and supersymmetry \cite{deWit:2004yr} and hence a further truncation down to two real supercharges could be a very valuable venue for string model-building. 
Three-dimensional theories with minimal supersymmetry can be  obtained by compactifying M-theory on Joyce 8-manifolds of $\textrm{Spin}(7)$ holonomy, and internal manifolds of this type admit an orbifold limit where they are described as $T^{8}/\Gamma$, where the discrete symmetry $\Gamma$ can be \emph{e.g.} $\mathbb{Z}_{2}^{4}$, so that our techniques can be applied to this case. One can also use the methods presented in this paper to investigate the duality relations with perturbative corners given by type IIA orientifolds of Joyce 7-manifolds of $\textrm{G}_{2}$ holonomy \cite{Majumder:2001dx}, or heterotic strings on such $\textrm{G}_{2}$-manifolds \cite{Acharya:1996ef}.
We leave a careful analysis of all these features of the three-dimensional case as a future project.

Finally, we stress again that the vast majority of the branes discussed in this paper are exotic, in the sense that they do not have a clear higher dimensional origin. It would be of extreme interest to get any understanding of the dynamics of these objects, that in our analysis must be included for symmetry arguments. This would dramatically improve our understanding of string theory and our ability to construct models.

\vskip .7cm

\section*{Acknowledgments}

We would like to thank G. Pradisi for carefully reading and suggesting corrections to the manuscript, and in particular N. Gubernari who contributed during his master thesis project to the derivation of the $D=8$ truncations. 
The authors would like to thank the Galileo Galilei Institute (GGI) in Florence for hosting the workshop ``Supergravity: what next?'' where this project was conceived.  
We furthermore respectively acknowledge the hospitality of the theory group of the University of Uppsala and La Sapienza University in Rome, where different parts of this project were completed. The work of GD is funded by the Swedish Research Council (VR).

\vskip 1.5cm

\appendix

\section{\label{appendix1} $D=8$ spinor conventions}

In this appendix we discuss in detail all the spinor conventions that we have adopted in section 2. All the fermions of the eight-dimensional maximal theory are spinors of ${\rm SO}(3)\times {\rm SO}(2)$.  We have denoted with $\sigma_m$ ($m=1,2,3$) the Pauli matrices of ${\rm SO}(3)$. The gamma matrices of ${\rm SO}(2)$ are the first two Pauli matrices, that we denote with $\tau_a$ ($a=1,2$), while the third Pauli matrix is the chiraly matrix of ${\rm SO}(2)$, 
\begin{equation}
\tau_a \tau_b = i \epsilon_{ab} \tau_3 \quad .\label{tau3definition}
\end{equation}
As eq. \eqref{sortofchirality} shows, all the spinors are chiral with respect to $\gamma_9 \tau_3$. They also satisfy the Majorana condition in eq. \eqref{symplMajorana}, with $C = C_8 \sigma_2 \tau_1$ as in eq. \eqref{definitionofC}. The matrix $C_8$ is defined in eq. \eqref{CdaggergammaCD=8}, and commutes with $\gamma_9$. On the other hand, the matrix $\tau_1$ anticommutes with $\tau_3$, and hence the matrix $C$ and $\gamma_9 \tau_3$ anticommute, so that the chirality conditions in eq. \eqref{sortofchirality} and the Majorana condition in eq. \eqref{symplMajorana} are compatible. We call the Majorana conditions in eq. \eqref{symplMajorana} `symplectic' because $C_8$ is symmetric and satisfies the condition in eq. \eqref{CdaggergammaCD=8} with the minus sign, so that the condition $\Psi = C_8 \overline{\Psi}^T$ would not be consistent. This is standard in the supergravity literature. 

We now want to discuss the reality properties and the properties under Majorana flip of the various fermionic bilinears that  can be constructed, and in particular of the ones that occur in section 2. We make use of the identities
\begin{align}
& C^\dagger \gamma_\mu C = - \gamma_\mu^T \nonumber \\
& C^\dagger \sigma_m C = - \sigma_m^T \nonumber \\
& C^\dagger \tau_a C = \tau_a^T\nonumber \\
& C^\dagger \tau_3 C = -\tau_3 \quad ,
\end{align}
the first of which is the same as eq. \eqref{CdaggergammaCD=8}. A Majorana spinor $\Psi$, satisfying eq. \eqref{symplMajorana},  also satisfies $\overline{\Psi} = -\Psi^T C^\dagger$. As a consequence, for instance the bilinear $\overline{\psi} \sigma_m \tau_a \chi$ is equal to $-\overline{\chi} \sigma_m \tau_a \psi$, which means that the bilinear is odd under Majorana flip. By complex conjugation the bilinear goes to $\overline{\chi} \sigma_m \tau_a \psi$, which is minus the bilinear itself because of Majorana flip. As a conseguence, the bilinear is purely imaginary. One can easily generalise this to get the reality conditions and the Majorana-flip properties for all the bilinears. We give a summary of the properties of the various bilinears in Table \ref{fermibilinearsD=8}.

\begin{table}[t!]
\begin{center}

\scalebox{1}{
\begin{tabular}{|c|c|c|}
\hline \rule[-1mm]{0mm}{6mm}  bilinear & reality property & flip\\
\hline
\hline \rule[-1mm]{0mm}{6mm}  $\overline{\psi} \chi$ & real & even\\
\hline \rule[-1mm]{0mm}{6mm}
 $\overline{\psi}\gamma_\mu \chi$ & imaginary & odd\\
 $\overline{\psi} \sigma_m \chi$ & imaginary & odd\\
$\overline{\psi} \tau_a \chi$ & real & even\\
$\overline{\psi} \tau_3 \chi$ & imaginary & odd\\
\hline \rule[-1mm]{0mm}{6mm} $\overline{\psi} \gamma_{\mu\nu} \chi$ & real & odd  \\
$\overline{\psi} \gamma_{\mu}  \sigma_m \chi$ & real & even  \\
$\overline{\psi} \gamma_{\mu}  \tau_a \chi$ & imaginary & odd  \\
$\overline{\psi} \gamma_{\mu}  \tau_3 \chi$ & real & even  \\
$\overline{\psi} \sigma_m  \tau_a \chi$ & imaginary & odd  \\
$\overline{\psi} \sigma_m  \tau_3 \chi$ & real & even  \\
\hline \rule[-1mm]{0mm}{6mm} $\overline{\psi} \gamma_{\mu\nu\rho} \chi$ & imaginary & even  \\
$\overline{\psi} \gamma_{\mu\nu} \sigma_m \chi$ & imaginary & even  \\
$\overline{\psi} \gamma_{\mu\nu} \tau_a \chi$ & real & odd  \\
$\overline{\psi} \gamma_{\mu\nu} \tau_3 \chi$ & imaginary & even  \\
$\overline{\psi} \gamma_{\mu}\sigma_m \tau_a \chi$ & real & even  \\
$\overline{\psi} \gamma_{\mu}\sigma_m \tau_3 \chi$ & imaginary & odd  \\
\hline \rule[-1mm]{0mm}{6mm} $\overline{\psi} \gamma_{\mu\nu\rho\sigma} \chi$ & real &  even \\
$\overline{\psi} \gamma_{\mu\nu\rho} \sigma_m \chi$ & real &   odd\\
$\overline{\psi} \gamma_{\mu\nu\rho} \tau_a \chi$ & imaginary &   even\\
$\overline{\psi} \gamma_{\mu\nu\rho} \tau_3 \chi$ & real &  odd \\
$\overline{\psi} \gamma_{\mu\nu}\sigma_m \tau_a \chi$ & imaginary & even \\
$\overline{\psi} \gamma_{\mu\nu}\sigma_m \tau_3 \chi$ & real  &  odd \\
\hline \rule[-1mm]{0mm}{6mm} 
$\overline{\psi} \gamma_{\mu\nu\rho\sigma} \sigma_m \chi$ & imaginary  & odd  \\
$\overline{\psi} \gamma_{\mu\nu\rho\sigma} \tau_a \chi$ & real & even  \\
$\overline{\psi} \gamma_{\mu\nu\rho}\sigma_m \tau_a \chi$ &  real & odd \\
$\overline{\psi} \gamma_{\mu\nu\rho}\sigma_m \tau_3 \chi$ & imaginary &  even \\
\hline \rule[-1mm]{0mm}{6mm} 
$\overline{\psi} \gamma_{\mu\nu\rho\sigma} \sigma_m \tau_a\chi$ & imaginary & odd  \\
\hline

\end{tabular}
}
\end{center}
  \caption{\sl The reality properties and the properties under Majorana flip of various fermionic bilinears. The analogous properties for all the other bilinears can be derived from those in this table using the duality relations in eq. \eqref{dualityrelations}.
  \label{fermibilinearsD=8}}
\end{table}

We now want to derive the duality relations among different bilinears. Starting from the definition of the $\gamma_9$ matrix given in eq. \eqref{definitiongammanine}, by multiple contractions from the left with  gamma matrices one gets
\begin{equation}
\gamma_{\mu_1 ...\mu_m} = -\frac{ (-)^{\left[ \tfrac{n+1}{2} \right]} i}{n!} \epsilon_{\mu_1 ...\mu_m \nu_1 ...\nu_n} \gamma^{\nu_1 ...\mu_n} \gamma_9 \qquad m+n=8 \quad . \label{dualityrelations}
\end{equation}
Using this relation and the chirality properties in eq. \eqref{sortofchirality}, the properties of the bilinears that contain the matrix $\gamma_{\mu_1 ...\mu_m}$ are related to those of the bilinears that contain the matrix $\gamma_{\nu_1 ...\nu_n} \tau_3$. In particular from the bilinears given in Table \ref{fermibilinearsD=8} one can derive all the others.

Given a spinor $\Psi$ satisfying $\gamma_9 \tau_3 \Psi = \Psi$, for the particular case of $m=n=4$ one gets
 \begin{equation}
\gamma_{\mu_1 ...\mu_4}\Psi = -\frac{i}{4!} \epsilon_{\mu_1 ...\mu_4 \nu_1 ...\nu_4} \gamma^{\nu_1 ...\mu_4} \tau_3 \Psi  \quad . 
\end{equation}
Contracting from the left with $\tau_a $ and using eq. \eqref{tau3definition}, one then obtains  the self-duality condition
 \begin{equation}
\gamma_{\mu_1 ...\mu_4}\tau_a\Psi = -\frac{1}{4!} \epsilon_{\mu_1 ...\mu_4 \nu_1 ...\nu_4} \epsilon_{ab} \gamma^{\nu_1 ...\mu_4} \tau_b \Psi  \quad . \label{gamma4psiselfduality}
\end{equation}

In general, in eight dimensions one can impose on a doublet of 4-forms $X_{\mu_1 ...\mu_4 \, a}$ the self-duality condition 
\begin{equation}
X_{\mu_1 ...\mu_4 \, a} = \frac{\alpha}{4!} \epsilon_{\mu_1 ...\mu_4 \nu_1 ...\nu_4} \epsilon_{ab} X^{\nu_1 ...\nu_4}{}_b \quad ,\label{selfdualityinD=8}
\end{equation}
where $\alpha$ can be either $1$ or $-1$. In particular, eq. \eqref{gamma4psiselfduality} corresponds to the case $\alpha=-1$. On the other hand, in section 2 we have shown that from the field-strength $F_{\mu\nu\rho\sigma}^A$ of the 3-form potential $A_{\mu\nu\rho}^A $ one can construct the composite quantity   $F_{\mu\nu\rho\sigma}^A V_{Aa}$ that satisfies the self-duality relation in eq. \eqref{selfdualityF4}, corresponding again to the case $\alpha=-1$ in eq. \eqref{selfdualityinD=8}. In general, if $X_{\mu_1 ...\mu_4 \, a}$ and $Y_{\mu_1 ...\mu_4 \, a}$ satisfy eq. \eqref{selfdualityinD=8} with the same $\alpha$, one can prove that the following identities hold:
\begin{equation}
X_{\mu\nu\rho\sigma \, a} Y^{\mu\nu\rho\sigma}{}_a = 0 \qquad \quad X_{\mu\nu\rho\sigma \, a} Y^{\mu\nu\rho\sigma}{}_b \epsilon_{ab} = 0 \quad .
\end{equation}
These relations have been used to prove the closure of the supersymmetry algebra on the 3-forms in section 2.

\end{document}